\documentclass[prd,aps,showpacs,preprintnumbers,amsmath,amssymb,nofootinbib,preprintnumbers,onecolumn]{revtex4}

\usepackage{amsfonts}
\usepackage{amsmath}
\usepackage{amssymb,epsf}
\usepackage{eurosym}
\usepackage{amsfonts}
\usepackage{amsmath}
\usepackage{amssymb,epsf}
\usepackage{color}
\usepackage{graphicx}
\usepackage{natbib}
\usepackage{float}
\usepackage{caption}
\usepackage{subfig}
\usepackage{epstopdf}

\begin{document}

\title{Rotating Lifshitz-like black holes in F(R) gravity}
\author{
        Kh. Jafarzade$^{1,2,3}$\footnote{email address: khadije.jafarzade@gmail.com},
        E. Rezaei$^{1,2}$\footnote{email address: rezaielham02@gmail.com } and
        S. H. Hendi$^{1,2,4}$\footnote{email address: hendi@shirazu.ac.ir}}
\affiliation{
        $^1$Department of Physics, School of Science, Shiraz University, Shiraz 71454, Iran \\
        $^2$Biruni Observatory, School of Science, Shiraz University, Shiraz 71454, Iran \\
        $^3$ICRANet-Mazandaran, University of Mazandaran, P. O. Box 47415-416, Babolsar, Iran\\
        $^4$Canadian Quantum Research Center 204-3002 32 Ave Vernon, BC V1T 2L7 Canada}

\begin{abstract}
One of the alternative theories of gravitation with a possible UV
completion of general relativity is Horava-Lifshitz gravity.
Regarding a particular class of pure $F(R)$ gravity in three
dimensions, we obtain an analytical rotating Lifshitz-like black
hole solution. We first investigate some geometrical properties of
the obtained solution that reduces to a charged rotating BTZ black
hole in a special limit. Then, we study the optical features of
such a black hole like the photon orbit and the energy emission
rate and discuss how electric charge, angular momentum, and
exponents affect them. In order to have an acceptable optical
behavior, we should apply some constraints on the exponents. We
continue our investigation with the study of the thermodynamic
behavior of the solutions in the extended phase space and examine
the validity of the first law of thermodynamics besides local
thermal stability by using the heat capacity. Evaluating the
existence of van der Waals-like phase transition, we obtain
critical quantities and show how they change under the variation
of black hole parameters. Finally, we construct a holographic heat
engine of such a black hole and obtain its efficiency in a cycle.
By comparing the obtained efficiency with the Carnot one, we
examine the second law of thermodynamics.
\end{abstract}

\maketitle

\section{Introduction}

$F(R)$ theory of gravity is one of the straightforward
generalization of Einstein's theory of general relativity (GR),
where the Ricci scalar of GR Lagrangian is replaced with an
arbitrary function of $R$ \cite{Akbar:1a,deSouza:1a,Cognola:1a}.
Unlike Einstein's gravity, $F(R)$ theory can explain the
accelerated expansion as well as structure formation of the
Universe without considering dark sectors
\cite{Perlmutter:1a,Riess:1a,Riess:1b}. Other motivations to
consider $F(R)$ gravity are including i) this theory seems to be
the only one that can avoid the long-known and fatal Ostrogradski
instability \cite{Woodard:1b}. ii) $F(R)$ theories of gravitation
can be compatible with Newtonian and post-Newtonian approximations
\cite{Capozziello:1a,Capozziello:1b}. iii) some viable $F(R)$
models have no ghosts ( $dF/dR>0 $), and the stability condition
$d^{2}F/dR^{2}\geq 0 $ of essentially amounts to guarantee that
the scalaron is not a tachyon \cite{Dolgov:1b,Faraon:1c}. iv)
although $F(R)$ theory is the simplest modification of the
gravitational interaction to higher-order known so far, its action
is sufficiently general to encapsulate some of the basic
characteristics of higher-order gravity.

From the geometrical point of view, Einstein's gravity cannot
explain the non-relativistic scale-invariant theory. To describe a
non-relativistic scale-invariant system that enjoys Galilean
symmetry, one can employ Horava-Lifshitz
\cite{Horava:1a,Horava:1b} approach. In this approach, an
anisotropic scale invariant between time and space directions is
considered, i.e. $(t,x) \rightarrow (\lambda^{z}t,\lambda x)$,
where the degree of anisotropy is measured by the dynamical
exponent $z$. Theories with $z\neq 1$ are invariant under
non-relativistic transformations \cite{Kachru:1ab} while for
$z=1$, the theory reduces to the relativistic isotropic scale
invariance model corresponding to the AdS spacetime. Systems with
such a Lifshitz scaling appear frequently in quantum/statistical
field theory of condensed matter physics and ultra-cold atomic
gases (see \cite{Cardy:1ab} for more details). Motivated by what
was mentioned above, here, we are going to consider a
Lifshitz-like geometry in a class of three dimensional $F(R)$
gravity model.

The study of three dimensional black holes known as BTZ
(Banados-Teitelboim-Zanelli) solutions \cite{Banados:1ab} has
opened different aspects of physics in low dimensional spacetimes.
The geometry of $(2+1)-$dimensional manifold has various
interesting properties such as the existence of specific relations
between the BTZ black holes and an effective action in string
theory \cite{Lee:1ab,Larranaga:1ac}, developing our
understanding of gravitational interaction in low dimensional manifolds \cite%
{Witten:1abc}, improvement in the quantum theory of gravity and
gauge field theory \cite{Witten:1abd,Carlip:1abx}, the possibility
of the existence of
gravitational Aharonov Bohm effect in the noncommutative spacetime \cite%
{Anacleto:1mn}, and so on. Therefore, the study of three-dimensional black
holes has attracted physicists not only in the context of Einstein's
gravity, but also in modified theories such as massive gravity \cite%
{Hendi:1az}, dilaton gravity \cite{Chan:1az}, gravity's rainbow \cite%
{Hendi:1ax} and also massive gravity's rainbow \cite{Hendi:1abx}.
Besides, the static vacuum solutions of a Lifshitz model in
$(2+1)-$dimensions have been investigated in \cite{Shu:1abv}. In
addition, three dimensional Lifshitz-like charged black hole
solutions in $F(R)$ gravity have been also studied in
\cite{Hendi:1abs}. In this paper, we introduce a new Lifshitz-like
charged rotating black hole solution in three-dimensional $F(R)$
gravity and investigate its optical and thermodynamical
properties.

From the theoretical viewpoint, one of the challenging subjects of
black hole physics is the thermodynamic behavior of a typical
black hole. The possible identification of a black hole as a
thermodynamic object was first realized by Bardeen, Carter, and
Hawking \cite{Bardeen:1973}. They clarified the laws of black hole
mechanics and showed that these laws are corresponding to ordinary
thermodynamics. Thereafter, various thermodynamic properties of
black holes have been widely studied, and we now have a
considerable understanding of the microscopic origin of these
properties due to a pioneering work by Strominger and Vafa
\cite{Strominger:1996}. The investigation of black hole
thermodynamics in anti-de Sitter (AdS) spacetime provided us with
a deep insight into understanding the quantum nature of
gravity which has been one of the open problems of physical communities \cite%
{Zeng:2020,Caldarelli:2000,Hertog:2005,Lu:2015}. Besides, the
existence of the cosmological constant can change both the
geometry and thermodynamic properties of spacetime. Notably, the
study in the context of black hole thermodynamics showed that the
correspondence between black hole mechanics and ordinary
thermodynamic systems is completed by considering the cosmological
constant as a variable parameter \cite{Kubiznak:2012}. To complete
the thermodynamic properties of a system, it is inevitable to
investigate the existence of phase transition and thermal
stability.  The investigation of phase transition has a crucial
role in exploring the critical behavior of a system near its
critical point. The black hole phase transition was first studied
by Hawking and Page who demonstrated the existence of a certain
phase transition (so-called Hawking-Page) between thermal AdS and
Schwarzschild-AdS BH which corresponds to the
confinement/deconfinement phase transition in the dual strongly
coupled gauge theory \cite{Hawking:1983}. The discovery of the
first-order phase transition in the charged AdS black hole
spacetime has gained a lot of attention in recent years
\cite{Chamblin:1999a,Chamblin:1999b}. This transition displays a
classical critical behavior and is superficially analogous to a
van der Waals liquid-gas phase transition. Especially, considering
the cosmological constant as a thermodynamical variable and
working in the extended phase space led to finding the additional
analogy between the black holes and the behavior of the van der
Waals liquid/gas system
\cite{Gibbons:1996,Breton:2005,Kastor:2009,Hendi:2013}. In this
regard, some efforts have been made in the context of $P-V$
criticality of black holes in modified theories of gravitation,
such as Horava-Lifshitz
gravity \cite{Haldar:2018,Mo:2015,Jafarzade:2017}, Gauss-Bonnet gravity \cite%
{Cai:2013,Mo:2014,Miao:2018}, Lovelock gravity
\cite{Mo:2014a,Hendi:2015},
dilaton gravity \cite{Zhao:2013,Dehghani:2014}, $F(R)$ gravity \cite%
{Wu:2016,Ovgun:2018}, massive gravity \cite%
{Mirza:2014,Hendi:2016xz,Chabab:2019}, gravity's rainbow \cite%
{Hendi:2016nm,Feng:2017}, and massive gravity's rainbow
\cite{Hendi:2017fs}. In addition, from the thermodynamics point of
view, one may consider a black hole as a heat engine in the
extended phase space. Indeed, the mechanical term PdV in the first
law provides the possibility of extracting the mechanical work and
consequently calculating the efficiency of a typical heat engine.
The concept of the holographic heat engine was first proposed by
Johnson in Ref.\cite{CVJohnson}. He creatively employed the
charged AdS black hole as a heat engine working substance to
construct a holographic heat engine and calculated the heat engine
efficiency. Afterward, holographic heat engines were investigated
in different black hole backgrounds, such as the rotating BHs
\cite{RAHennigar,Jafarzade1}, Horava-Lifshitz BHs
\cite{Jafarzade2}, Born-Infeld BHs \cite{CVJohnson2}, charged BTZ
BHs \cite{JXMo}, accelerating BHs \cite{Zhang1}, black holes in
massive gravity \cite{Meng} and gravity's rainbow \cite{Panah}.

This paper is organized in the following manner: In Sec. \ref{SecII}, we
consider three-dimensional Lifshitz-like background spacetime and obtain
charged rotating black hole solutions in a special class of $F(R)$ gravity.
We also determine the null geodesics equations as well as the radius of the
photon orbit, and explore the conditions to find an acceptable
optical behavior. The energy emission rate for these black holes and the
influence of the model's parameters on the emission of particles are
investigated. In Sec. \ref{SecIII}, we study the thermodynamic properties of
the corresponding black hole. With thermodynamic quantities in hand, we
study the thermal stability of these black holes in the context of the
canonical ensemble. We also investigate the critical behavior of the system
and discuss how the parameters of the black holes affect critical
quantities. The heat engine efficiency is the other interesting quantity
that we will evaluate in an independent subsection. Finally, we present our
conclusions in the last section.

\section{Geometrical properties}\label{SecII}

Our line of work in this paper includes investigating three-dimensional
rotating Lifshitz-like black holes in $F(R)$ gravity and studying their
geometrical and thermodynamical properties. In this section, we first
introduce the action of $3-$dimensional $F(R)$ gravity, and then we obtain
the metric function and confirm the existence of black holes. At the end of
this section, we investigate optical features of the black hole including
the photon orbit and energy emission rate, and
examine the effects of electric charge, angular momentum and dynamical
exponents on the optical properties.

\subsection{ Constructing the solutions}\label{SubSecIIA}

Here, we intend to construct three-dimensional rotating Lifshitz-like black
holes in $F(R)$ gravity and study their geometrical properties. To do so, we
consider the following action
\begin{equation}
S=\int_{\mathcal{M}} d^{2+1}x\sqrt{-g}F(R),  \label{Eqaction}
\end{equation}
in which $\mathcal{M}$ is a $3-$dimensional spacetime and $F(R)$ is an
arbitrary function of Ricci scalar $R$. Variation with respect to metric
tensor, $g_{\mu\nu}$, leads to the following field equation
\begin{equation}
G_{\mu\nu}F_{R}-\frac{1}{2}g_{\mu\nu}[F(R)-RF_{R}]-[\nabla_{\mu}\nabla_{%
\nu}-g_{\mu\nu}\square]F_{R}=0,  \label{field-equation}
\end{equation}
where $G_{\mu\nu}$ is the Einstein tensor and $F_{R}\equiv dF(R)/dR $. Now,
we want to obtain the rotating Lifshitz-like solutions of Eq. (\ref%
{field-equation}). For this purpose, we assume that the metric has the
following ansatz
\begin{equation}
ds^{2}=-\left( \frac{r}{r_{0}}\right) ^{z}f\left( r\right) dt^{2}+\frac{%
dr^{2}}{f\left( r\right) }+r^{2}\left( d\varphi -\frac{J\left( \frac{r}{r_{0}%
}\right) ^{w}}{2r^{2}}dt\right) ^{2},  \label{metric 3-d Lif}
\end{equation}
where $z$ and $w$ play the role of dynamical exponents and $r_{0}$ is an
arbitrary (positive) length scale. Here, we study black hole solutions with
constant Ricci scalar with the condition of $F(R_{0})=F_{R}=0 $, and
therefore, it is easy to show that the equation of motion reduces to the
following differential equation
\begin{equation}
R=R_{0}=\frac{J^{2}}{2r^{4}}\left( \frac{r}{r_{0}}\right) ^{2w-z}\left(
\frac{w^{2}}{4}-w+1\right) -\frac{z^{2}f}{2r^{2}}-\left( 2+\frac{3z}{2}
\right)\frac{f^{\prime}}{r}-f^{\prime\prime},  \label{Ricci scalar}
\end{equation}
with the following metric function as the solution
\begin{equation}
f\left( r\right) =-\Lambda r^{2}-\frac{m}{r^{\gamma }}+\frac{2^{-\frac{1}{4}%
}q^{\frac{3}{2}}}{r^{\delta }}+\frac{\left( w-2\right) ^{2}\left( \frac{r}{%
r_{0}}\right) ^{\left( 2w-z\right) }J^{2}}{8r^{2}\left[ 4w^{2}-(z+6)w +2%
\right] },  \label{metric function F(R)}
\end{equation}%
where $m$ and $q$ are two integration constants related to the total mass
and electric charge of the black hole, respectively. It is worth mentioning
that these two integration constants are set in such a way that for the case
of $w=z=0$, the solution reduces to the rotating BTZ black hole solution in
the presence of a special model of the Power-Maxwell field. Here, $\Lambda$
is a (positive/negative or zero) constant which depends on the sign/value of
$R_{0}$ as
\begin{equation}
\Lambda =\frac{2R_{0}}{z^{2}+6z+12}.  \label{cos cons}
\end{equation}

Besides, $\gamma$ and $\delta$ are defined as
\begin{eqnarray}
\gamma &=&\frac{1}{4}\left( 3z+2-\sqrt{z^{2}+12z+4}\right) ,
\label{gamma delta} \\
\delta &=&\frac{1}{4}\left( 3z+2+\sqrt{z^{2}+12z+4}\right) .
\nonumber
\end{eqnarray}

\begin{figure}[!htb]
\centering
\subfloat[$ z=w=0$]{
        \includegraphics[width=0.31\textwidth]{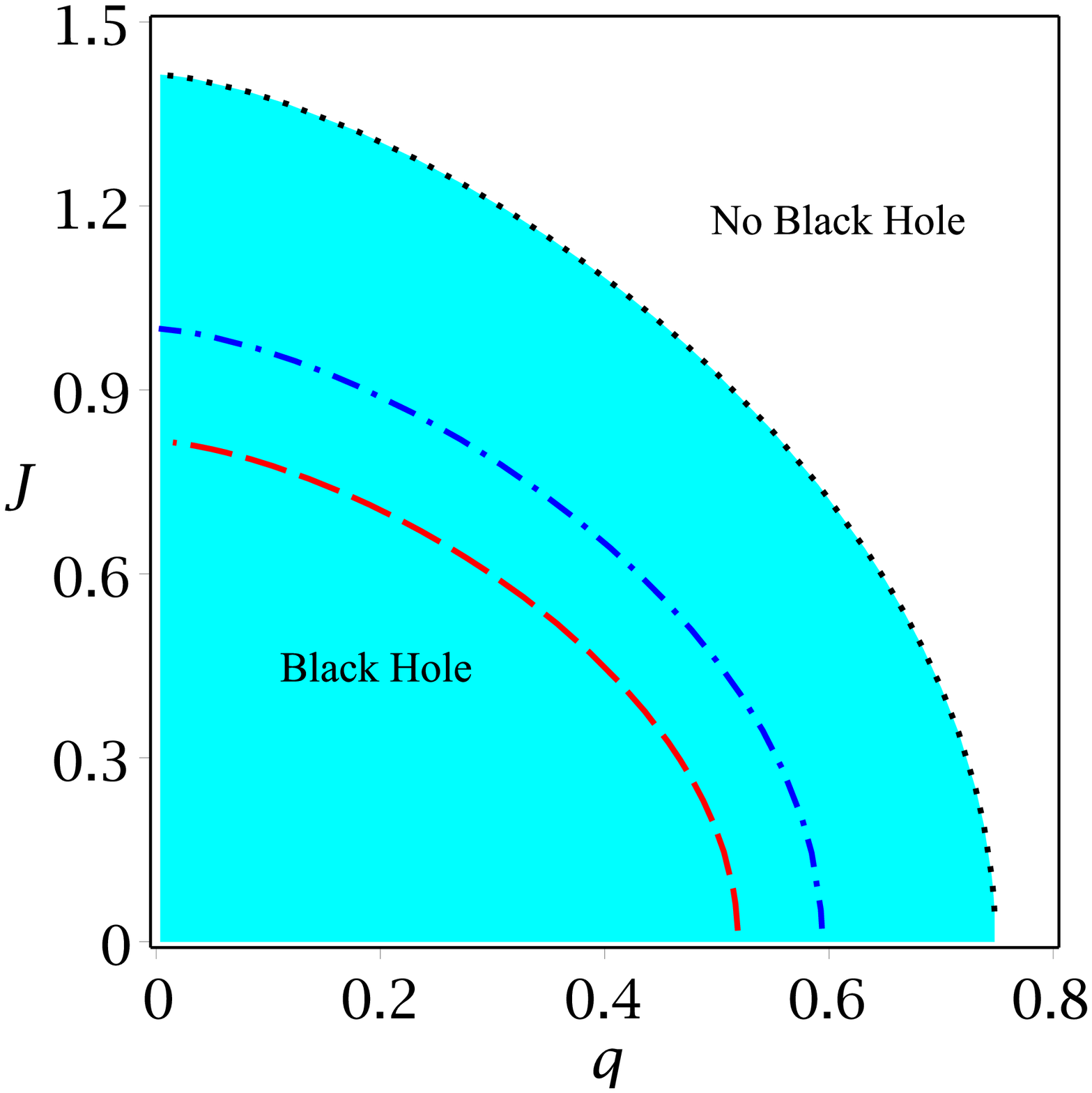}}
\subfloat[$ z=0$ and $w=3$]{
        \includegraphics[width=0.31\textwidth]{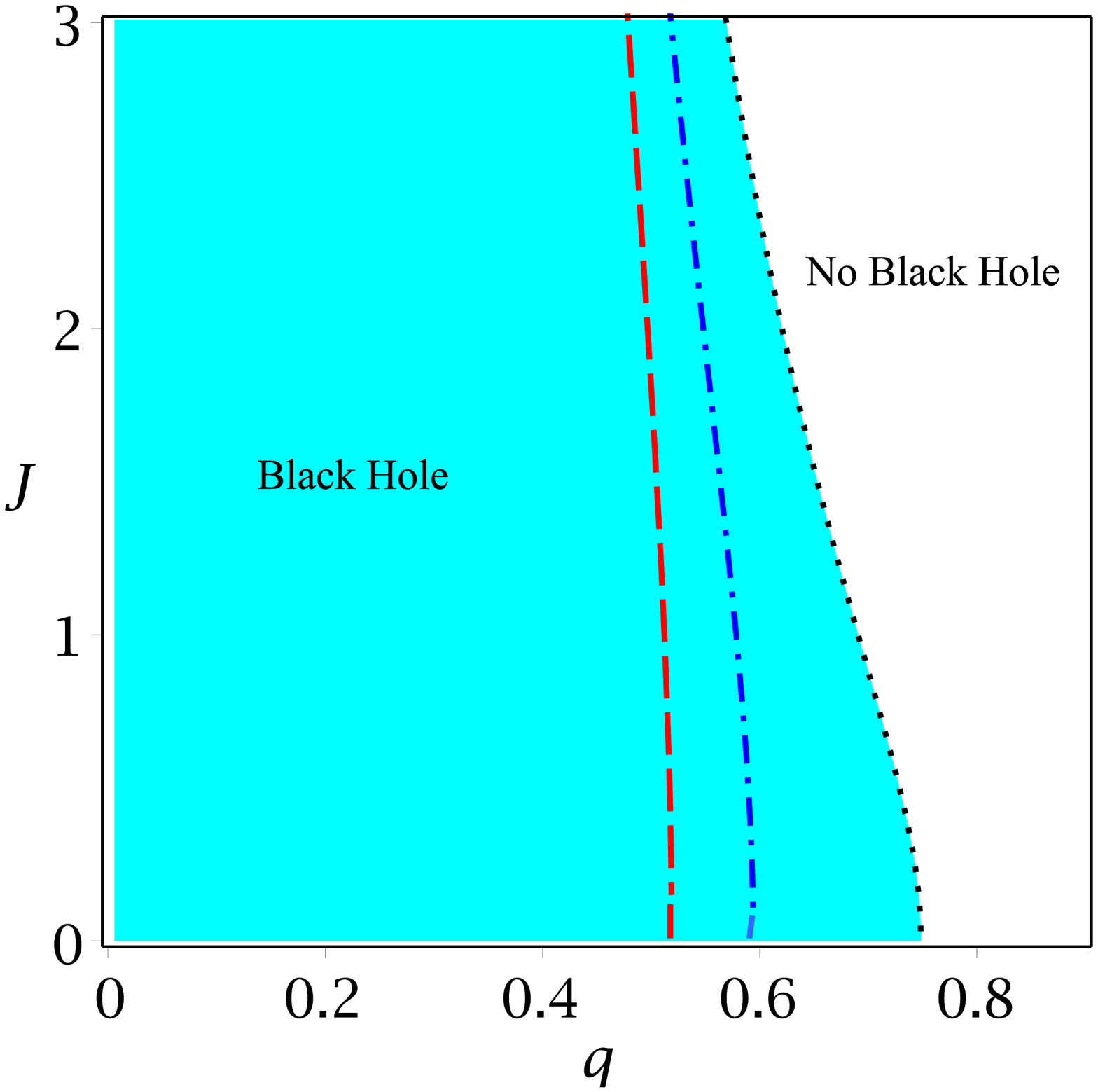}}
\newline
\subfloat[$ z=1$ and $w=0$]{
        \includegraphics[width=0.315\textwidth]{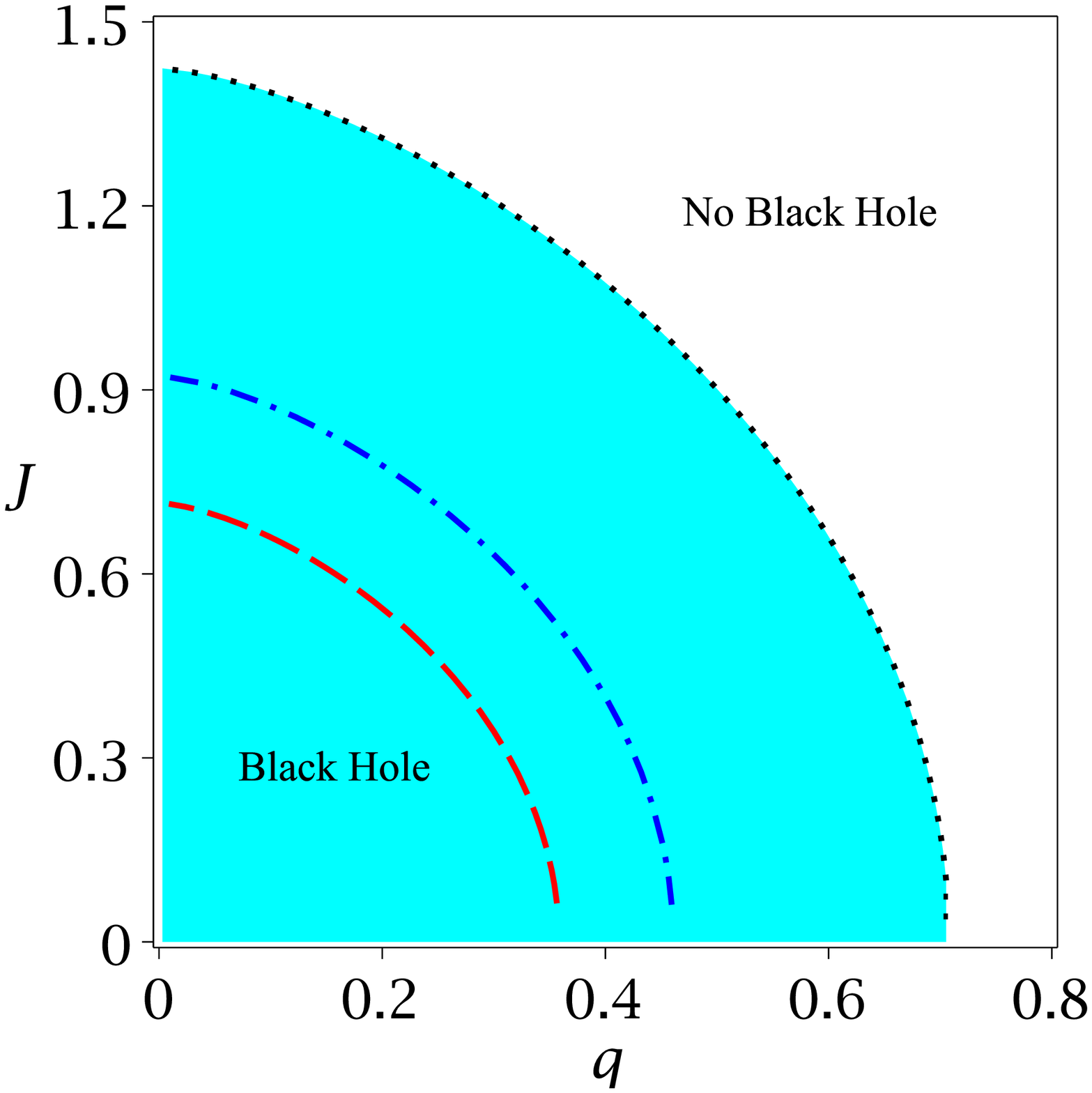}}
\subfloat[$ z=3$ and $w=0$]{
        \includegraphics[width=0.32\textwidth]{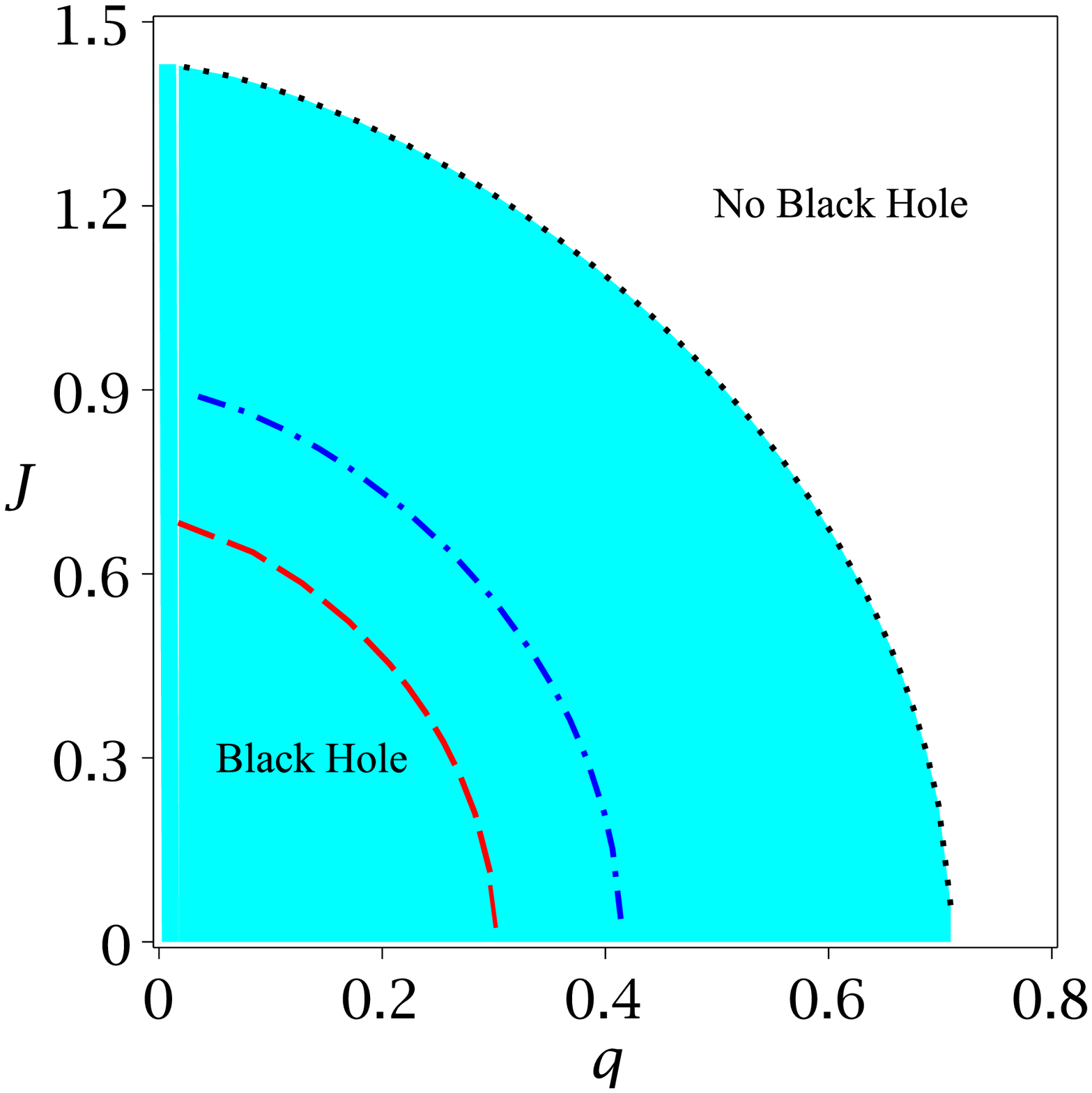}}
\newline
\caption{ The admissible parameter space to have a physical solution for $%
m=1 $, $r_{0}=1 $, $\Lambda=-0.5 $ (dotted line), $\Lambda=-1$ (dash-dotted
line) and $\Lambda=-1.5 $ (dashed line).}
\label{Fig1}
\end{figure}

\subsection{Singularity and Event horizon}\label{SubSecIIB}

With the exact solution in hand, we examine whether the obtained solution
could be interpreted as a black hole. The interpretation of solution as a
(singular) black hole has two criteria: I) Presence of singularity, II)
Existence of an event horizon covering the singularity. In order to look for
the singularity, we use the Kretschmann scalar as%
\begin{eqnarray}
K &=&f^{\prime \prime ^{2}}+\left( \frac{3zf^{\prime }}{r}+\frac{z\left(
z-2\right) f}{r^{2}}-\frac{3J^{2}\left( \frac{r}{r_{0}}\right) ^{2w-z}\left(
w-2\right) ^{2}}{4r^{2}}\right) f^{\prime \prime }+  \nonumber \\
&&  \nonumber \\
&&\frac{(\frac{9z^{2}}{4}+2)}{r^{2}}f^{\prime ^{2}}+\left( \frac{3\left(
z^{2}-2z+\frac{4}{3}\right) zf}{2r^{3}}-\frac{J^{2}\left( \frac{r}{r_{0}}%
\right) ^{2w-z}\left( w-2\right) ^{2}\left( \frac{9z}{4}-1\right) }{2r^{5}}%
\right) f^{\prime }+  \nonumber \\
&&  \nonumber \\
&&\frac{z^{2}\left( \frac{z^{2}}{4}-z+2\right) }{r^{4}}f^{2}-\frac{%
J^{2}\left( \frac{r}{r_{0}}\right) ^{2w-z}\left( w-2\right) ^{2}\left[
z^{2}-z\left( w+2\right) +w^{2}\right] }{2r^{6}}f+\frac{11J^{2}\left( \frac{r%
}{r_{0}}\right) ^{4w-2z}\left( w-2\right) ^{4}}{64r^{8}}.
\label{Kretschmann}
\end{eqnarray}

Inserting the metric function $f(r)$ into Eq. (\ref{Kretschmann}),
one finds
\begin{eqnarray*}
K&=&\Upsilon_{1}r^{2(w-z-4)}+(\Upsilon_{2} r^{\gamma}+
\Upsilon_{3} r^{\gamma+\delta +2}+ \Upsilon_{4}r^{\delta})r^{2w-2.5z-7}\nonumber \\
&&  \nonumber \\
&&+\left[ (\Upsilon_{5}r^{\gamma +2}+\Upsilon_{6})r^{\gamma+\delta}+(\Upsilon_{7}
r^{2\gamma +4}+\Upsilon_{8}r^{\gamma +2}+\Upsilon_{9})r^{2\delta}+\Upsilon_{10}r^{2\gamma}\right] r^{-3(z+2)},
\end{eqnarray*}
where $\Upsilon_{i}$'s are functions of $z$, $w$, $J$, $r_{0}$,
$\Lambda $, $m$ and $q$. According to our analysis, the scalar
curvature diverges in the limit of $r\rightarrow 0 $ which
confirms that there is a curvature singularity at $r=0$.

\begin{figure}[!htb]
\centering
\subfloat[$ z=1$ and $w=0$]{
        \includegraphics[width=0.31\textwidth]{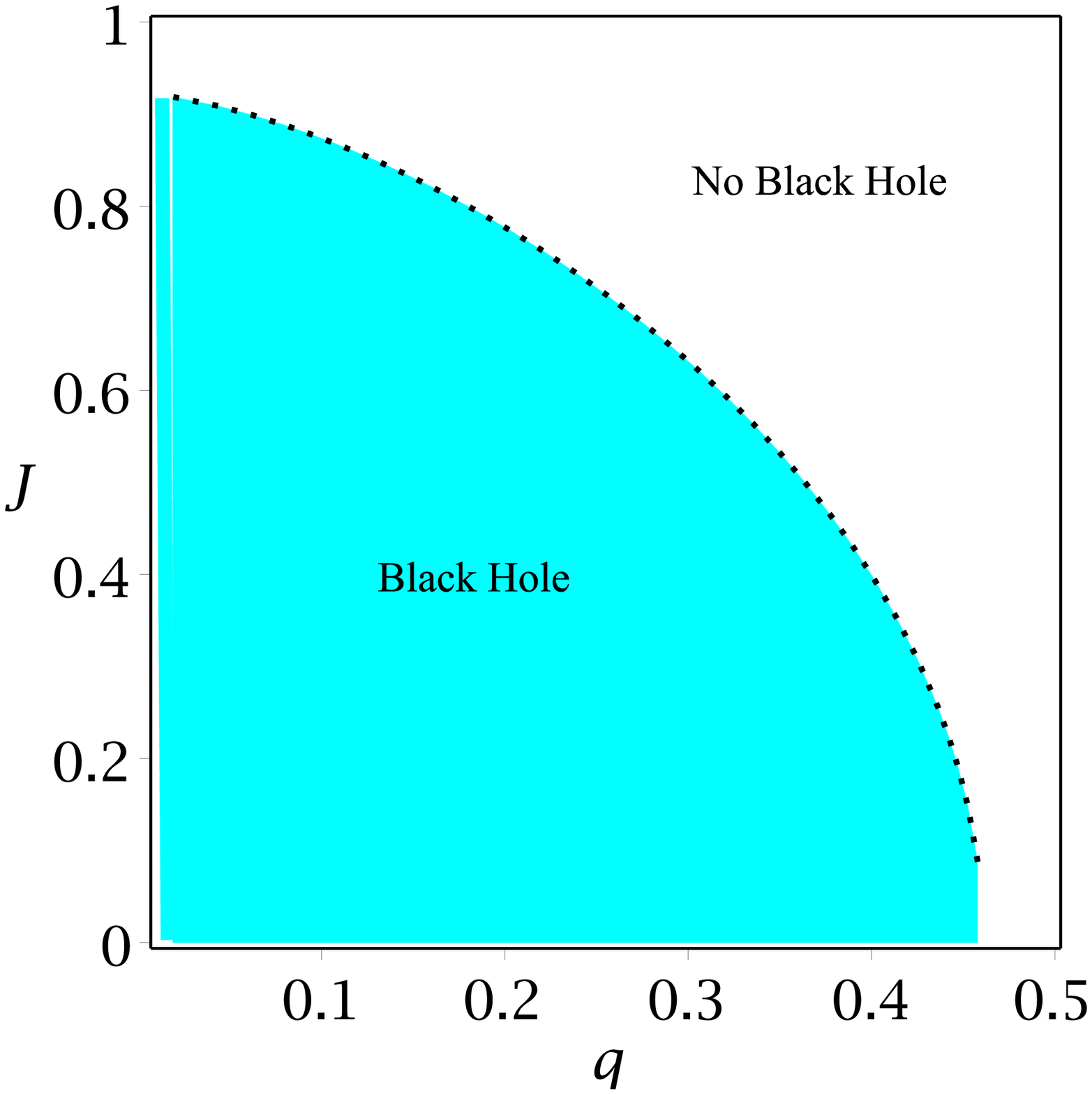}}
\subfloat[$ z=1$ and $w=1$]{
        \includegraphics[width=0.31\textwidth]{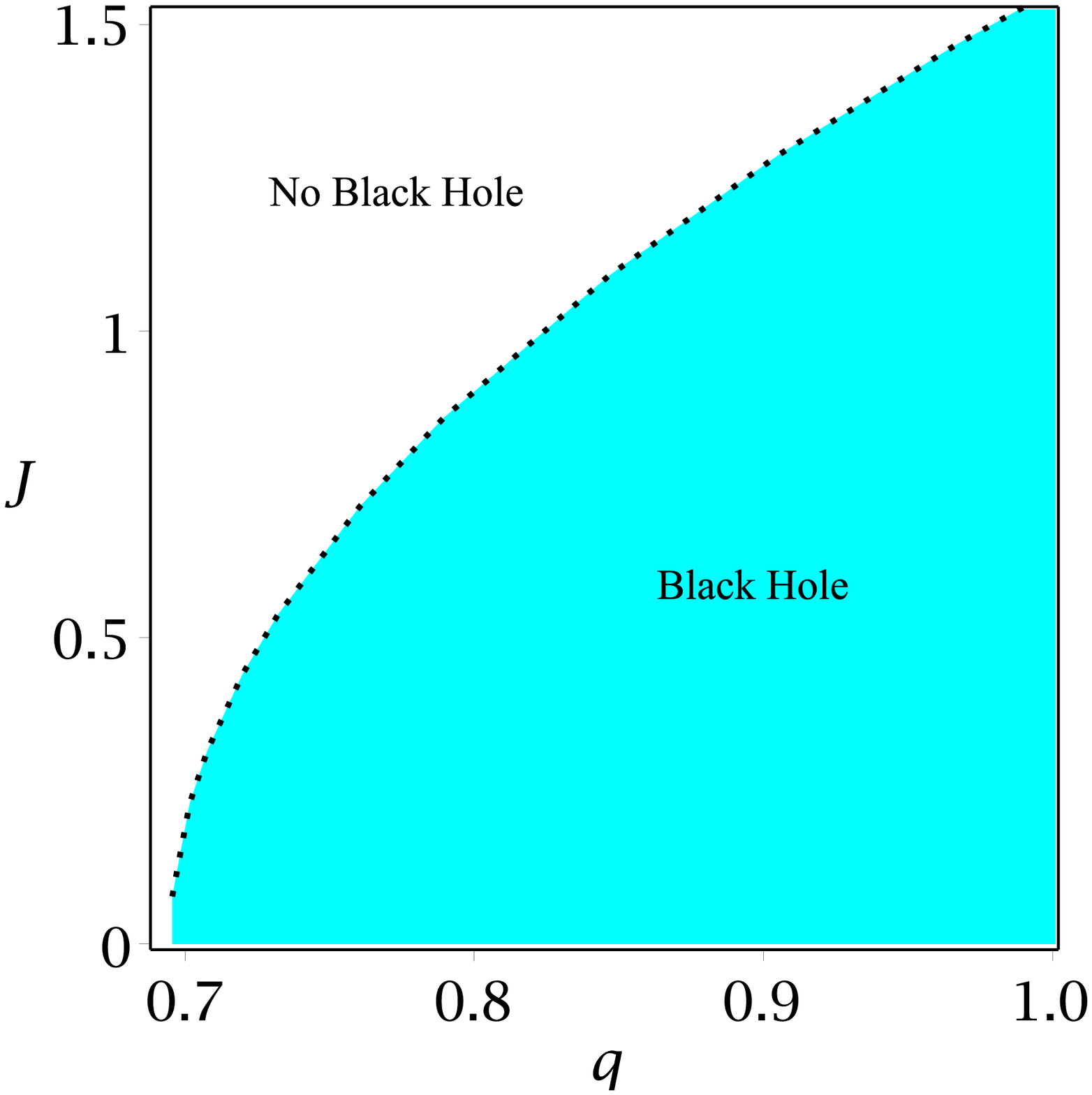}}
\subfloat[$ z=1$ and $w=3$]{
        \includegraphics[width=0.31\textwidth]{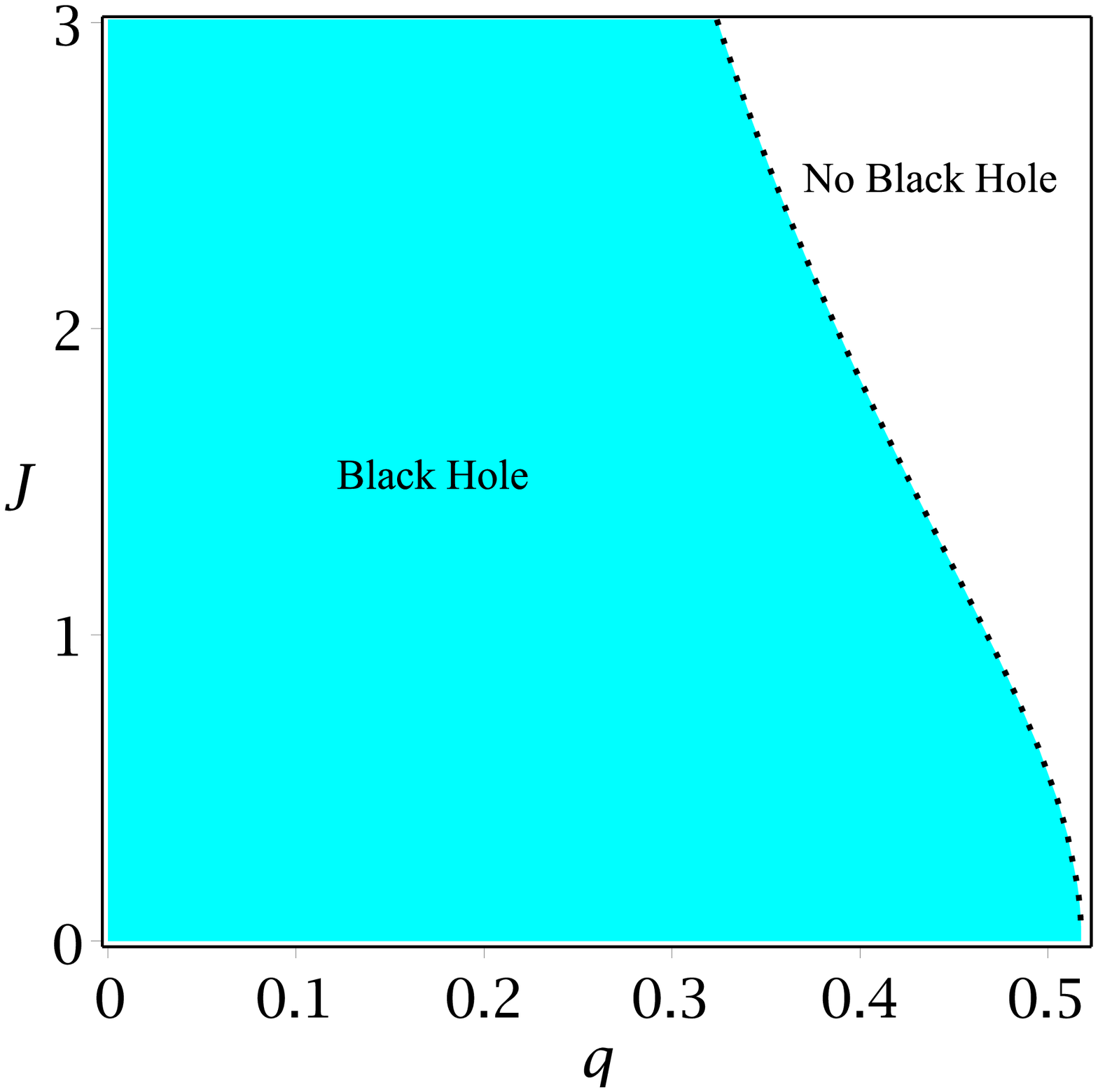}}\newline
\subfloat[$ z=2$ and $w=0$]{
        \includegraphics[width=0.31\textwidth]{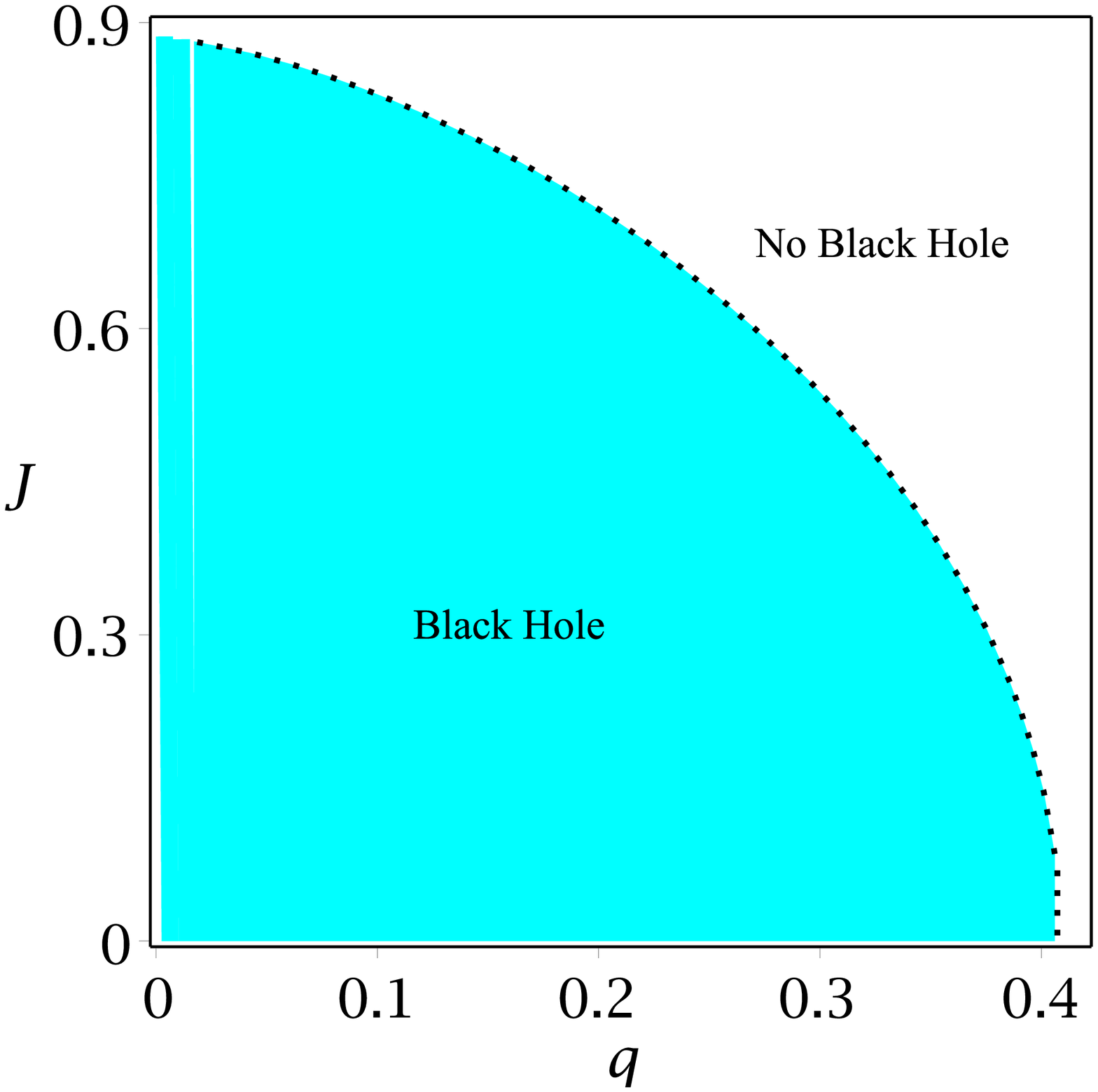}}
\subfloat[$ z=2$ and $w=1$]{
        \includegraphics[width=0.31\textwidth]{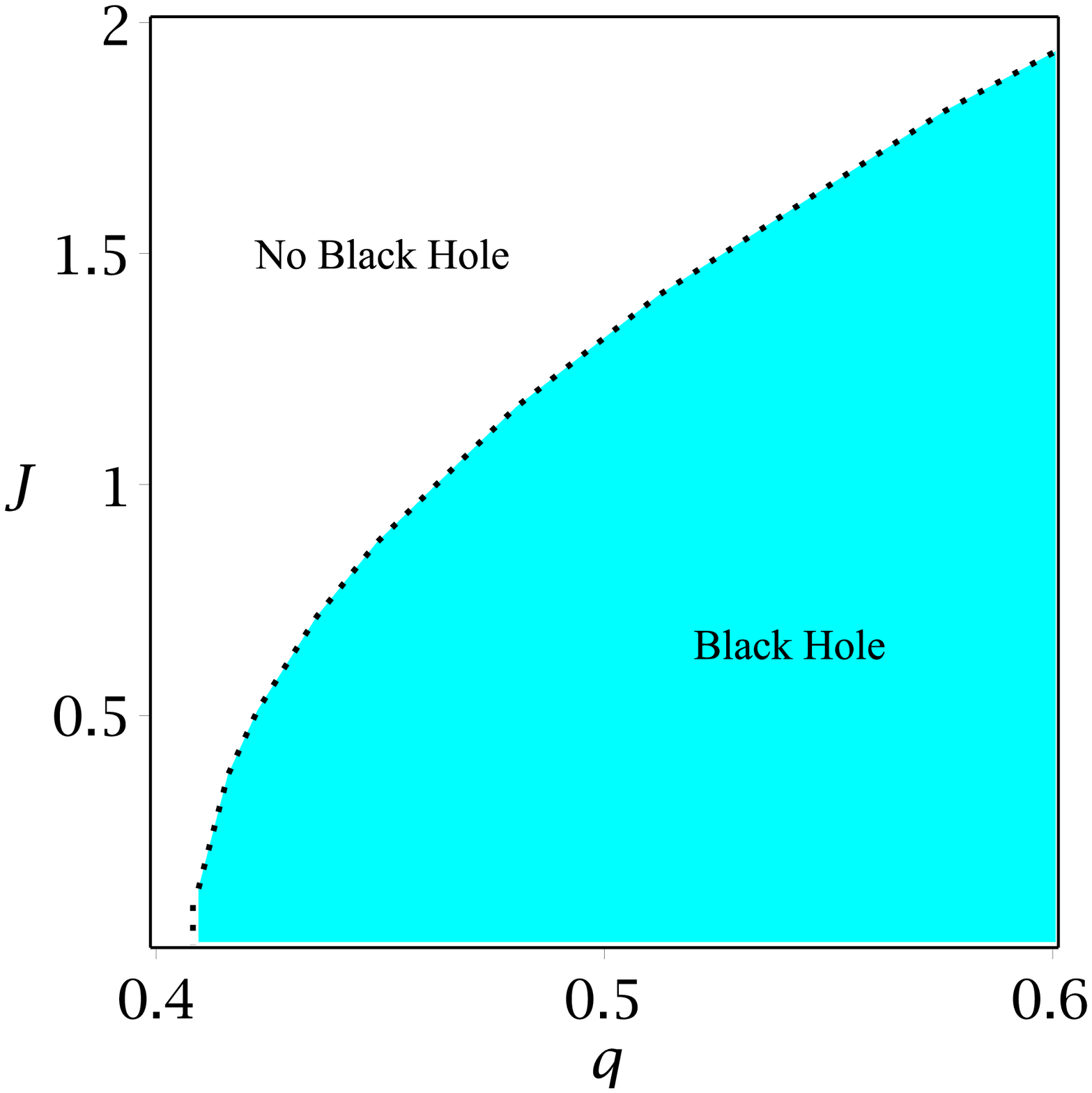}}
\subfloat[$ z=2$ and $w=3$]{
        \includegraphics[width=0.31\textwidth]{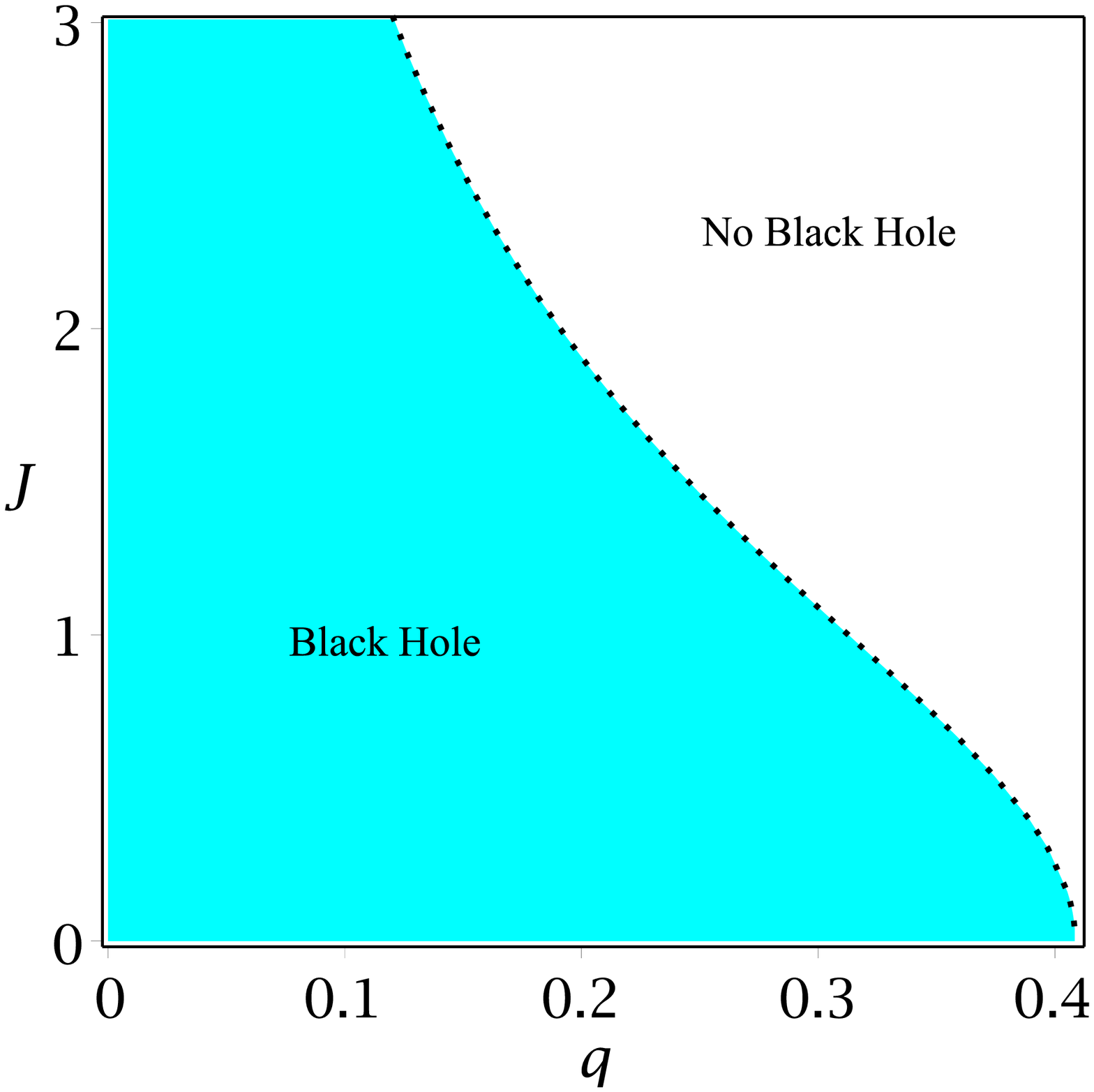}}\newline
\caption{ The admissible parameter space to have a physical solution for $%
m=1 $, $r_{0}=1 $ and $\Lambda=-1 $.}
\label{Fig2}
\end{figure}

Now, we require to investigate the second condition (existence of
a horizon(s)). Strictly speaking, roots of the metric function,
$g^{rr}=f(r)$, are where black hole's horizons are. The absence of
a root for the metric function indicates that the solution is not
a black hole but a naked singularity. Due to the fact that the
metric function goes to $+\infty$ for spatial infinity and also
near the origin, one can find that such a function has a minimum
($r_{min}$). Depending the sign of $f(r_{min})$, one may find a
black hole with two horizons ($f(r_{min})<0$), an extreme black
hole ($f(r_{min})=0$) or naked singularity ($f(r_{min})>0$). We
examine the condition of extreme BH by studying the following
criteria
\begin{equation}
f(r_{min})=0=f^{\prime}(r_{min}).  \label{Root}
\end{equation}

Equation (\ref{Root}) shows that the metric function has one
degenerate horizon at $r_{min} $ which corresponds to the radius
of extremal black holes (the coincidence of the inner and outer BH
horizons). Solving these two equations, simultaneously, leads to

\begin{eqnarray}
q &=& \left( \frac{\left[ (2w-z-4)\Lambda r_{min}^{\gamma
+2}+m(\gamma +2w-z-2)\right]
r_{min}^{\delta}2^{\frac{1}{4}}}{r_{min}^{\gamma}\left(
2w-z-2+\delta \right)}\right)^{\frac{2}{3}},  \label{q-j} \\
J &=& \left( \frac{8\left(m(\gamma - \delta) -\Lambda (\delta
+2)r_{min}^{\gamma +2}\right)\left(w(z+6)-2-4w^{2} \right)
}{r_{min}^{\gamma
-2}\left( 2w-z-2+\delta \right)\left(w-2 \right)^{2} \left( \frac{r_{min}}{%
r_{0}}\right)^{2w-z} }\right) ^{\frac{1}{2}}.  \nonumber
\end{eqnarray}

The resultant curve provides a lower bound for the existence of
the black hole, and is depicted in Fig. \ref{Fig1} see dotted,
dash-dotted and dashed lines and Fig. \ref{Fig2} by the dotted
line, denoting the extremal limit. Below this line, black holes
(with two horizons) are present, whereas no black hole exists
above it. As it was observed, the second condition for the
existence of roots (horizons) for the metric function can be
satisfied. In other words, the curvature singularity can be
covered by an event horizon. So, the obtained solution can be
interpreted as a black hole solution.

A significant point regarding these figures is that the
admissible parameter space is highly affected by values of the
exponent $w$. For case of $z=0$, there is no physical solution for
values of $w=1 $ and $w=2$. For case of $z\neq 0 $, a physical
solution cannot be observed for small values of the electric
charge for $w=1$ (see the middle panels of Fig. \ref{Fig2}). For
the same $w$, the admissible parameter space decreases with
increase of $z$ (compare left panels of these two figures with
each other). Also, from Fig. \ref{Fig1}, one can find that the
cosmological constant has a decreasing effect on admissible
parameter space.

\subsection{Optical features}\label{SubSecIIC}

Here, we are going to investigate another geometric property of
the black hole, photon orbit. To do so, we investigate the photon orbit radius
of the black hole and explore the effect of the exponents $z$ and
$w$ and parameters $q$ and $J$ on the radii size.
At the first step, we employ the Hamilton-Jacobi equation for null
curves as \cite{Carter1a}
\begin{equation}
\frac{\partial S}{\partial \lambda}=-\frac{1}{2}g^{\mu\nu}\frac{\partial S}{%
\partial x^{\mu}}\frac{\partial S}{\partial x^{\nu}},  \label{HJ1}
\end{equation}
where $S $ and $\lambda$ denote, respectively, the Jacobi action
of the photon and the affine parameter of the null geodesic. Using
known constants of the motion, one can separate the Jacobi
function as follows
\begin{equation}
S=-Et+L\phi+S_r(r),  \label{HJ2}
\end{equation}
where $E$ and $L$ are, respectively, the energy and angular
momentum of the photon in the direction of rotation axis. By
inserting the Jacobi action (\ref{HJ2}) into the Hamilton-Jacobi
equation (\ref{HJ1}), and using also the metric components, we
acquire
\begin{equation}
r^{2}f(r)\left(\frac{r}{r_{0}} \right) ^{z}\left(\frac{d S_{r}}{dr} \right)
^{2}-\frac{L^{2}J^{2}}{4r^{2}f(r)}\left(\frac{r}{r_{0}} \right)
^{2w}+L^{2}\left(\frac{r}{r_{0}} \right) ^{z}-\frac{E^{2}r^{2}}{f(r)}-\frac{%
E L J}{f(r)}\left(\frac{r}{r_{0}} \right) ^{w}=0.  \label{HJ3}
\end{equation}
Considering $S_{r}^{\prime}(r)=\frac{\sqrt{\mathcal{R}(r)}}{f(r)}$
and inserting it  into Eq. (\ref{HJ3}), one finds
\begin{equation}
\mathcal{R}(r)=\frac{f(r)}{r^{2}\left(\frac{r}{r_{0}} \right) ^{z}}\left( \frac{%
L^{2}J^{2}}{4r^{2}f(r)}\left(\frac{r}{r_{0}} \right) ^{2w}-L^{2}\left(\frac{r%
}{r_{0}} \right) ^{z}+\frac{E^{2}r^{2}}{f(r)}+\frac{E L J}{f(r)}\left(\frac{r%
}{r_{0}} \right) ^{w}\right) .  \label{R(r)}
\end{equation}

Thus, the photon propagation obeys the following three equations
of motion, obtained from the variation of the Jacobi action with
respect to the affine parameter $\lambda$
\begin{align}  \label{HJ6}
&\frac{dt}{d\lambda}=\frac{E}{f(r)\left(\frac{r}{r_{0}} \right) ^{z}}-\frac{%
LJ}{2r^{2}f(r)}\left(\frac{r}{r_{0}} \right) ^{w-z} \\
&\frac{dr}{d\lambda}=\sqrt{\mathcal{R}(r)}, \\
&\frac{d\varphi}{d\lambda}=\frac{L}{r^{2}}-\frac{LJ^{2}}{4r^{4}f(r)}\left(%
\frac{r}{r_{0}} \right) ^{2w-z}-\frac{EJ}{2r^{2}f(r)}\left(\frac{r}{r_{0}}
\right) ^{w-z}.
\end{align}
\begin{figure}[!htb]
\centering
\subfloat[$ z=w=0 $]{
        \includegraphics[width=0.31\textwidth]{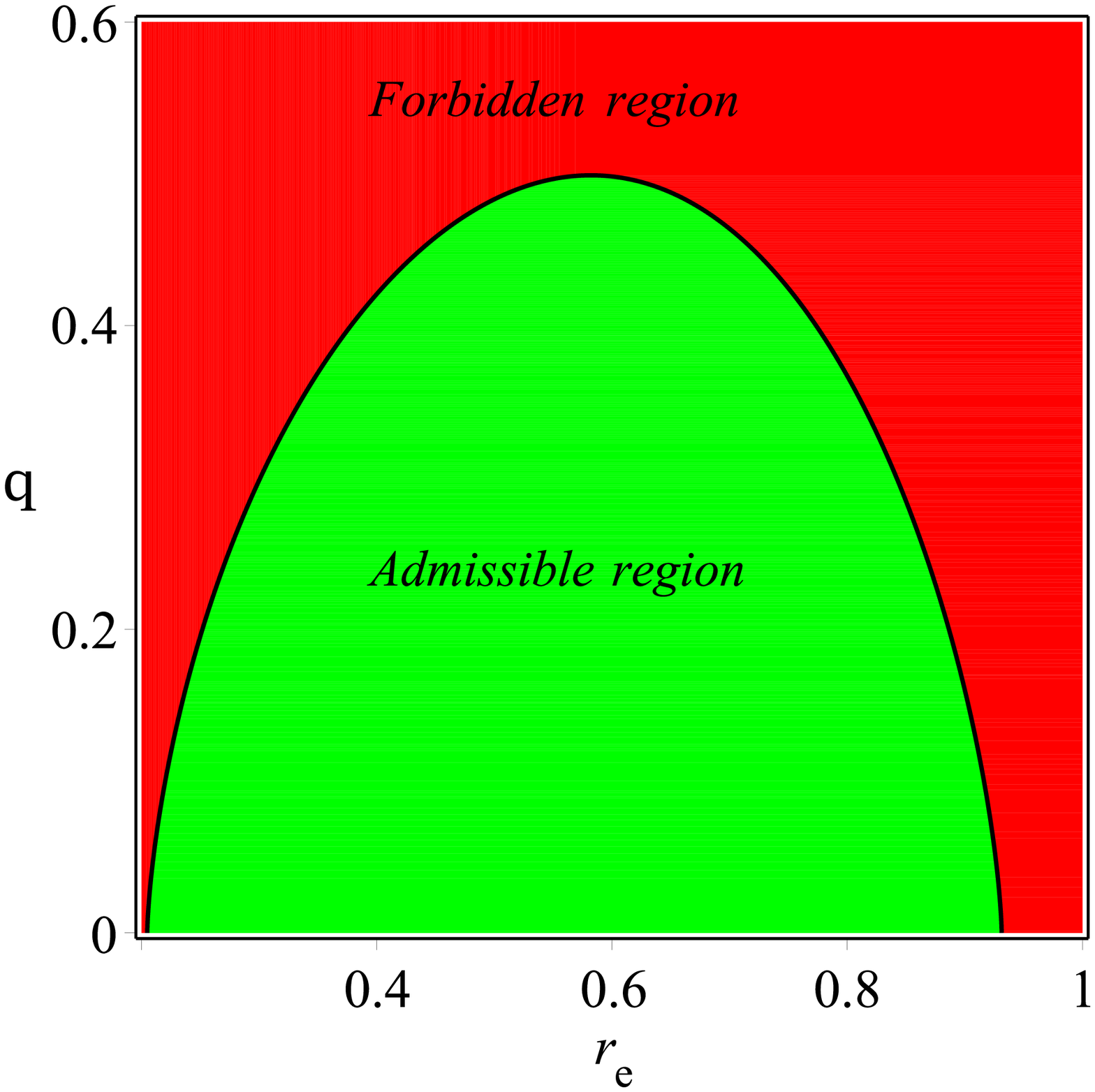}}
\subfloat[$ z=w=0 $]{
        \includegraphics[width=0.31\textwidth]{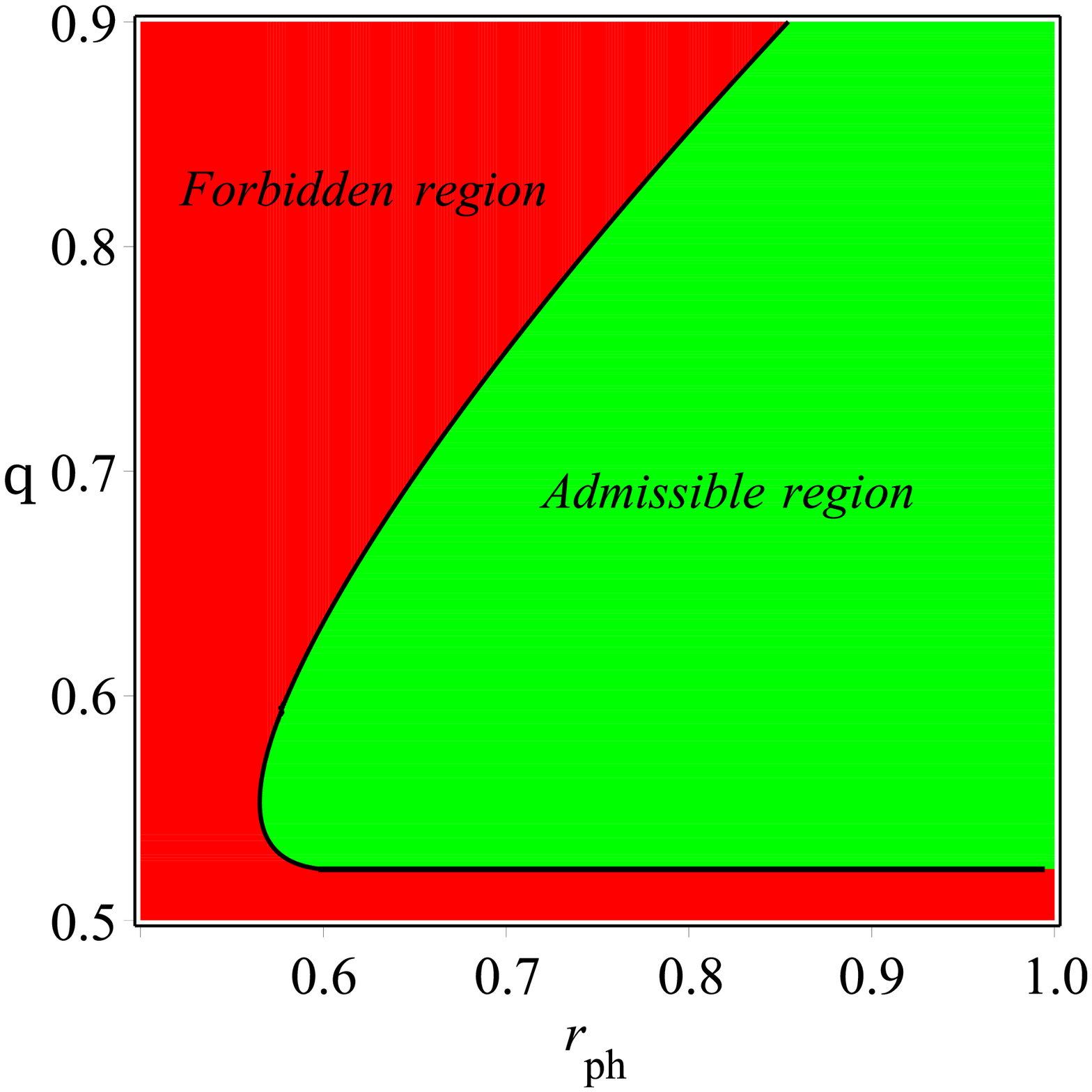}}\newline
\subfloat[$ z=0 $ and $ w=2 $]{
        \includegraphics[width=0.31\textwidth]{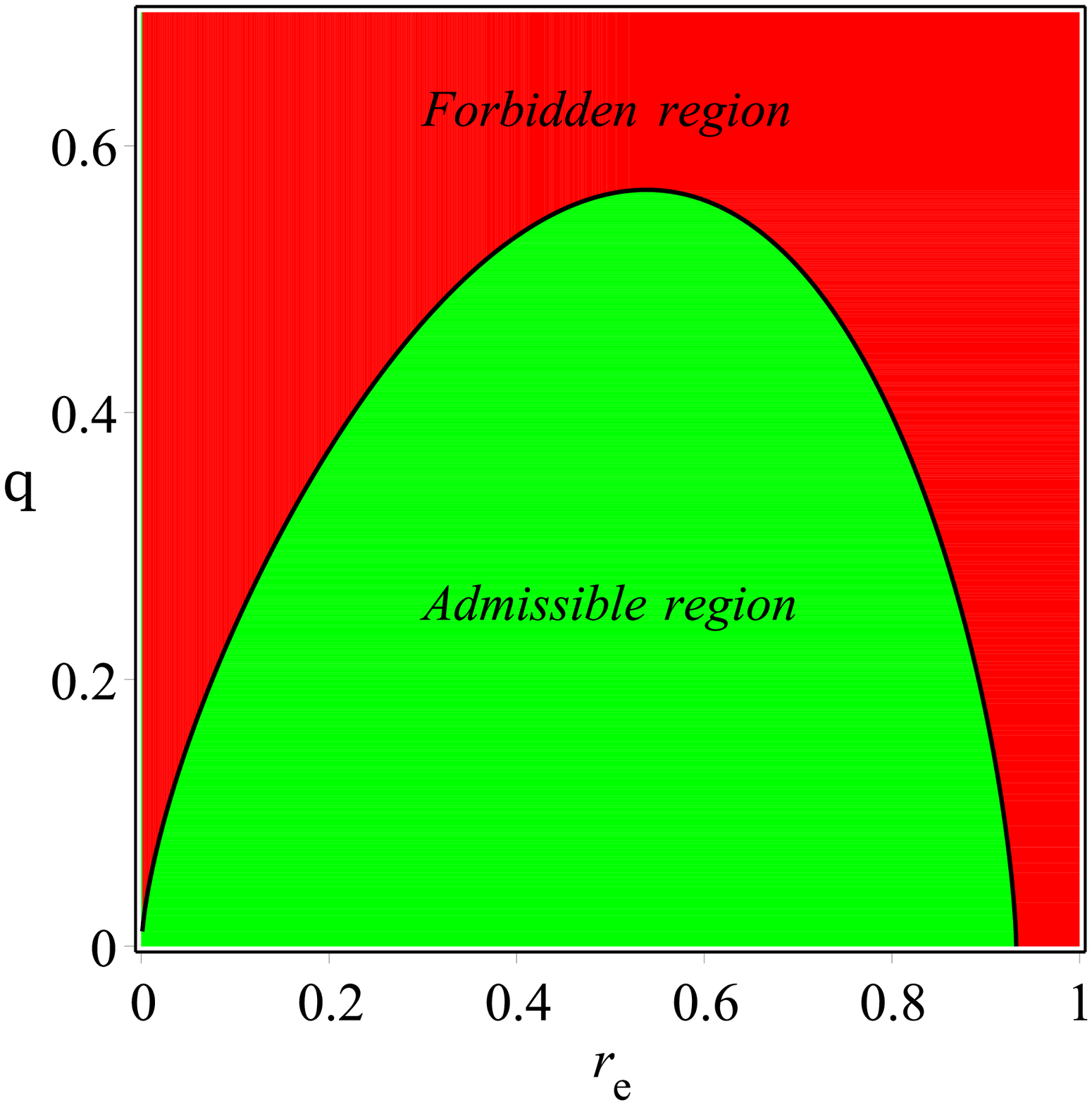}}
\subfloat[$ z=0 $ and $ w=2 $]{
        \includegraphics[width=0.315\textwidth]{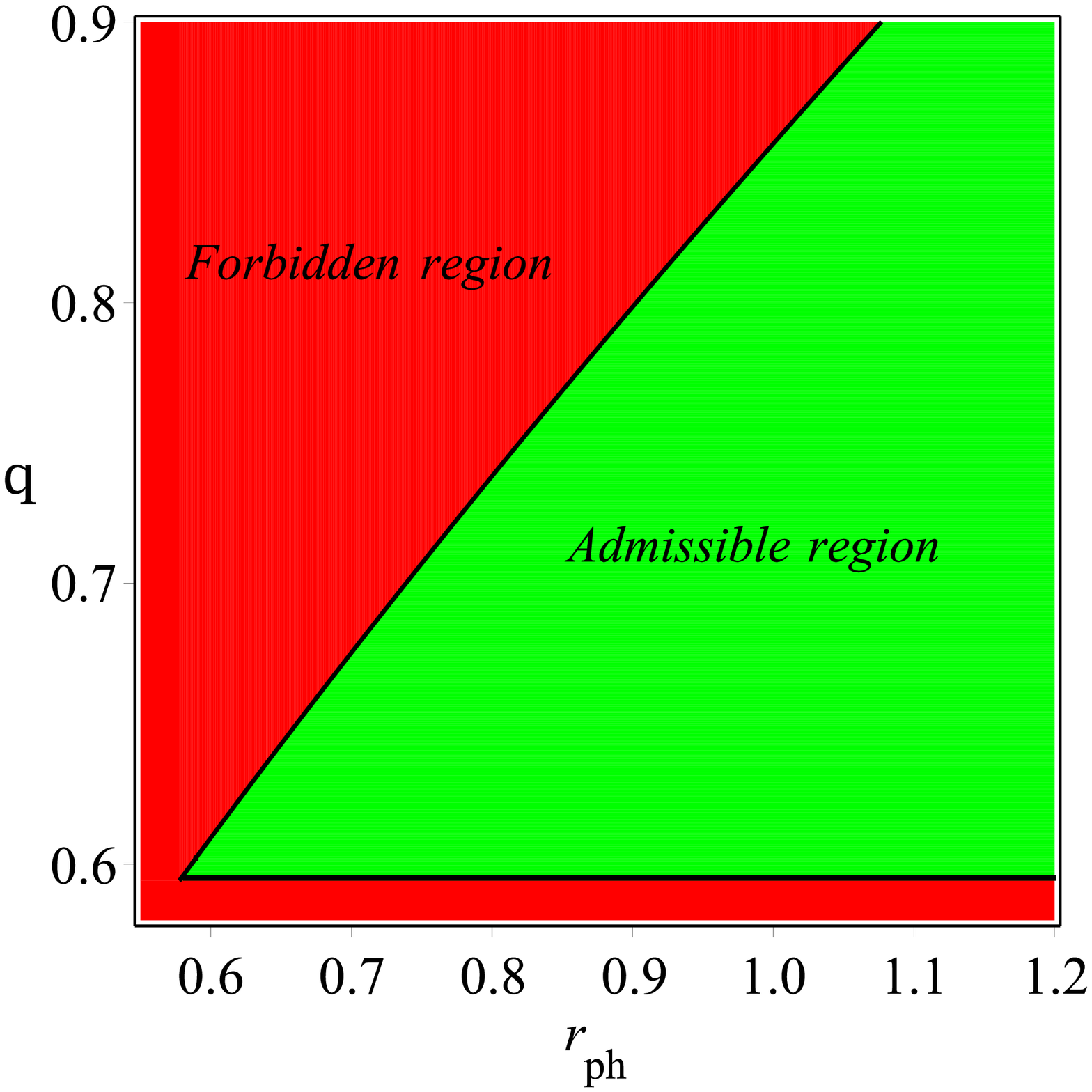}}\newline
\caption{ The admissible parameter space to have a real horizon (left
panels) and a real photon sphere (right panels) for $m=1 $, $\Lambda=-1 $, $%
r_{0}=1 $ and $J=0.4 $.}
\label{Fig3}
\end{figure}

In order to investigate the photon trajectories, one usually
expresses the radial geodesics in terms of the effective potential
$V_{\text{eff}}$ as
\begin{equation*}
\left(\frac{dr}{d\lambda}\right)^2+V_{\text{eff}}=0,
\end{equation*}
with
\begin{equation}
V_{\text{eff}}=\frac{f(r)}{r^{2}\left(\frac{r}{r_{0}} \right) ^{z}}\left(
L^{2}\left(\frac{r}{r_{0}} \right) ^{z}-\frac{E^{2}r^{2}}{f(r)}-\frac{E L J}{%
f(r)}\left(\frac{r}{r_{0}} \right) ^{w}-\frac{L^{2}J^{2}}{4r^{2}f(r)}\left(%
\frac{r}{r_{0}} \right) ^{2w}\right) .  \label{Veff}
\end{equation}

Now, we are in a position to obtain the photon critical circular orbit.
Therefore, the following unstable conditions should be satisfied,
simultaneously
\begin{equation}  \label{cond}
V_{\text{eff}}(r_{ph})=0,\quad~~~\frac{dV_{\text{eff}}(r_{ph})}{dr}=0, ~~~~
\frac{d^{2}V_{\text{eff}}(r_{ph})}{dr^{2}}<0.
\end{equation}

\begin{figure}[!htb]
\centering
\subfloat[$ z=w=2 $]{
        \includegraphics[width=0.31\textwidth]{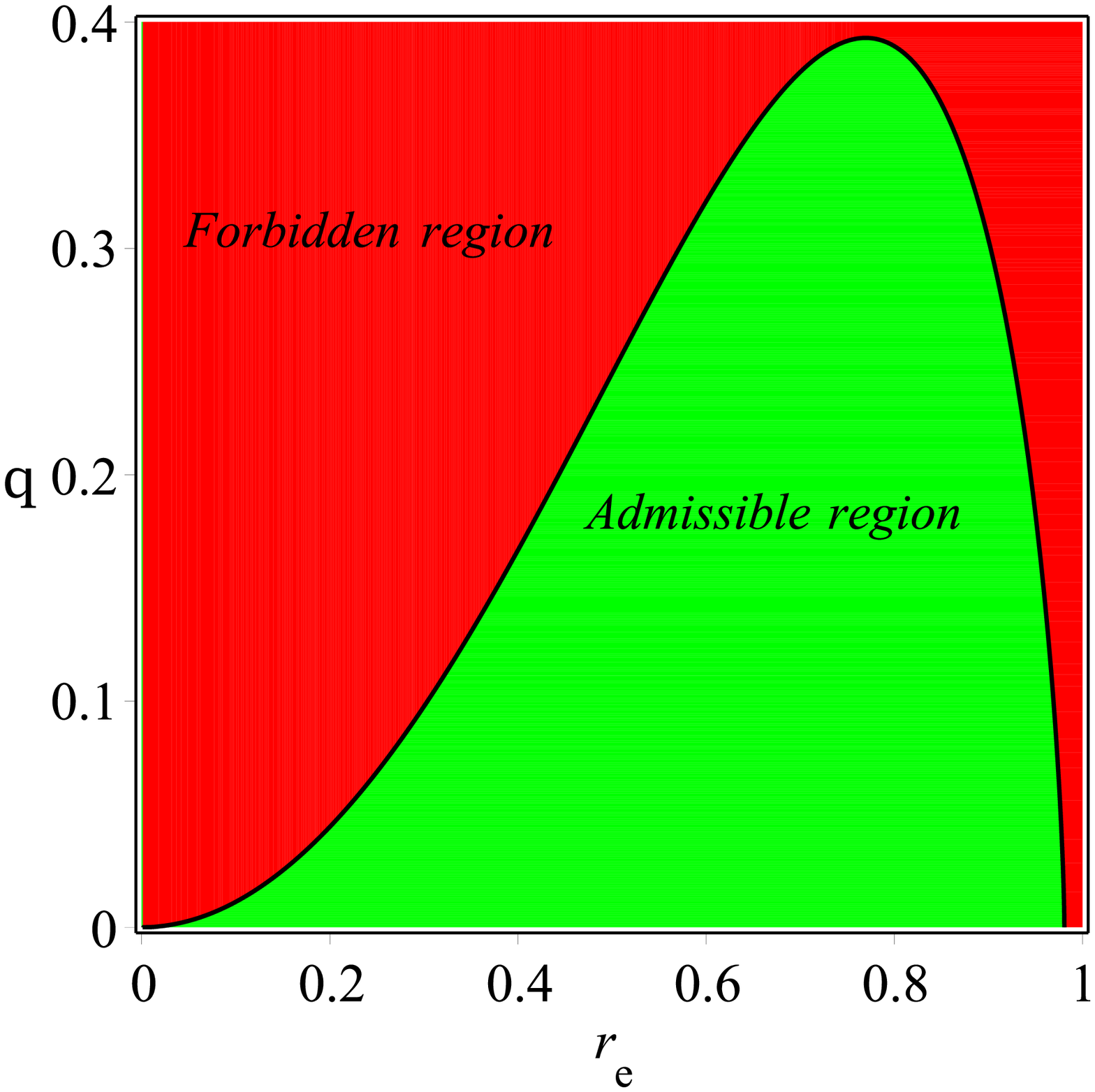}}
\subfloat[$ z=w=2 $]{
        \includegraphics[width=0.31\textwidth]{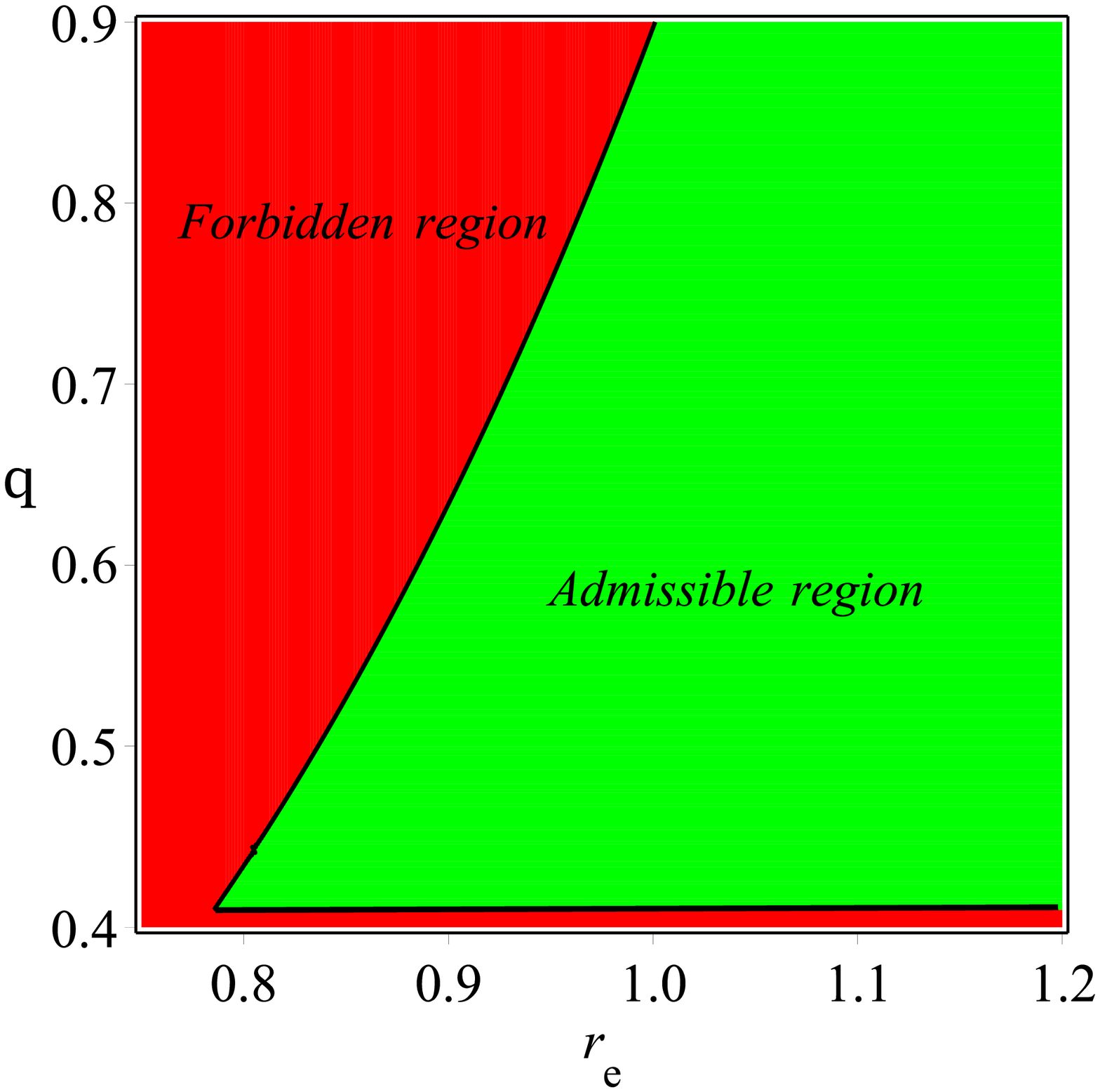}}\newline
\subfloat[$ z=2 $ and $ w=4$]{
        \includegraphics[width=0.31\textwidth]{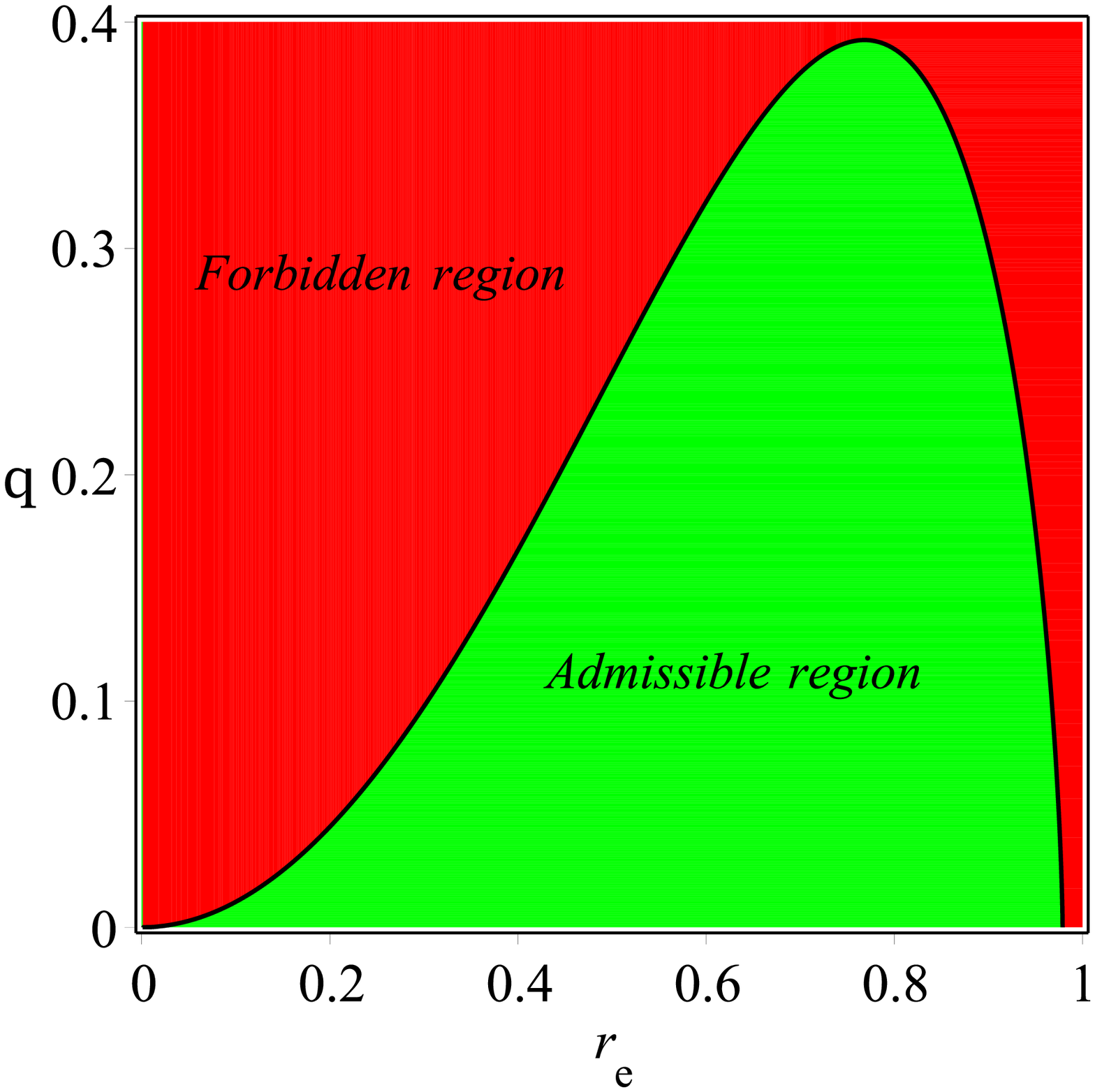}}
\subfloat[$ z=2 $ and $ w=4$]{
        \includegraphics[width=0.312\textwidth]{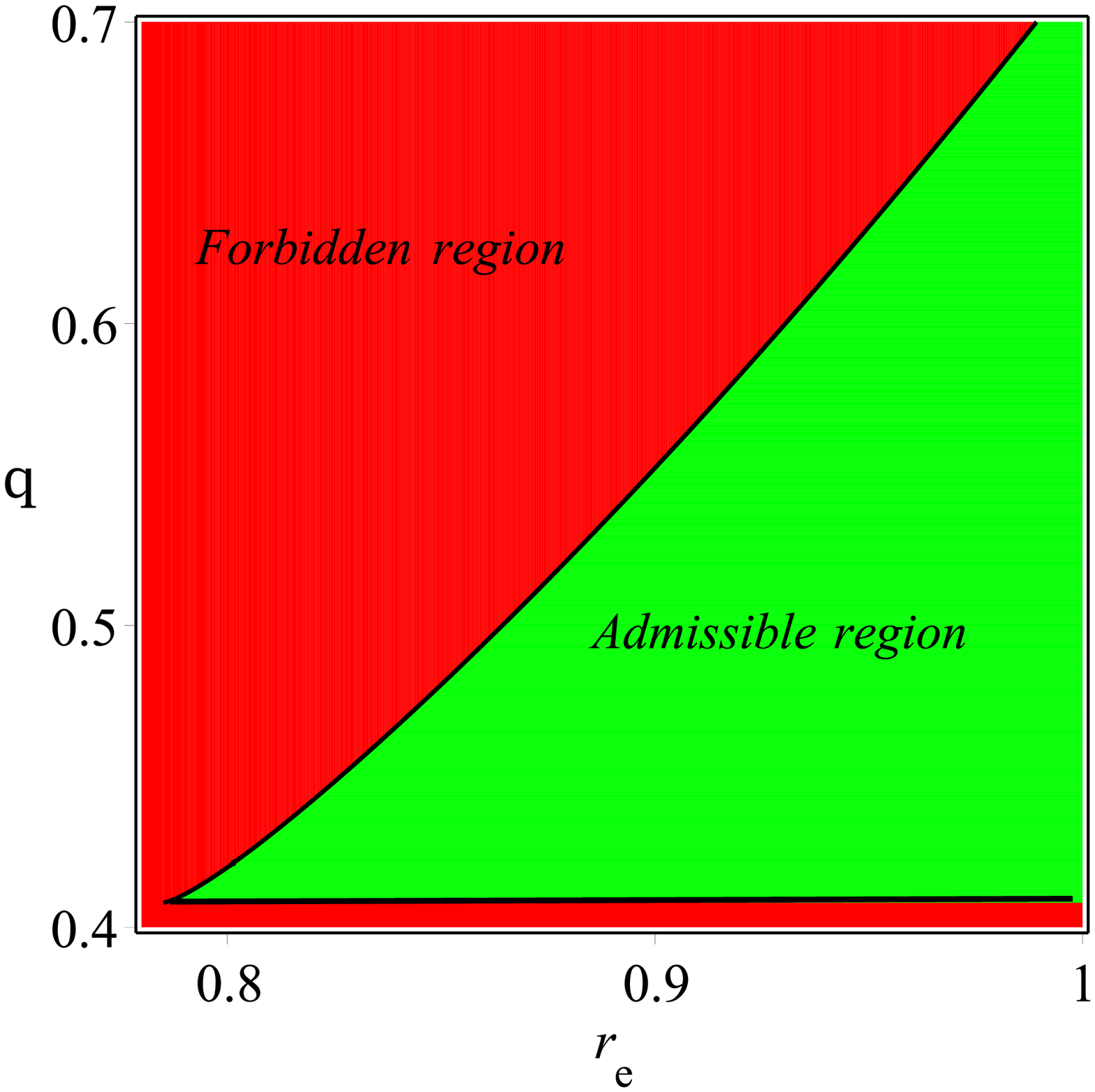}}\newline
\caption{ The admissible parameter space to have a real horizon (left
panels) and a real photon sphere (right panels) for $m=1 $, $\Lambda=-1 $, $%
r_{0}=1 $ and $J=0.5 $.}
\label{Fig4}
\end{figure}

To have an acceptable optical behavior, we need to examine the
condition $r_{e}<r_{ph} $ where $r_{ph}$ and $r_{e}$ are the
radius of photon orbit and event horizon radius, respectively.
Figures (\ref{Fig3}) and (\ref{Fig4}) display the admissible
parameter space to have a real horizon and real photon orbit. From
these two figures, one can find that an acceptable optical
behavior cannot be observed for $w-z<3$. In fact, by comparing the
left panels with the right panels of each figure, one can see that
the photon orbit radius will be imaginary in the region where the
event horizon is real. In other words, in a given region one
cannot observe the real horizon and real photon orbit,
simultaneously.

As was already mentioned, in order to have an acceptable optical
result, the condition $w-z\geq 3$ should be satisfied which is
quite evident in the tables \ref{table1}-\ref{table3}. Since
relation (\ref{cond}) leads to a complicated equation, it is not
possible to solve the equation analytically. Thus, we employ
numerical methods to obtain the radius of photon orbit. In this
regard, several values of the event horizon and the photon orbit
radius are listed in tables \ref{table1} ($z=0 $), \ref{table2}
($z=1 $), and \ref{table3} ($z=2 $). According to these tables,
only for limited regions of the electric charge and cosmological
constant, one can observe acceptable optical results. As one can
see, the increase of $q$ and $\vert \Lambda \vert $ leads to an
imaginary event horizon which is not a physical consequence. From
these three tables, it can also be seen that the electric charge,
angular momentum and the absolute value of the cosmological
constant have decreasing effects on the event horizon and the
radius of photon orbit. Taking a close look at the tables, one can
notice that the effect of parameter $r_{0}$ is opposite of that of
electric charge and angular momentum. Studying the effect of the
exponent $w$ shows that increasing this parameter leads to
increasing (decreasing) the event horizon (photon orbit).
Comparing these three tables to each other, one can examine the
effect of exponent $z$. Our analysis shows that as the parameter
$z$ increases the size of the event horizon and photon orbit
radius increase.
\begin{table*}[htb!]
\caption{The event horizon ($r_{e}$)and photon sphere radius ($r_{ph}$) for the variation of $q$, $J$, $r_{0}$, $w $ and $%
\Lambda$ for $m =1$ and $z=0 $.}
\label{table1}\centering
\begin{tabular}{||c|c|c|c|c||}
\hline
{\footnotesize $q$ \hspace{0.3cm}} & \hspace{0.3cm}$0.3$ \hspace{0.3cm} &
\hspace{0.3cm} $0.4$\hspace{0.3cm} & \hspace{0.3cm} $0.5$\hspace{0.3cm} &
\hspace{0.3cm}$0.6$\hspace{0.3cm} \\ \hline
$r_{e}$ ($\Lambda =-1 $, $J=0.4 $, $r_{0} =1$, $w=3$) & $0.9216$ & $0.8684$
& $0.7890$ & $0.57+0.05I$ \\ \hline
$r_{ph}$ ($\Lambda =-1 $, $J=0.4 $, $r_{0} =1$, $w=3$) & $1.7455 $ & $1.6989$
& $1.6385$ & $1.5576$ \\ \hline
$r_{ph}>r_{e}$ & \checkmark & \checkmark & \checkmark & $\times$  \\
\hline\hline
&  &  &  &  \\
{\footnotesize $J$ \hspace{0.3cm}} & \hspace{0.3cm} $0.3$ \hspace{0.3cm} &
\hspace{0.3cm} $0.4$\hspace{0.3cm} & \hspace{0.3cm} $0.5$\hspace{0.3cm} &
\hspace{0.3cm} $0.6$\hspace{0.3cm} \\ \hline
$r_{e}$ ($\Lambda =-1 $, $q=0.3 $, $r_{0} =1$, $w=3$) & $0.9217$ & $0.9216$
& $0.9213$ & $0.9210$ \\ \hline
$r_{ph}$ ($\Lambda =-1 $, $q=0.3 $, $r_{0} =1$, $w=3$) & $1.9067 $ & $1.7455$
& $1.6325$ & $1.5474$ \\ \hline
$r_{ph}>r_{e}$ & \checkmark & \checkmark & \checkmark & \checkmark \\
\hline\hline
&  &  &  &  \\
{\footnotesize $r_{0}$ \hspace{0.3cm}} & \hspace{0.3cm}$0.7$ \hspace{0.3cm}
& \hspace{0.3cm} $0.9$\hspace{0.3cm} & \hspace{0.3cm} $1$\hspace{0.3cm} &
\hspace{0.3cm}$1.1$\hspace{0.3cm} \\ \hline
$r_{e}$ ($\Lambda =-1 $, $q=0.3 $, $J =0.4$, $w=3$) & $0.9204 $ & $0.9212$ &
$0.9216$ & $0.9217$ \\ \hline
$r_{ph}$ ($\Lambda =-1 $, $q=0.3 $, $J =0.4$, $w=3$) & $1.4354 $ & $1.5883$
& $1.7455$ & $1.9057$ \\ \hline
$r_{ph}>r_{e}$ & \checkmark & \checkmark & \checkmark & \checkmark \\
\hline\hline
&  &  &  &  \\
{\footnotesize $\Lambda$ \hspace{0.3cm}} & \hspace{0.3cm}$-0.5$ \hspace{0.3cm%
} & \hspace{0.3cm} $-0.8$\hspace{0.3cm} & \hspace{0.3cm} $-1$\hspace{0.3cm}
& \hspace{0.3cm}$-1.2$\hspace{0.3cm} \\ \hline
$r_{e}$ ($r_{0}=1 $, $q=0.3 $, $J =0.4$, $w=3$) & $1.3367 $ & $1.0404$ & $%
0.9216$ & $0.83+0.06I$ \\ \hline
$r_{ph}$ ($r_{0}=1 $, $q=0.3 $, $J =0.4$, $w=3$) & $2.0506 $ & $1.8348$ & $%
1.7455$ & $1.6777$ \\ \hline
$r_{ph}>r_{e}$ & \checkmark & \checkmark & \checkmark & $\times$ \\
\hline\hline
&  &  &  &  \\
{\footnotesize $w$ \hspace{0.3cm}} & \hspace{0.3cm}$4$ \hspace{0.3cm} &
\hspace{0.3cm} $5$\hspace{0.3cm} & \hspace{0.3cm} $6$\hspace{0.3cm} &
\hspace{0.3cm}$7$\hspace{0.3cm} \\ \hline
$r_{e}$ ($r_{0}=1 $, $q=0.2 $, $J =0.4$, $\Lambda=-1.5$) & $0.77572 $ & $%
0.77576$ & $0.77581$ & $0.77584$ \\ \hline
$r_{ph}$ ($r_{0}=1 $, $q=0.2 $, $J =0.4$, $\Lambda=-1.5$) & $1.2488 $ & $%
1.1189$ & $1.0566$ & $1.0213$ \\ \hline
$r_{ph}>r_{e}$ & \checkmark & \checkmark & \checkmark & \checkmark \\
\hline\hline
\end{tabular}%
\end{table*}

\begin{table*}[htb!]
\caption{The event horizon ($r_{e}$) and photon sphere radius ($r_{ph}$) for the variation of $q$, $J$, $r_{0}$, $w $ and $%
\Lambda$ for $m =1$ and $z=1 $.}
\label{table2}\centering
\begin{tabular}{||c|c|c|c|c||}
\hline
{\footnotesize $q$ \hspace{0.3cm}} & \hspace{0.3cm}$0.3$ \hspace{0.3cm} &
\hspace{0.3cm} $0.35$\hspace{0.3cm} & \hspace{0.3cm} $0.4$\hspace{0.3cm} &
\hspace{0.3cm}$0.45$\hspace{0.3cm} \\ \hline
$r_{e}$ ($\Lambda =-1 $, $J=0.5 $, $r_{0} =1$, $w=4$) & $0.9216$ & $0.8932$
& $0.8588$ & $0.72+0.06I$ \\ \hline
$r_{ph}$ ($\Lambda =-1 $, $J=0.5 $, $r_{0} =1$, $w=4$) & $1.5012 $ & $1.4898$
& $1.4768$ & $1.4552$ \\ \hline
$r_{ph}>r_{e}$ & \checkmark & \checkmark & \checkmark & $\times$ \\
\hline\hline
&  &  &  &  \\
{\footnotesize $J$ \hspace{0.3cm}} & \hspace{0.3cm} $0.2$ \hspace{0.3cm} &
\hspace{0.3cm} $0.4$\hspace{0.3cm} & \hspace{0.3cm} $0.6$\hspace{0.3cm} &
\hspace{0.3cm} $0.8$\hspace{0.3cm} \\ \hline
$r_{e}$ ($\Lambda =-1 $, $q=0.4 $, $r_{0} =1$, $w=4$) & $0.8564$ & $0.8540$
& $0.8530$ & $0.8515$ \\ \hline
$r_{ph}$ ($\Lambda =-1 $, $q=0.4 $, $r_{0} =1$, $w=4$) & $2.2107 $ & $1.6150$
& $1.3783$ & $1.2453$ \\ \hline
$r_{ph}>r_{e}$ & \checkmark & \checkmark & \checkmark & \checkmark \\
\hline\hline
&  &  &  &  \\
{\footnotesize $r_{0}$ \hspace{0.3cm}} & \hspace{0.3cm}$0.8$ \hspace{0.3cm}
& \hspace{0.3cm} $1$\hspace{0.3cm} & \hspace{0.3cm} $1.2$\hspace{0.3cm} &
\hspace{0.3cm}$1.3$\hspace{0.3cm} \\ \hline
$r_{e}$ ($\Lambda =-1 $, $q=0.4 $, $J =0.5$, $w=4$) & $0.8487 $ & $0.8588$ &
$0.8597$ & $0.8599$ \\ \hline
$r_{ph}$ ($\Lambda =-1 $, $q=0.4 $, $J =0.5$, $w=4$) & $1.1135 $ & $1.4768$
& $1.5482$ & $2.1824$ \\ \hline
$r_{ph}>r_{e}$ & \checkmark & \checkmark & \checkmark & \checkmark \\
\hline\hline
&  &  &  &  \\
{\footnotesize $\Lambda$ \hspace{0.3cm}} & \hspace{0.3cm}$-0.5$ \hspace{0.3cm%
} & \hspace{0.3cm} $-0.8$\hspace{0.3cm} & \hspace{0.3cm} $-1$\hspace{0.3cm}
& \hspace{0.3cm}$-1.5$\hspace{0.3cm} \\ \hline
$r_{e}$ ($r_{0}=1 $, $q=0.4 $, $J =0.5$, $w=4$) & $1.2808 $ & $0.9879$ & $%
0.8588$ & $0.61+0.09I$ \\ \hline
$r_{ph}$ ($r_{0}=1 $, $q=0.4 $, $J =0.5$, $w=4$) & $1.6002 $ & $1.4973$ & $%
1.4768$ & $1.4681$ \\ \hline
$r_{ph}>r_{e}$ & \checkmark & \checkmark & \checkmark & $\times$ \\
\hline\hline
&  &  &  &  \\
{\footnotesize $w$ \hspace{0.3cm}} & \hspace{0.3cm}$5$ \hspace{0.3cm} &
\hspace{0.3cm} $6$\hspace{0.3cm} & \hspace{0.3cm} $7$\hspace{0.3cm} &
\hspace{0.3cm}$8$\hspace{0.3cm} \\ \hline
$r_{e}$ ($r_{0}=1 $, $q=0.2 $, $J =0.4$, $\Lambda=-1.5$) & $0.78831 $ & $%
0.78844$ & $0.78849$ & $0.78852$ \\ \hline
$r_{ph}$ ($r_{0}=1 $, $q=0.2 $, $J =0.4$, $\Lambda=-1.5$) & $1.2809 $ & $%
1.1480$ & $1.0831$ & $1.0455$ \\ \hline
$r_{ph}>r_{e}$ & \checkmark & \checkmark & \checkmark & \checkmark \\
\hline\hline
\end{tabular}%
\end{table*}
\begin{table*}[htb!]
\caption{The event horizon ($r_{e}$) and photon sphere radius ($r_{ph}$) for the variation of $q$, $J$, $r_{0}$, $w $ and $%
\Lambda$ for $m =1$ and $z=2 $.}
\label{table3}\centering
\begin{tabular}{||c|c|c|c|c||}
\hline
{\footnotesize $q$ \hspace{0.3cm}} & \hspace{0.3cm}$0.2$ \hspace{0.3cm} &
\hspace{0.3cm} $0.3$\hspace{0.3cm} & \hspace{0.3cm} $0.4$\hspace{0.3cm} &
\hspace{0.3cm}$0.5$\hspace{0.3cm} \\ \hline
$r_{e}$ ($\Lambda =-1 $, $J=0.5 $, $r_{0} =1$, $w=5$) & $0.9640$ & $0.9248$
& $0.8281$ & $0.81+0.15I$ \\ \hline
$r_{ph}$ ($\Lambda =-1 $, $J=0.5 $, $r_{0} =1$, $w=5$) & $1.4011 $ & $1.3891$
& $1.3733$ & $1.3525$ \\ \hline
$r_{ph}>r_{e}$ & \checkmark & \checkmark & \checkmark & $\times$ \\
\hline\hline
&  &  &  &  \\
{\footnotesize $J$ \hspace{0.3cm}} & \hspace{0.3cm} $0.2$ \hspace{0.3cm} &
\hspace{0.3cm} $0.4$\hspace{0.3cm} & \hspace{0.3cm} $0.6$\hspace{0.3cm} &
\hspace{0.3cm} $0.8$\hspace{0.3cm} \\ \hline
$r_{e}$ ($\Lambda =-1 $, $q=0.4 $, $r_{0} =1$, $w=5$) & $0.8302$ & $0.8290$
& $0.8270$ & $0.8257$ \\ \hline
$r_{ph}$ ($\Lambda =-1 $, $q=0.4 $, $r_{0} =1$, $w=5$) & $1.9762 $ & $1.4922$
& $1.2868$ & $1.2202$ \\ \hline
$r_{ph}>r_{e}$ & \checkmark & \checkmark & \checkmark & \checkmark \\
\hline\hline
&  &  &  &  \\
{\footnotesize $r_{0}$ \hspace{0.3cm}} & \hspace{0.3cm}$0.9$ \hspace{0.3cm}
& \hspace{0.3cm} $1$\hspace{0.3cm} & \hspace{0.3cm} $1.1$\hspace{0.3cm} &
\hspace{0.3cm}$1.2$\hspace{0.3cm} \\ \hline
$r_{e}$ ($\Lambda =-1 $, $q=0.4 $, $J =0.5$, $w=5$) & $0.8248 $ & $0.8281$ &
$0.8290$ & $0.8300$ \\ \hline
$r_{ph}$ ($\Lambda =-1 $, $q=0.4 $, $J =0.5$, $w=5$) & $1.1836 $ & $1.3733$
& $1.5908$ & $1.8348$ \\ \hline
$r_{ph}>r_{e}$ & \checkmark & \checkmark & \checkmark & \checkmark \\
\hline\hline
&  &  &  &  \\
{\footnotesize $\Lambda$ \hspace{0.3cm}} & \hspace{0.3cm}$-0.5$ \hspace{0.3cm%
} & \hspace{0.3cm} $-0.8$\hspace{0.3cm} & \hspace{0.3cm} $-1$\hspace{0.3cm}
& \hspace{0.3cm}$-1.1$\hspace{0.3cm} \\ \hline
$r_{e}$ ($r_{0}=1 $, $q=0.4 $, $J =0.5$, $w=5$) & $0.9872 $ & $0.8417$ & $%
0.8281$ & $0.76+0.07I$ \\ \hline
$r_{ph}$ ($r_{0}=1 $, $q=0.4 $, $J =0.5$, $w=5$) & $1.4472 $ & $1.3738$ & $%
1.3733$ & $1.3726$ \\ \hline
$r_{ph}>r_{e}$ & \checkmark & \checkmark & \checkmark & $\times$ \\
\hline\hline
&  &  &  &  \\
{\footnotesize $w$ \hspace{0.3cm}} & \hspace{0.3cm}$6$ \hspace{0.3cm} &
\hspace{0.3cm} $7$\hspace{0.3cm} & \hspace{0.3cm} $8$\hspace{0.3cm} &
\hspace{0.3cm}$9$\hspace{0.3cm} \\ \hline
$r_{e}$ ($r_{0}=1 $, $q=0.2 $, $J =0.4$, $\Lambda=-1.5$) & $0.81788 $ & $%
0.81794$ & $0.81799$ & $0.81800$ \\ \hline
$r_{ph}$ ($r_{0}=1 $, $q=0.2 $, $J =0.4$, $\Lambda=-1.5$) & $1.2768 $ & $%
1.1616$ & $1.1013$ & $1.0649$ \\ \hline
$r_{ph}>r_{e}$ & \checkmark & \checkmark & \checkmark & \checkmark \\
\hline\hline
\end{tabular}%
\end{table*}

\subsection{Energy emission rate}\label{SubSecIID}

Now, we are interested in studying the effect of the black hole
parameters on the emission of particles around the black hole. It
has been known that at very high energies, the
absorption cross-section for black holes oscillates around a
limiting constant value $\sigma_{lim}$ which is defined in the
following form for an arbitrary spacetime dimension \cite{Wei:2013}
\begin{equation}
\sigma _{lim}=\frac{\pi ^{\frac{d-2}{2}}b_{c}^{d-2}}{\Gamma (\frac{d}{2})},
\label{Eqsigma}
\end{equation}
where the critical impact parameter $ b_{c} $ is given by
\begin{equation}
b_{c}=\frac{r_{ph}}{\sqrt{f(r_{ph})}}.  \label{Eqemission1}
\end{equation}

The energy emission rate for three-dimensional spacetime is
obtained as \cite{Decanini}
\begin{equation}
\frac{d^{2}E(\omega )}{dtd\omega }=\frac{4\pi ^{2}\omega ^{2}b_{c}}{e^{%
\frac{\omega }{T}}-1},  \label{Eqemission}
\end{equation}%
where $\omega $ is the emission frequency and $T $ is the Hawking
temperature. For the corresponding black hole, the Hawking temperature is
given by
\begin{equation}
T=\frac{\kappa }{2\pi }=\frac{f^{\prime}(r) }{4\pi }\left( \frac{r}{r_{0}}%
\right) ^{\frac{z}{2}}\Bigg\vert _{r=r_{e}},  \label{temp1F(R)}
\end{equation}%
in which $\kappa $ is the surface gravity. Using Eqs. (\ref{temp1F(R)}) and (%
\ref{metric function F(R)}), one can find
\begin{equation}
T=\frac{\left( \frac{r_{e}}{r_{0}}\right) ^{\frac{z}{2}}}{4\pi }\left( -%
\frac{A_{1}\left( w-2\right) ^{2} J^{2}\left( \frac{r_{e}}{r_{0}}\right)
^{2w-z}}{32A_{2}r_{e}^{3}}-\Lambda \left( \gamma +2\right) r_{e}-\frac{2^{-%
\frac{1}{4}}q^{\frac{3}{2}}\left( \delta -\gamma \right) }{r_{e}^{\gamma +1}}%
\right) ,  \label{temp2F(R)}
\end{equation}
where
\begin{eqnarray}
A_{1}&=& 2\delta -2\gamma -8w+z+6, \\
A_{2}&=& 2-zw-6w+4w^{2},  \nonumber
\end{eqnarray}

\begin{figure}[!htb]
\centering
\subfloat[$ z=0 $, $ w=3 $, $ J=0.4 $, $ r_{0} =1$, $ \Lambda=-1 $]{
        \includegraphics[width=0.31\textwidth]{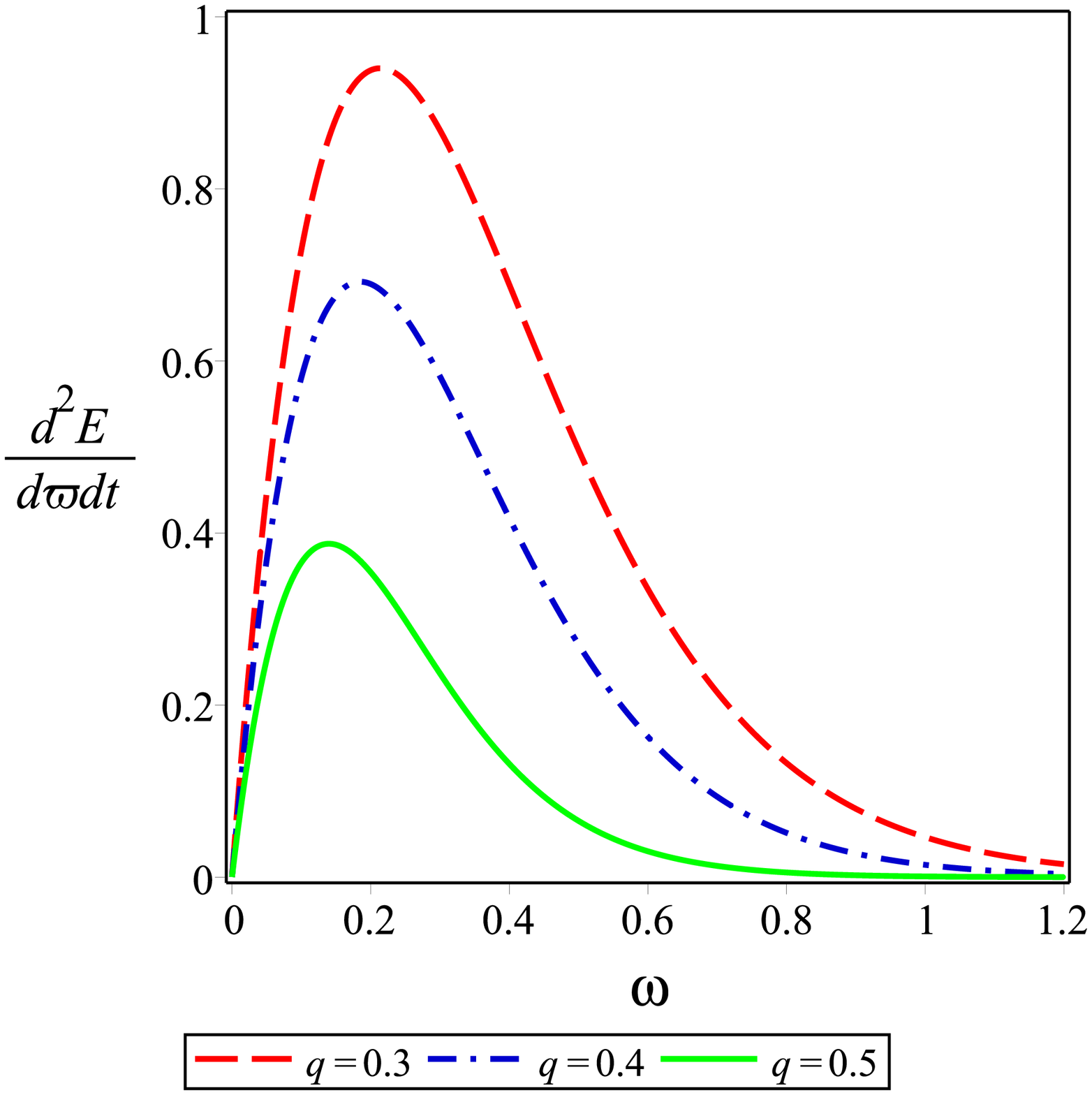}}
\subfloat[$ z=0 $, $ w=3 $, $ q=0.3 $, $ r_{0} =1$, $ \Lambda=-1 $]{
        \includegraphics[width=0.31\textwidth]{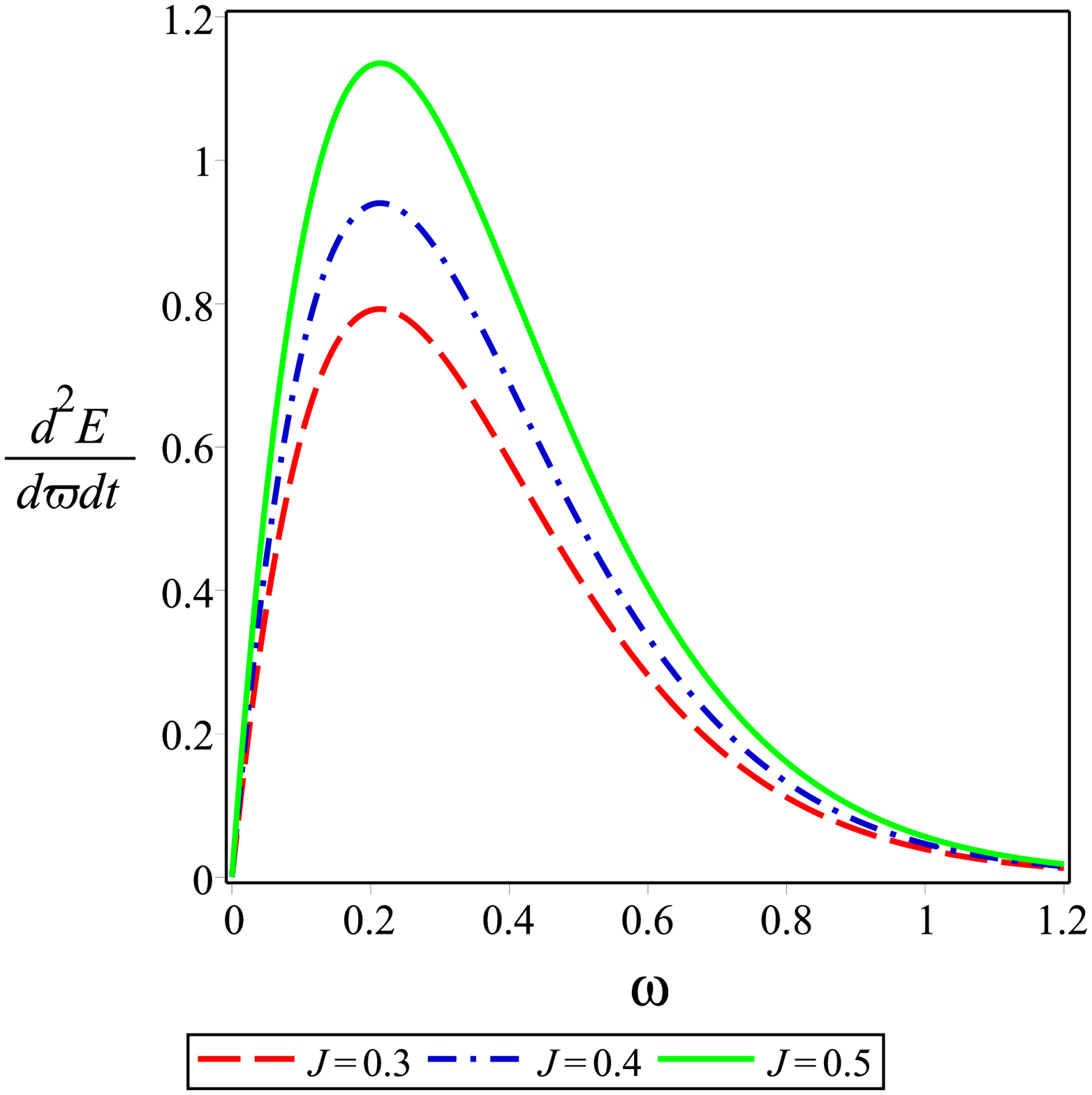}}
\subfloat[$ z=0 $, $ w=3 $, $ J=0.4 $, $ q=0.3$, $ \Lambda=-1 $]{
        \includegraphics[width=0.31\textwidth]{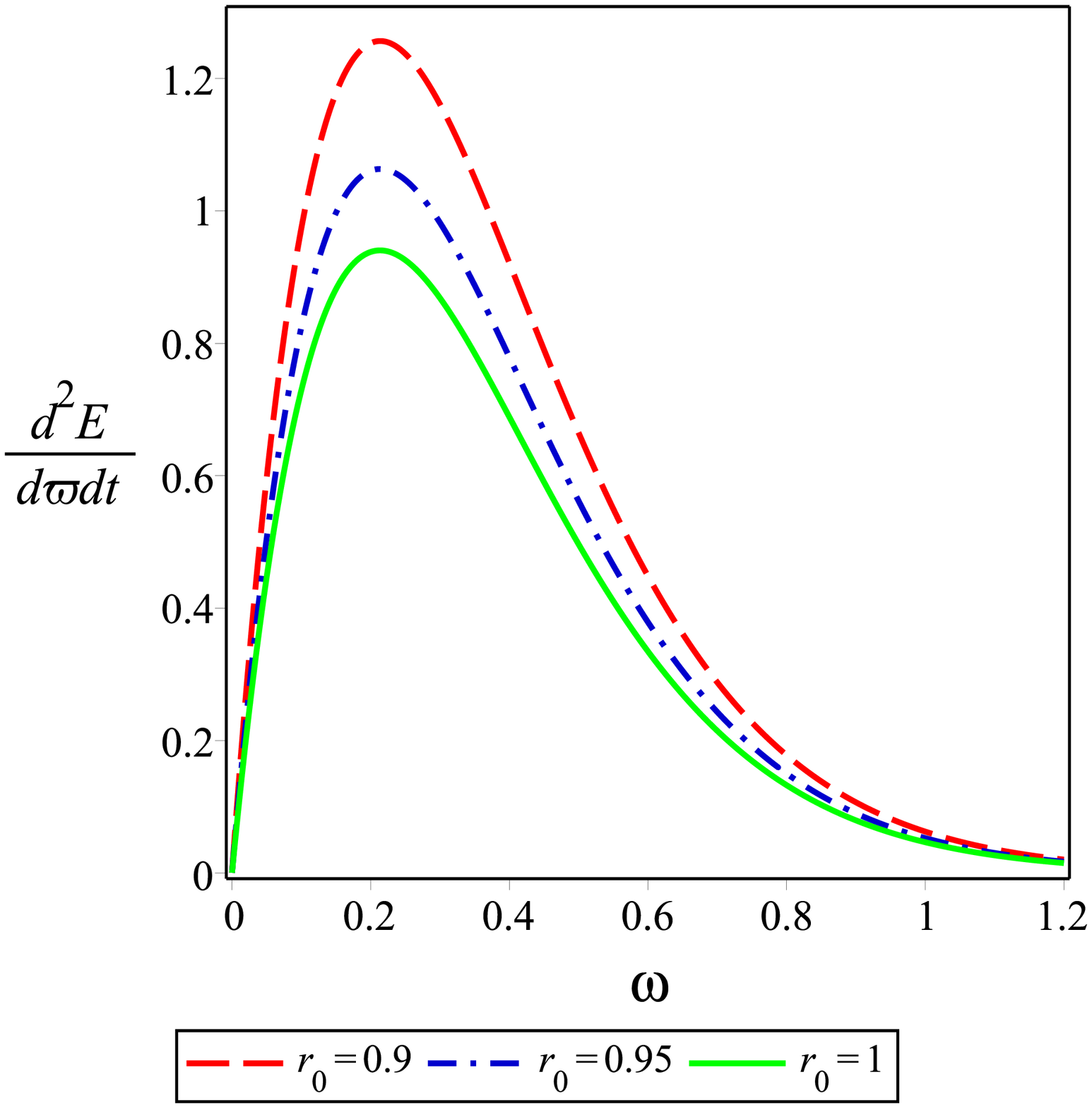}}\newline
\subfloat[$ z=0 $, $ w=3 $, $ J=0.4 $, $ q=0.3 $, $ r_{0} =1$]{
        \includegraphics[width=0.31\textwidth]{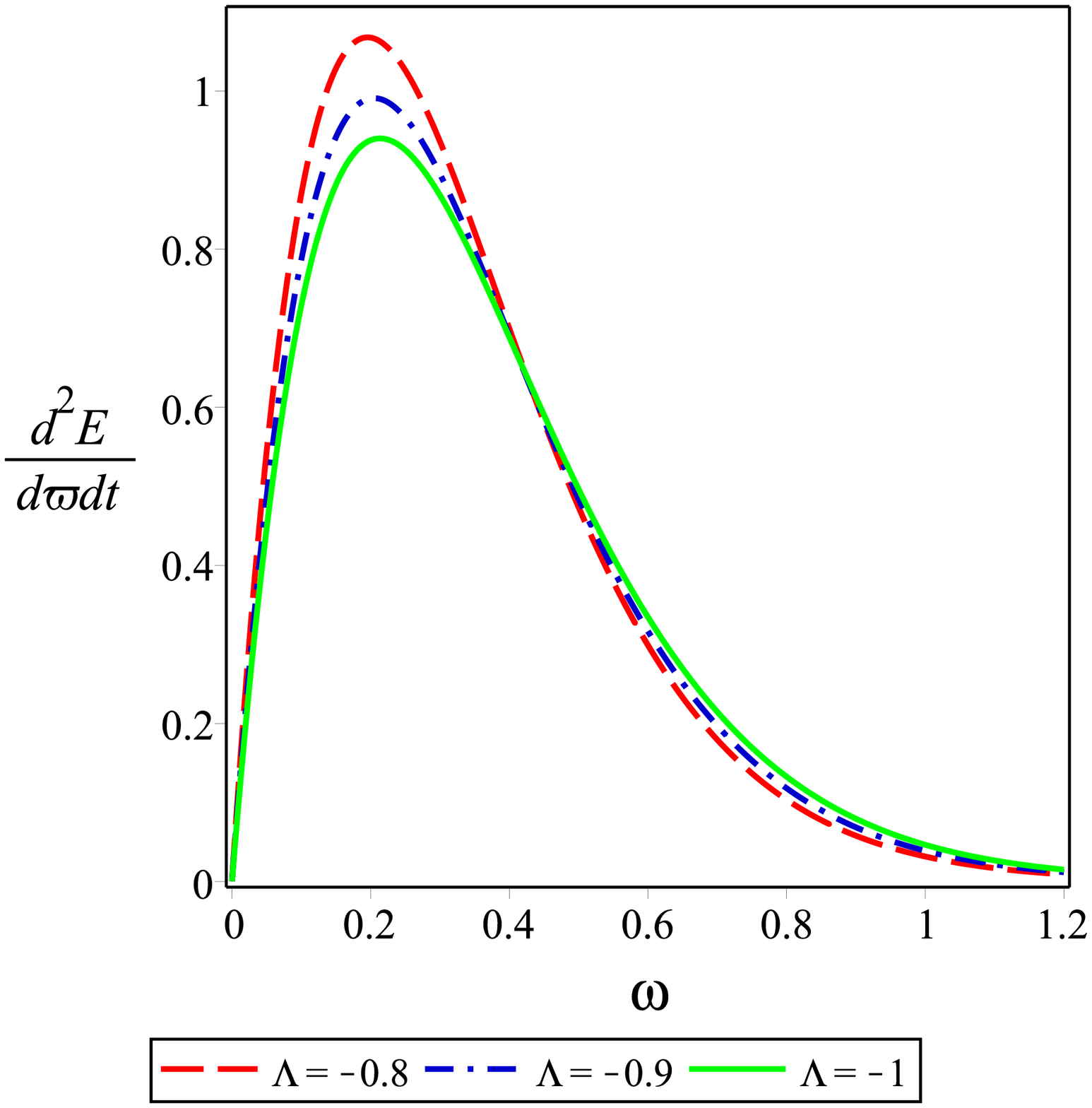}}
\subfloat[$ z=0 $, $ q=0.3 $, $ J=0.4 $, $ r_{0} =1$, $ \Lambda=-1 $]{
        \includegraphics[width=0.31\textwidth]{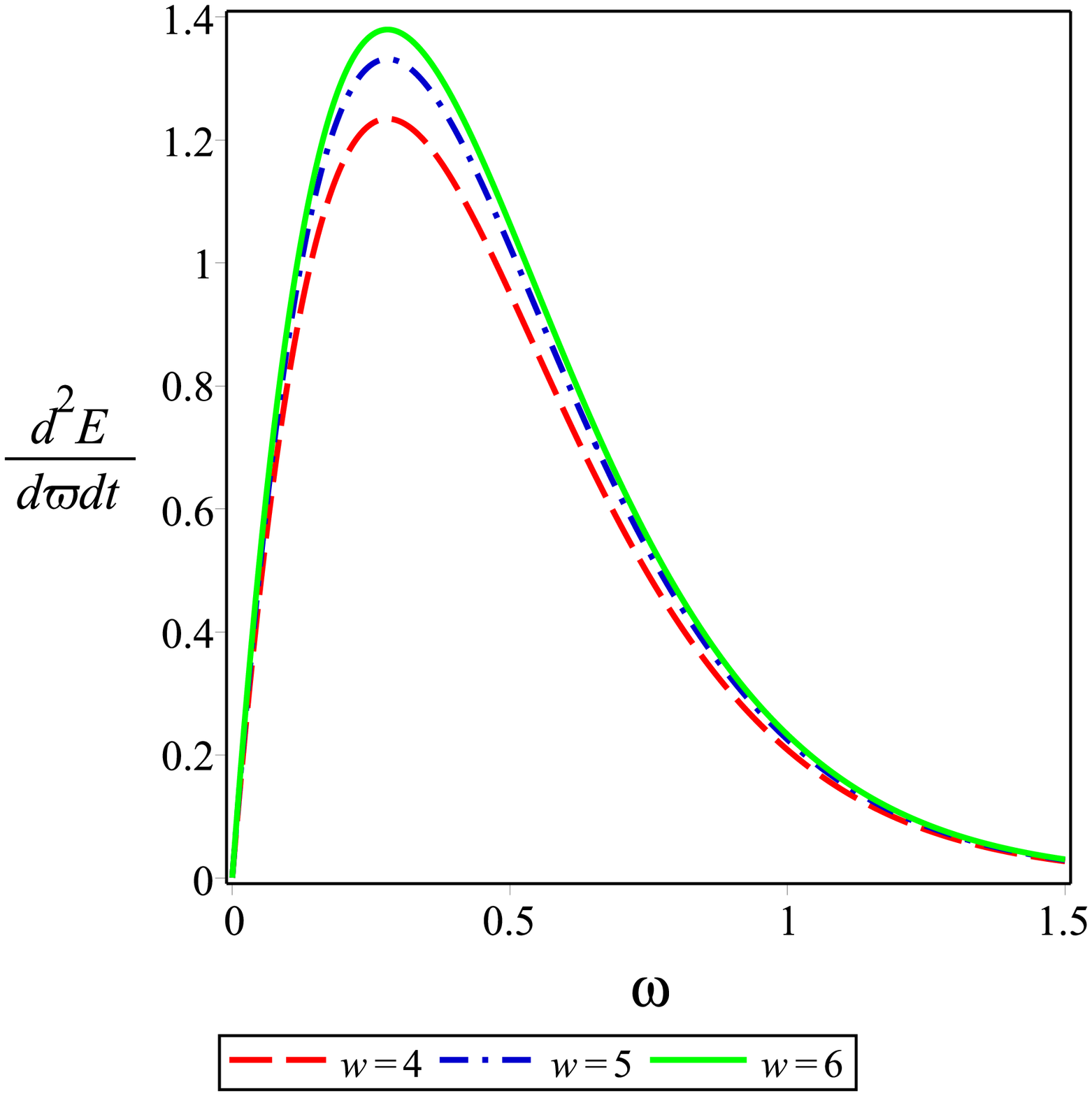}}
\subfloat[ $ w=6 $, $ J=0.4 $, $ q=0.2 $ ,$ r_{0} =1$, $ \Lambda=-1.5 $]{
        \includegraphics[width=0.312\textwidth]{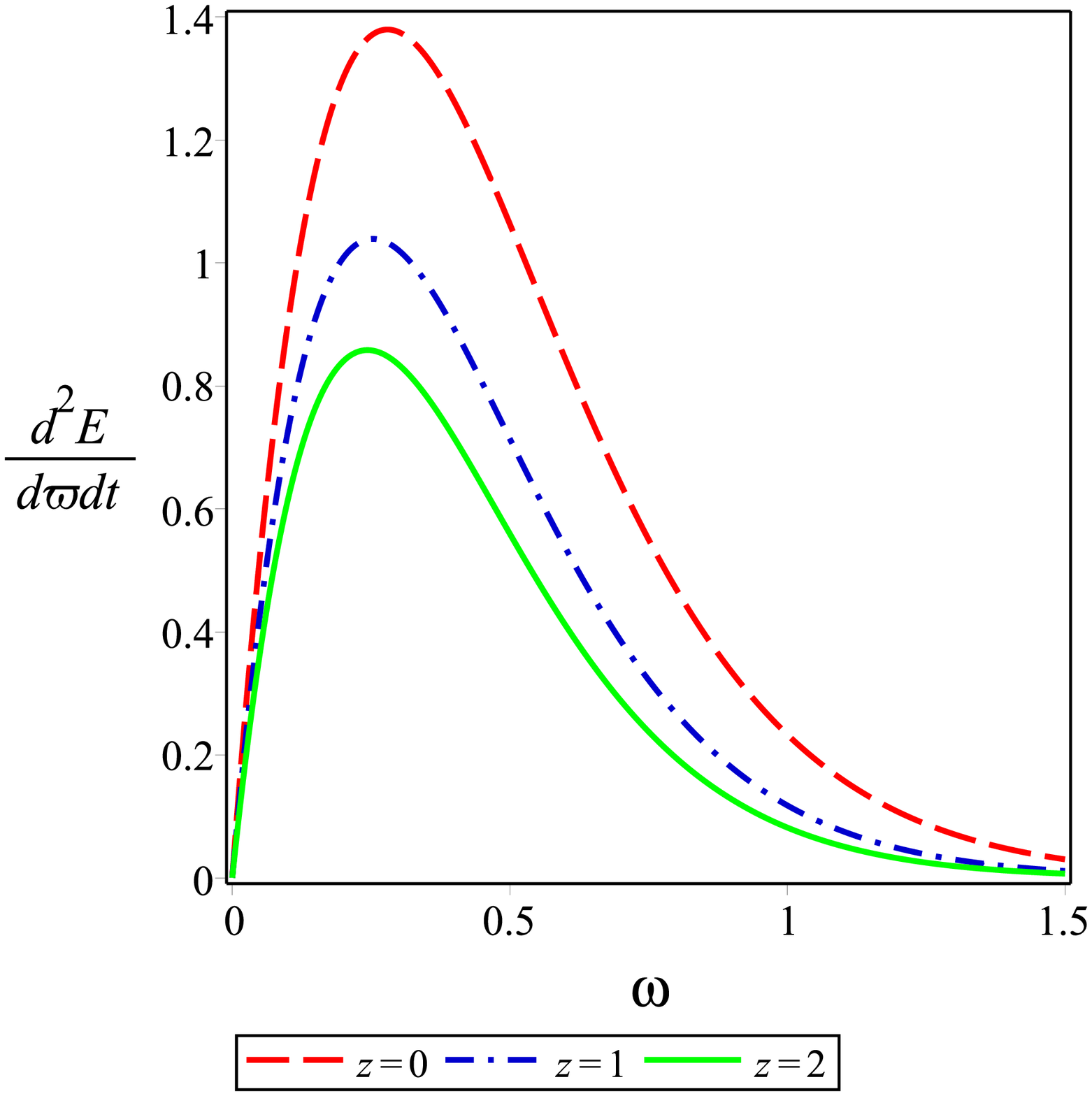}}\newline
\caption{Energy emission rate for the corresponding black hole with $m=1 $
and different values of black hole parameters.}
\label{FigEmr}
\end{figure}

To study the impact of black hole parameters on the energy
emission rate, we have plotted Fig. \ref{FigEmr}. Figure
\ref{FigEmr}(a) illustrates the influence of the electric charge
on the emission rate of a Lifshitz rotating black hole. As it is
clear, there exists a peak of the energy emission rate for the
black hole which shifts to the low frequency with the increase of
$q$. From this figure, one can also find that this parameter has a
decreasing contribution to the energy emission rate. This reveals
the fact that the evaporation process would be slow for a black
hole located in a powerful electric field. The effect of the
angular momentum on the emission rate is depicted in Fig.
\ref{FigEmr}(b), indicating that the impact of this parameter is
opposite of that of the electric charge. Studying the impact of
parameter $r_{0}$ and cosmological constant, we observe that both
parameters have a decreasing effect on this optical quantity
similar to the electric charge (see Figs. \ref{FigEmr}(c) and
\ref{FigEmr}(d)). This reveals the fact that as the effect of
these two parameters get weak the energy emission rate becomes
significant. To study the effects of two exponents, we plot Figs.
\ref{FigEmr}(e) and \ref{FigEmr}(f). From Fig. \ref{FigEmr}(e), it
is clear that the parameter $w$ has an increasing contribution to
the energy emission rate, while the effect of the exponent $z$ is
to decrease it (see Fig. \ref{FigEmr}(f)). From what was
expressed, one can find that the black hole has a longer lifetime
when it rotates slowly or when it is located in a high curvature
background or a powerful electric field.

\section{Thermodynamic properties}\label{SecIII}

In this section, we would like to study the thermodynamic
structure of the system. We first calculate the conserved and
thermodynamics quantities of the black hole solution and examine
the first law of thermodynamics. Then, we investigate phase
transition and thermal stability of the black hole in the context
of the canonical ensemble by calculating the heat capacity. We
also examine the effects of the black hole parameters on phase
transition and stability of the system and show that a certain
relation between the exponents $z$ and $w$ should be satisfied in
order to have a phase transition. We investigate the possibility
of the existence of van der Waals-like phase transition and
critical behavior for the solutions, and determine critical
values. Finally, we construct a heat engine by taking into account
this black hole as the working substance, and obtain the heat
engine efficiency. Comparing the engine efficiency with Carnot
efficiency, we investigate the criteria of having a consistent
thermodynamic second law.

\subsection{Thermodynamic quantities and the first law}\label{SubSecIIIA}

In this subsection, we obtain the thermodynamical quantities of
the solutions and check the validity of the first law of black
hole thermodynamics. Before we go on, we introduce a new notion
for our solutions. We consider the negative branch of cosmological
constant to be a thermodynamical quantity known as pressure.
Considering the cosmological constant as a thermodynamical
pressure and its conjugate quantity as a thermodynamical volume
leads to a new insight into thermodynamical structure of the black
holes, called extended phase space thermodynamics. From now on, we
replace the cosmological constant with the pressure using the
following relation \cite{Kubiznak:2012}
\begin{eqnarray}
\Lambda &=&-8\pi P  \nonumber
\end{eqnarray}

The finite mass is the first quantity that we would like to
calculate. In the non-extended phase space (the cosmological
constant is not allowed to vary), the total mass of the black
holes is depicted as internal energy. Considering the variable
cosmological constant, the role of the mass is changed to
enthalpy. There are several methods for calculating the mass.
Here, to calculate this property, we use ADM
(Arnowitt-Deser-Misner) approach which yields
\begin{equation}
M=\frac{m}{16\pi }r_{0}^{-\gamma}.  \label{M1F(R)}
\end{equation}

Evaluating the metric function on horizon $(f(r = r_{e}) = 0)$ and
solving it with respect to geometrical mass result into the
following relation for total mass of the black hole
\begin{equation}
M=\left( \frac{\left( w-2\right) ^{2}\left( \frac{r_{e}}{r_{0}}\right)
^{2w-z}}{128\pi A_{2}r_{e}^{2}}J^{2}+\frac{r_{e}^{2}P}{2}+\frac{2^{-\frac{1}{%
4}}q^{\frac{3}{2}}}{16\pi r_{e}^{\delta }}\right) \left( \frac{r_{e}}{r_{0}}%
\right) ^{\gamma }.  \label{M2F(R)}
\end{equation}

The temperature of the black hole has already been obtained from
Eq. (\ref{temp2F(R)}). Replacing cosmological constant with
pressure, it can be rewritten as
\begin{equation}
T=\frac{\left( \frac{r_{e}}{r_{0}}\right) ^{\frac{z}{2}}}{4\pi }\left( -%
\frac{A_{1}\left( w-2\right) ^{2} J^{2}\left( \frac{r_{e}}{r_{0}}\right)
^{2w-z}}{32A_{2}r_{e}^{3}}+8\pi P \left( \gamma +2\right) r_{e}-\frac{2^{-%
\frac{1}{4}}q^{\frac{3}{2}}\left( \delta -\gamma \right) }{r_{e}^{\gamma +1}}%
\right).  \label{temp3F(R)}
\end{equation}

As the next step, we calculate the entropy of the black hole. The
method for obtaining the entropy of black holes depends on
gravities under consideration and topological structure of the
black holes. In Einsteinian black holes, without higher curvature
terms, the entropy could be obtained by using the area law. But,
since our solutions are obtained in a class of $F(R)$ gravity with
$F_R=0$, we suppose the validity of the first law of
thermodynamics to calculate the entropy as
\begin{equation}
\delta S=\frac{1}{T}\delta M,
\end{equation}
yielding
\begin{equation}
S=\frac{\left( 4+2\delta -\gamma \right) r_{e}\left(
\frac{r_{e}}{r_{0}}\right) ^{\gamma -\frac{z}{2}}}{24}. \label{S}
\end{equation}

As we see, entropy depends only on the exponent $z$ and there is
no direct contributions of matter field. Besides, using the
concept of enthalpy, one can obtain the volume of these black
holes as
\begin{equation}
V=\left( \frac{\partial H}{\partial P}\right) \Bigg\vert _{S,Q,J}=\frac{
r_{e}^{2}\left( \frac{r_{e}}{r_{0}}\right) ^{\gamma }}{2}.  \label{Volume}
\end{equation}

Evidently, there is a direct relationship between the total volume of the
black holes and the horizon radius, indicating that one can use the horizon
radius instead of using volume in calculations.

The total electric charge of the black hole can be obtained from the power
Maxwell nonlinear electrodynamics as
\begin{equation}
Q=\frac{3\sqrt{q}2^{\frac{3}{4}}}{32\pi },  \label{Q U1}
\end{equation}%
and the electric potential is determined as
\begin{equation}
U=r_{e}^{-\delta }q\left( \frac{r_{e}}{r_{0}}\right) ^{\gamma }.
\label{Q U2}
\end{equation}

To obtain the angular velocity, we take advantage of the standard
equation as
\begin{equation}
\Omega =-\frac{g_{t\varphi }}{g_{\varphi \varphi }}=\frac{J\left( \frac{%
r_{e} }{r_{0}}\right) ^{w}}{2r_{e}^{2}}.  \label{angular velocity}
\end{equation}

Considering the above equation and the first law, the angular momentum can
be obtained as

\begin{equation}
\xi =\frac{\left( \frac{r_{e}}{r_{0}}\right) ^{w+\gamma -z}\left( w-2\right)
^{2}}{32\pi \left( 2-zw-6w+4w^{2}\right) }J.  \label{angular momentum}
\end{equation}

It is easy to show that the first law of thermodynamics is as follows
\begin{equation}
dM=TdS+UdQ+VdP+\Omega d\xi ,  \label{first law F(R)}
\end{equation}

Taking into account the scaling argument for our Lifshitz like solutions in
the extended phase space, one can find the following Smarr relation holds
\begin{equation}
\gamma M=\frac{\left( 4-\delta +2\gamma \right) }{3}TS+\frac{\delta }{3}%
QU-2PV-\frac{\left( 3w-\delta -\gamma \right) }{3}\Omega \xi .  \label{smarr}
\end{equation}

It is notable that for $z=w=0$, Eq. (\ref{smarr}) reduces to that
of nonlinearly charged rotating BTZ black holes in which mass term
has no scaling.

\begin{figure}[!htb]
\centering
\subfloat[$ z=0.5 $, $ w=1.5 $, $ J=0.5 $, $ r_{0} =0.1$, $ \Lambda=-0.01 $]{
        \includegraphics[width=0.31\textwidth]{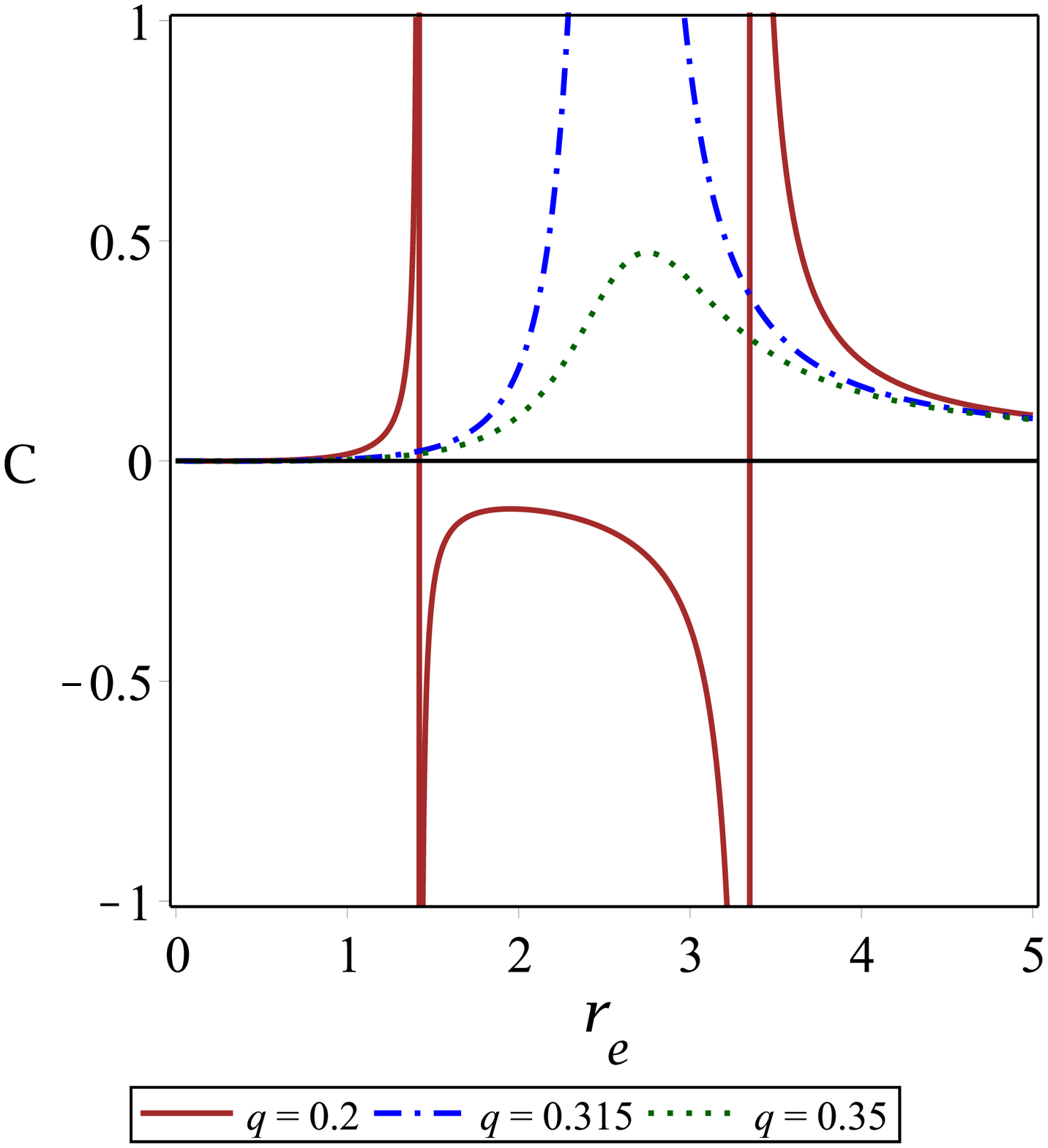}}
\subfloat[$ z=0.5 $, $ w=1.5 $, $ q=0.2 $, $ r_{0} =0.1$, $ \Lambda=-0.01 $]{
        \includegraphics[width=0.31\textwidth]{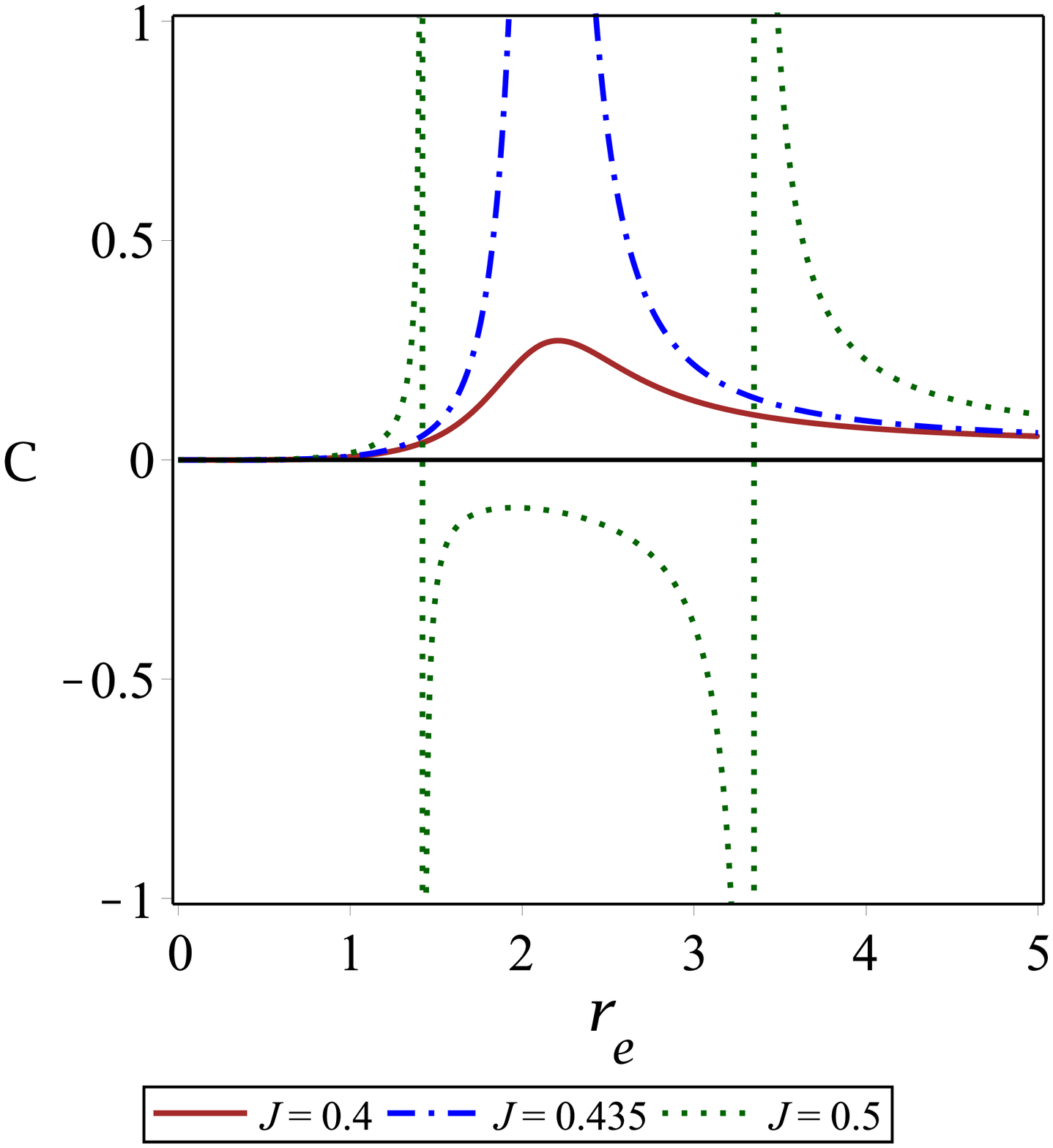}}
\subfloat[$ z=0.5 $, $ w=1.5 $, $ q=0.2 $, $ J =0.5$, $ \Lambda=-0.01 $]{
        \includegraphics[width=0.31\textwidth]{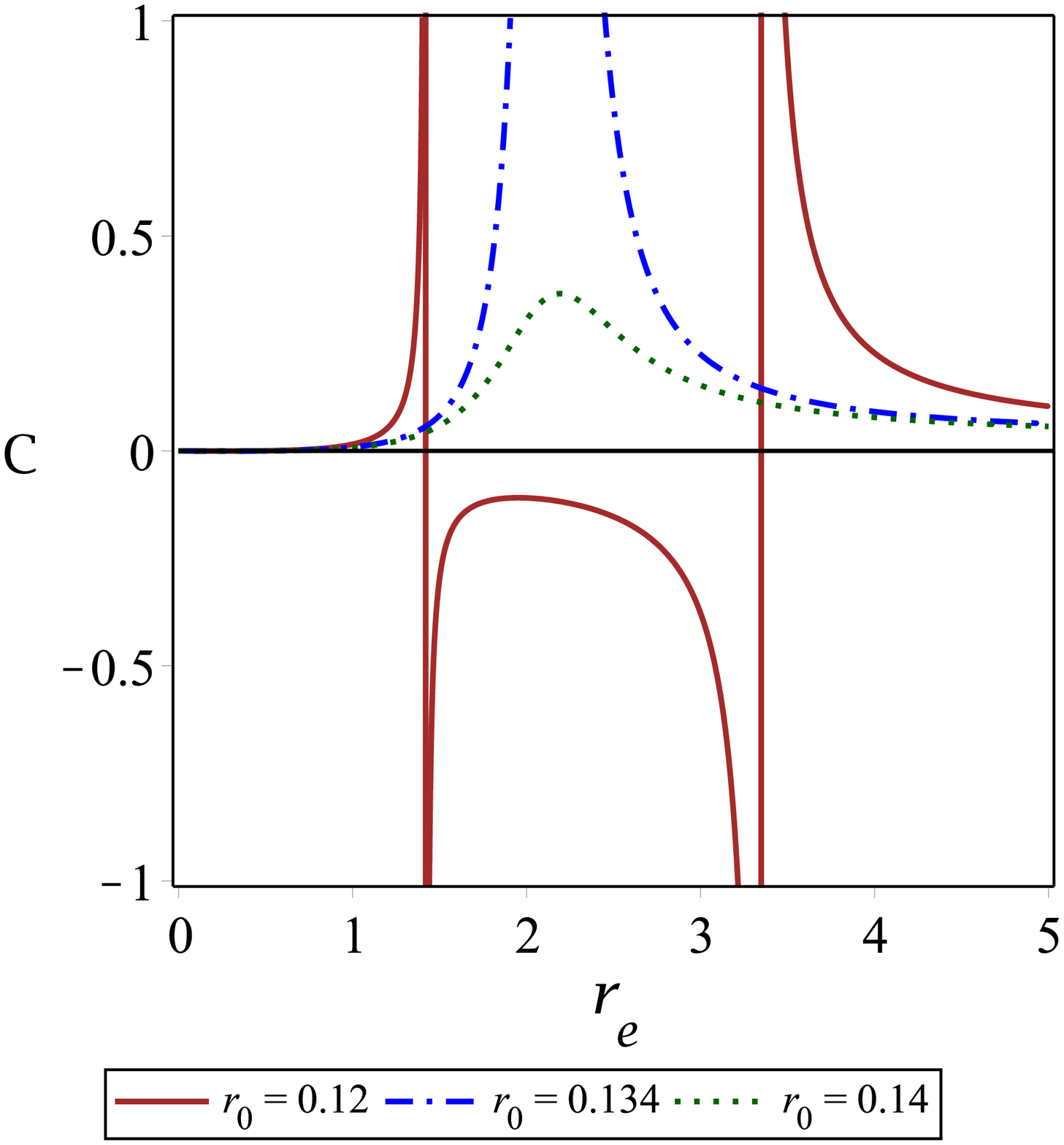}}\newline
\subfloat[$ z=0.5 $, $ J=0.5 $, $ q=0.2 $, $ r_{0} =0.1$, $ \Lambda=-0.01 $]{
        \includegraphics[width=0.31\textwidth]{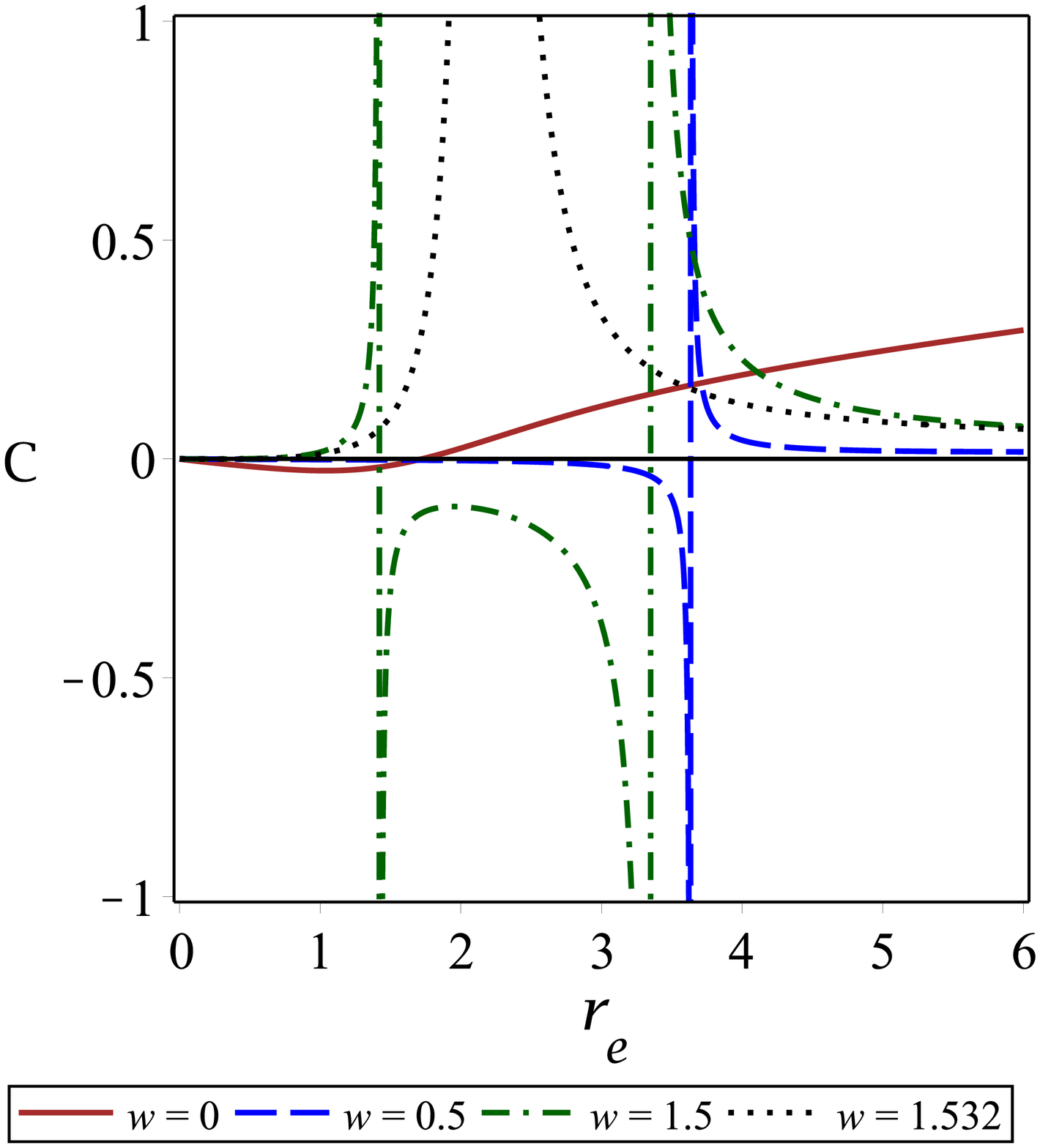}}
\subfloat[$w=1 $, $ J=0.3 $, $ q=0.2 $, $ r_{0} =0.1$, $ \Lambda=-0.01 $]{
        \includegraphics[width=0.32\textwidth]{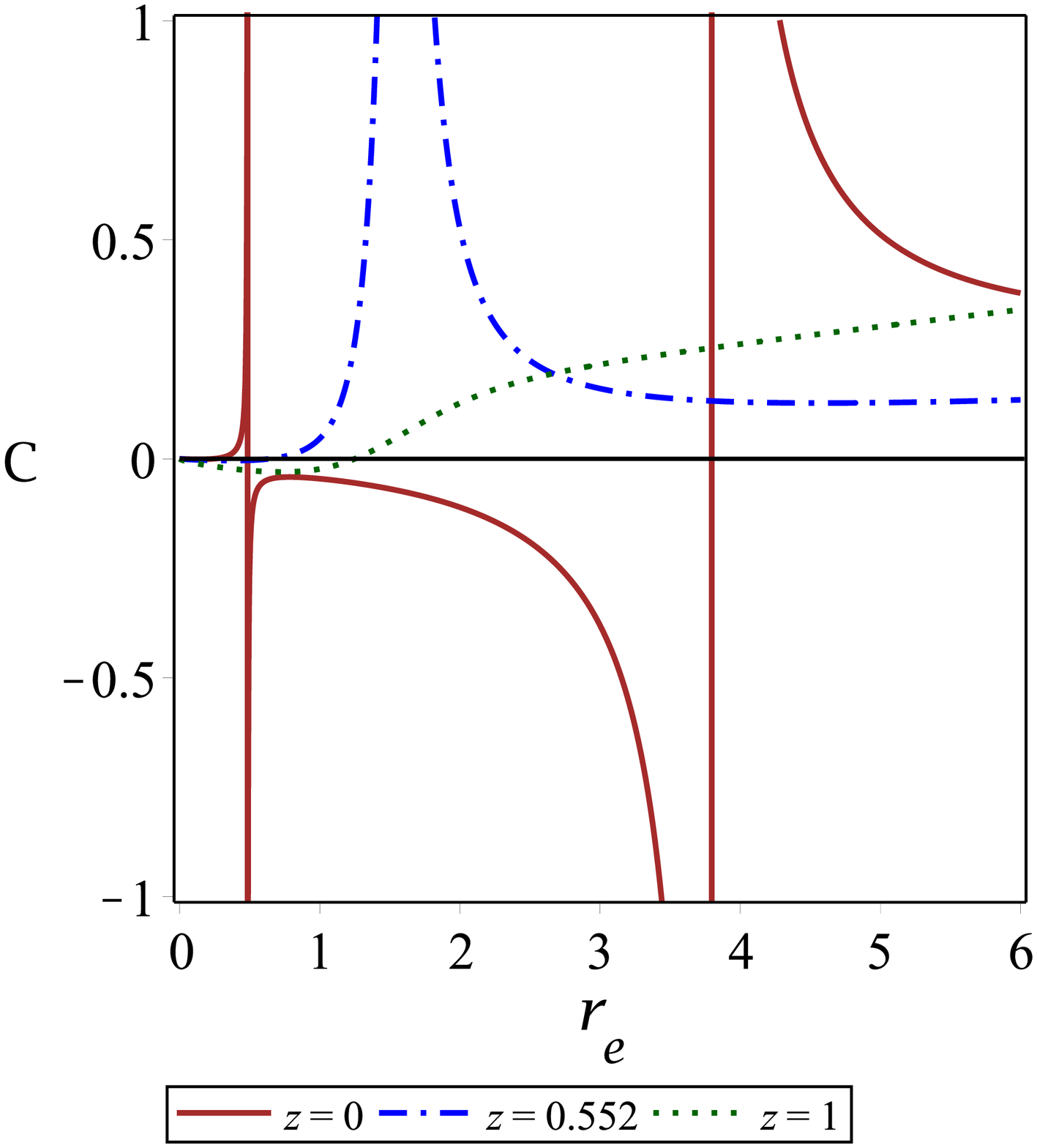}}
\subfloat[$ z=0.5 $, $ w=1.5 $, $ J=0.5 $, $ q=0.2 $, $ r_{0} =0.1$]{
        \includegraphics[width=0.312\textwidth]{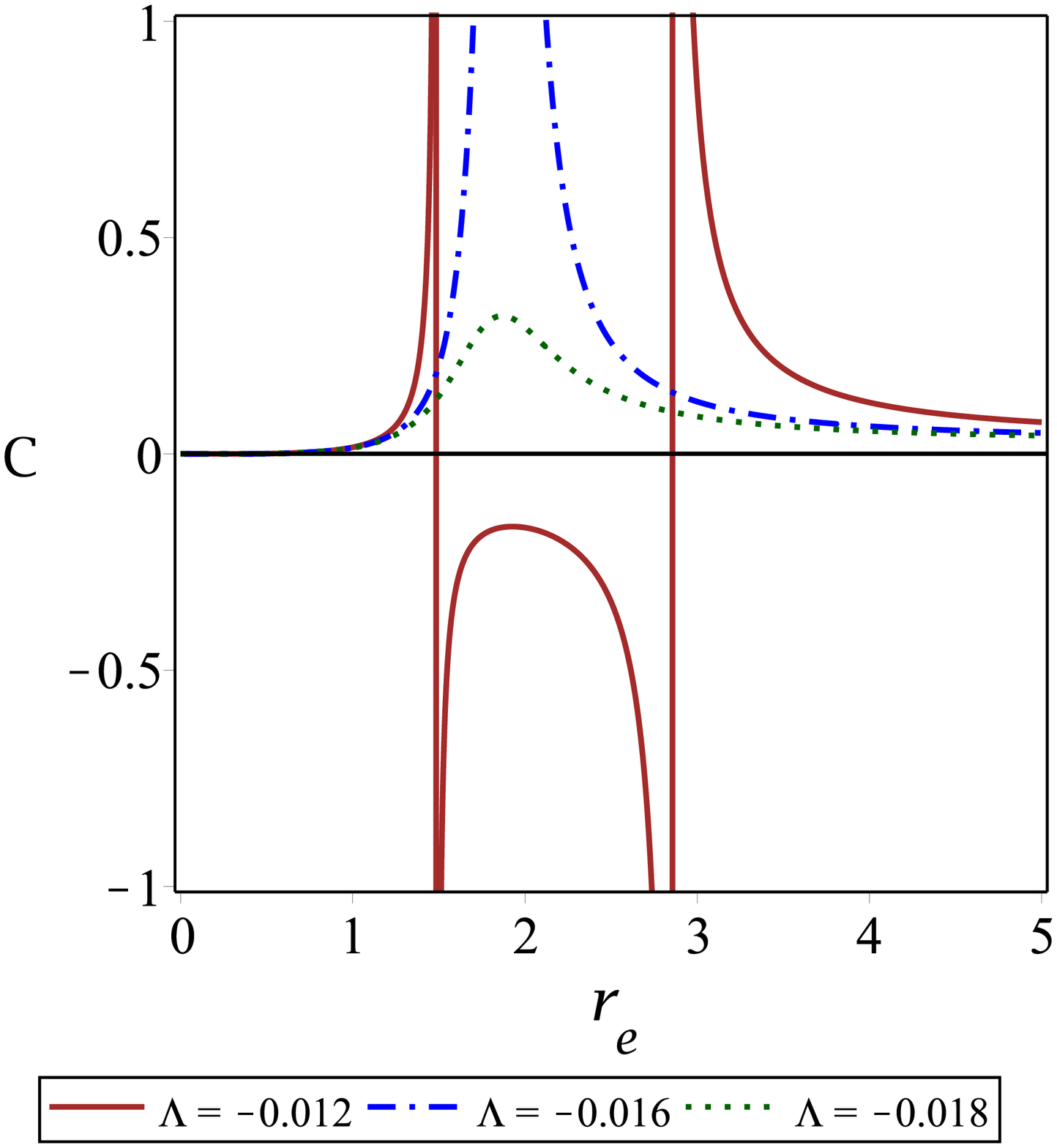}}\newline
\caption{Heat capacity versus $r_{e} $ for different values of parameters.}
\label{FigHeat}
\end{figure}


\subsection{Thermal stability and phase transition}\label{SubSecIIIB}

Heat capacity is one of the interesting thermodynamical quantities
which could be used to extract two important properties of the
solutions: I) Phase transition points. II) Thermal stability of
the solutions. The signature of heat capacity determines the
thermal stability/instability of the system. The positivity of
heat capacity represents the black hole being in thermally stable
state, while the opposite corresponds to thermally unstable case.
As was already mentioned, the heat capacity can provide a
mechanism to study the phase transition of the system. In fact,
this thermodynamic quantity can be employed to investigate two
distinctive points, bound and phase transition points. The bound
point is where the sign of temperature is changed. In other words,
the root of temperature (or heat capacity) indicates a limitation
point, which separates physical solutions (positive temperature)
from non-physical ones (negative temperature). The phase
transition point may be related to the divergence points of $C$.
Indeed, the divergencies of the heat capacity are where the system
goes under phase transition. The heat capacity is given by
\begin{eqnarray}
C_{P,Q,J} &=&T\left( \frac{\partial S}{\partial T}\right) _{P,Q,J}.
\label{C}
\end{eqnarray}

\begin{figure}[!htb]
\centering \subfloat[$ z=0.5 $, $ w=1.5 $, $ J=0.5 $, $ r_{0}
=0.1$, $ \Lambda=-0.01 $]{
        \includegraphics[width=0.31\textwidth]{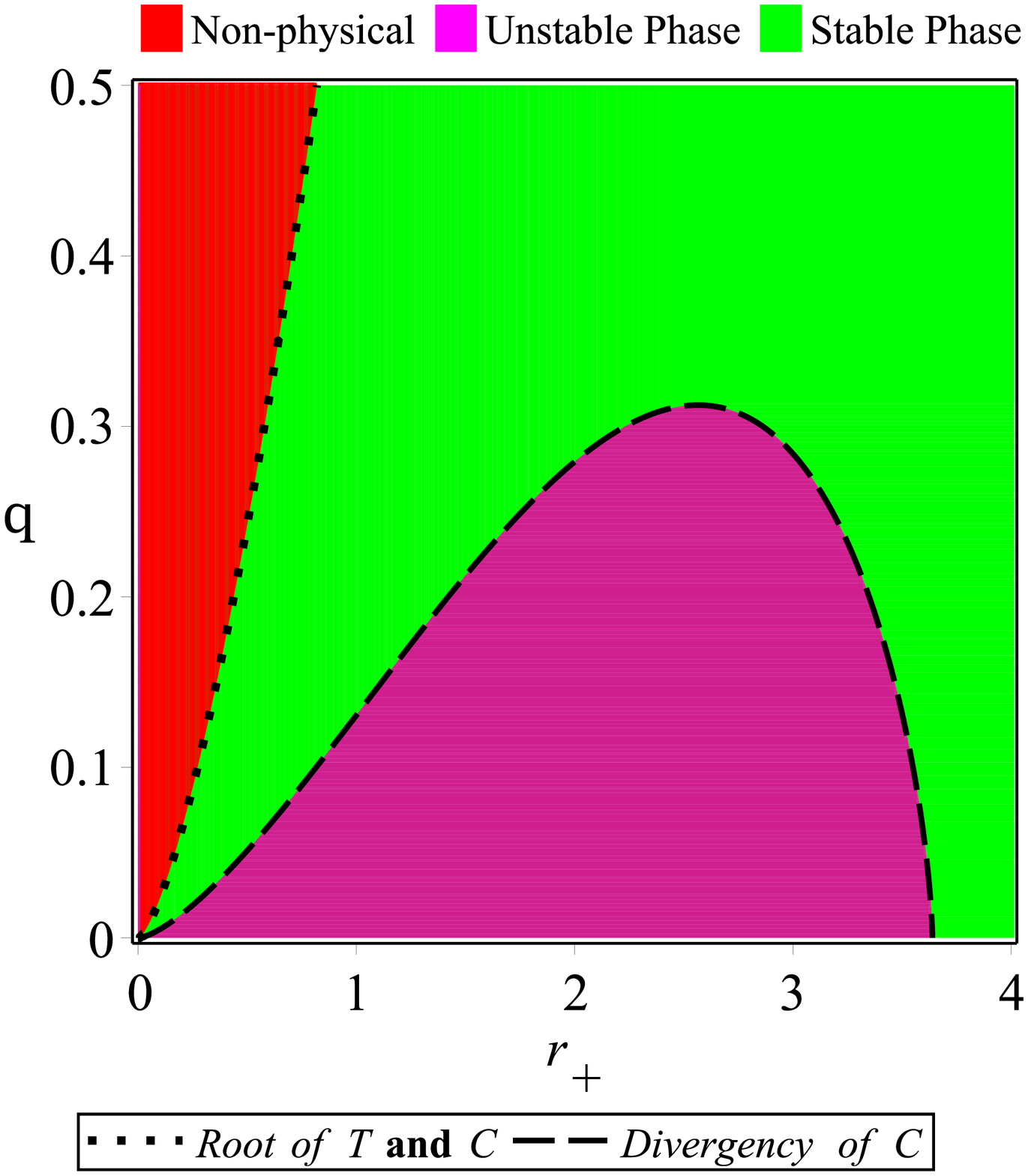}}
\subfloat[$ z=0.5 $, $ w=1.5 $, $ q=0.2 $, $ r_{0} =0.1$, $
\Lambda=-0.01 $]{
        \includegraphics[width=0.31\textwidth]{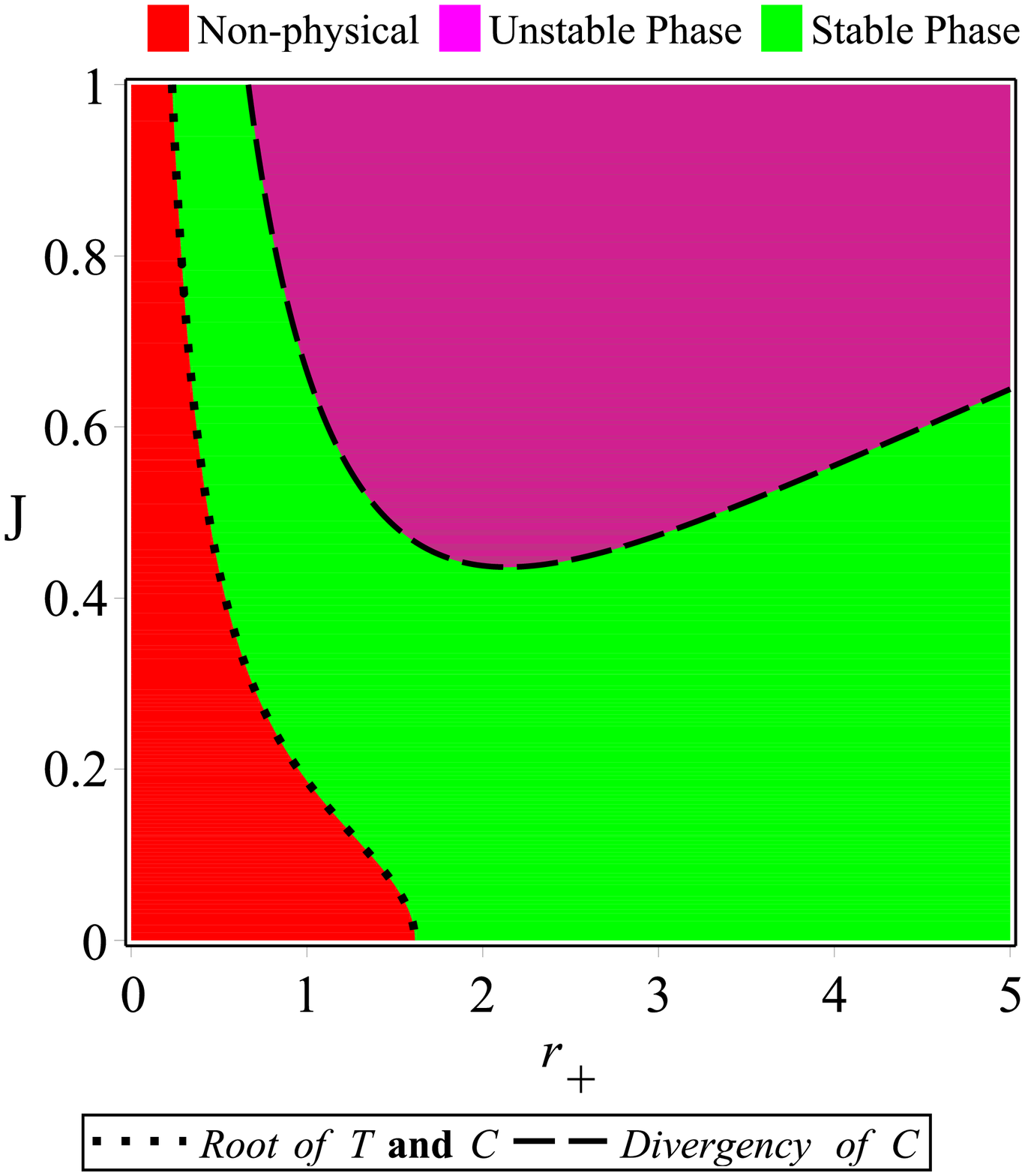}}
\subfloat[$ z=0.5 $, $ w=1.5 $, $ q=0.2 $, $ J =0.5$, $
\Lambda=-0.01 $]{
        \includegraphics[width=0.31\textwidth]{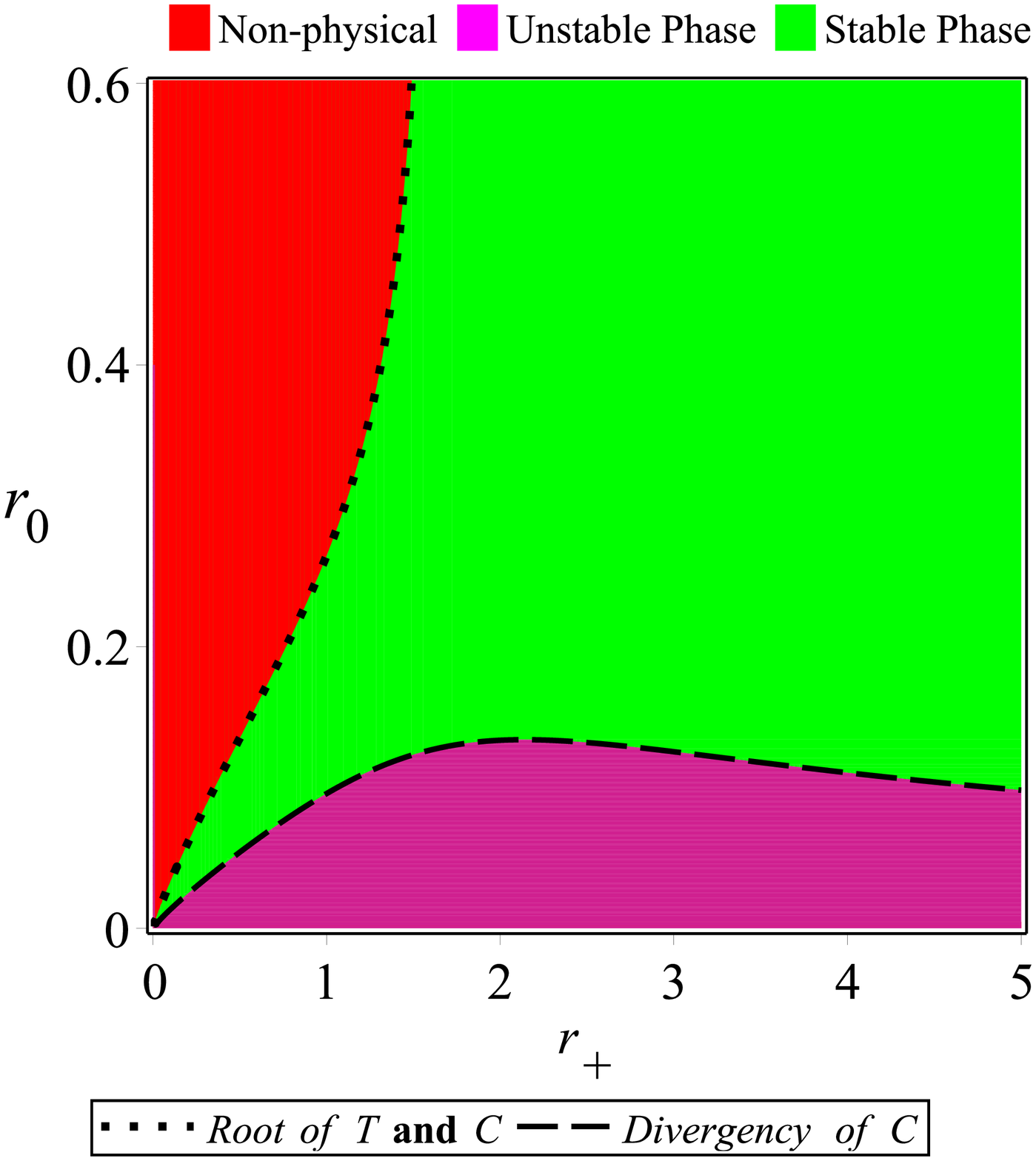}}\newline
\subfloat[$ z=0.5 $, $ J=0.5 $, $ q=0.2 $, $ r_{0} =0.1$, $
\Lambda=-0.01 $]{
        \includegraphics[width=0.305\textwidth]{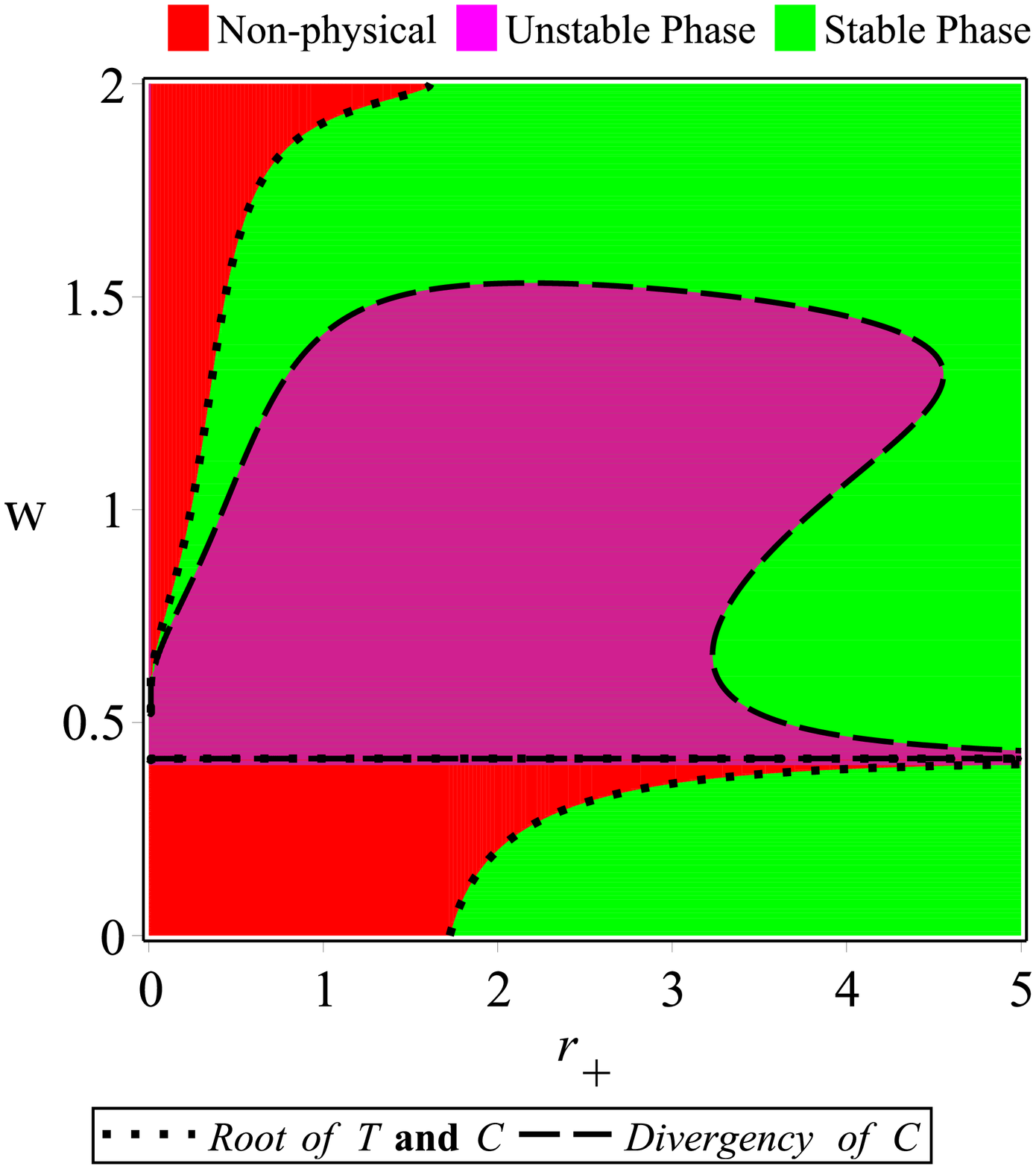}}
\subfloat[$w=1.5 $, $ J=0.5 $, $ q=0.2 $, $ r_{0} =0.1$, $
\Lambda=-0.01 $]{
        \includegraphics[width=0.31\textwidth]{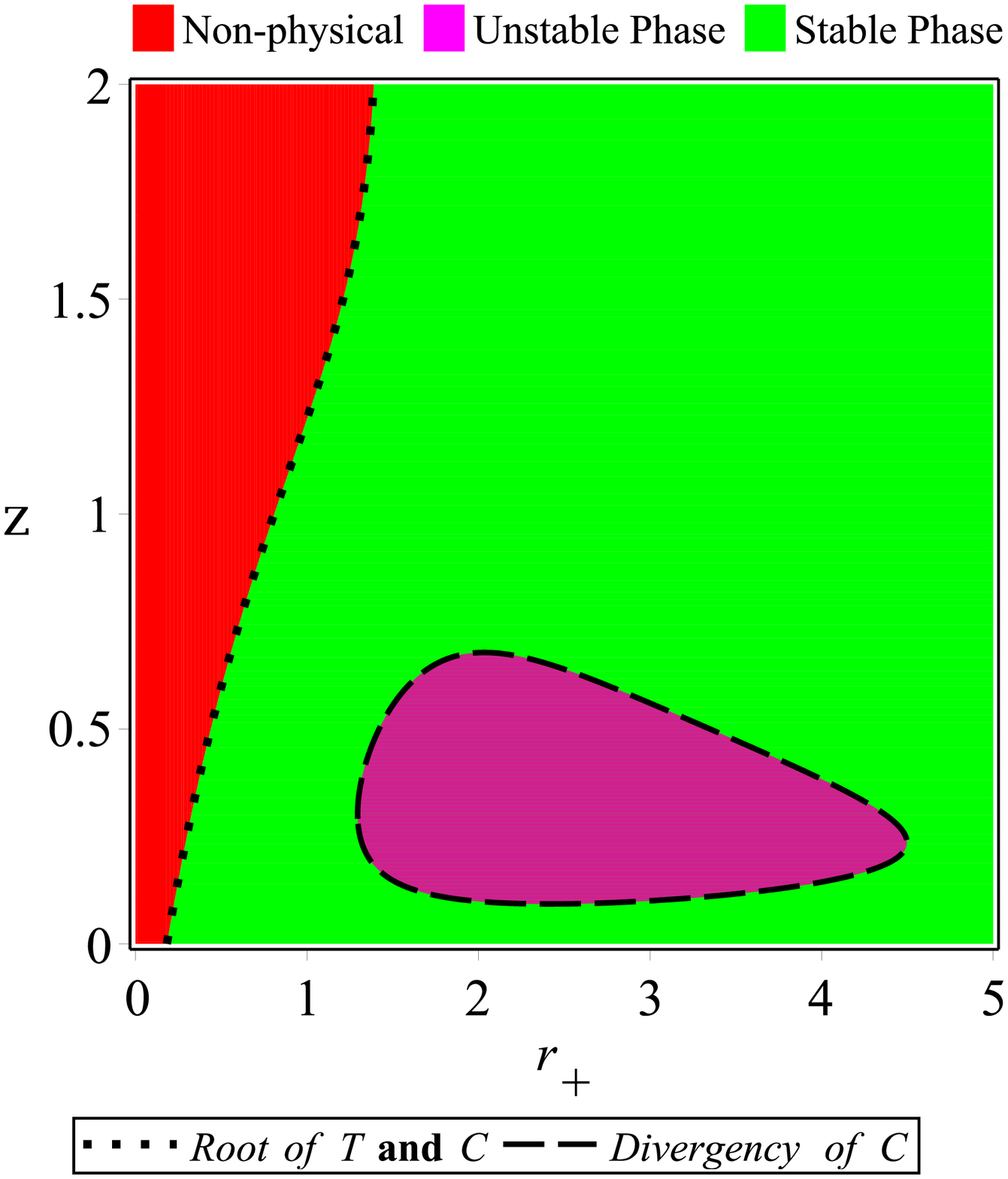}}
\subfloat[$ z=0.5 $, $ w=1.5 $, $ J=0.5 $, $ q=0.2 $, $ r_{0}
=0.1$]{
        \includegraphics[width=0.335\textwidth]{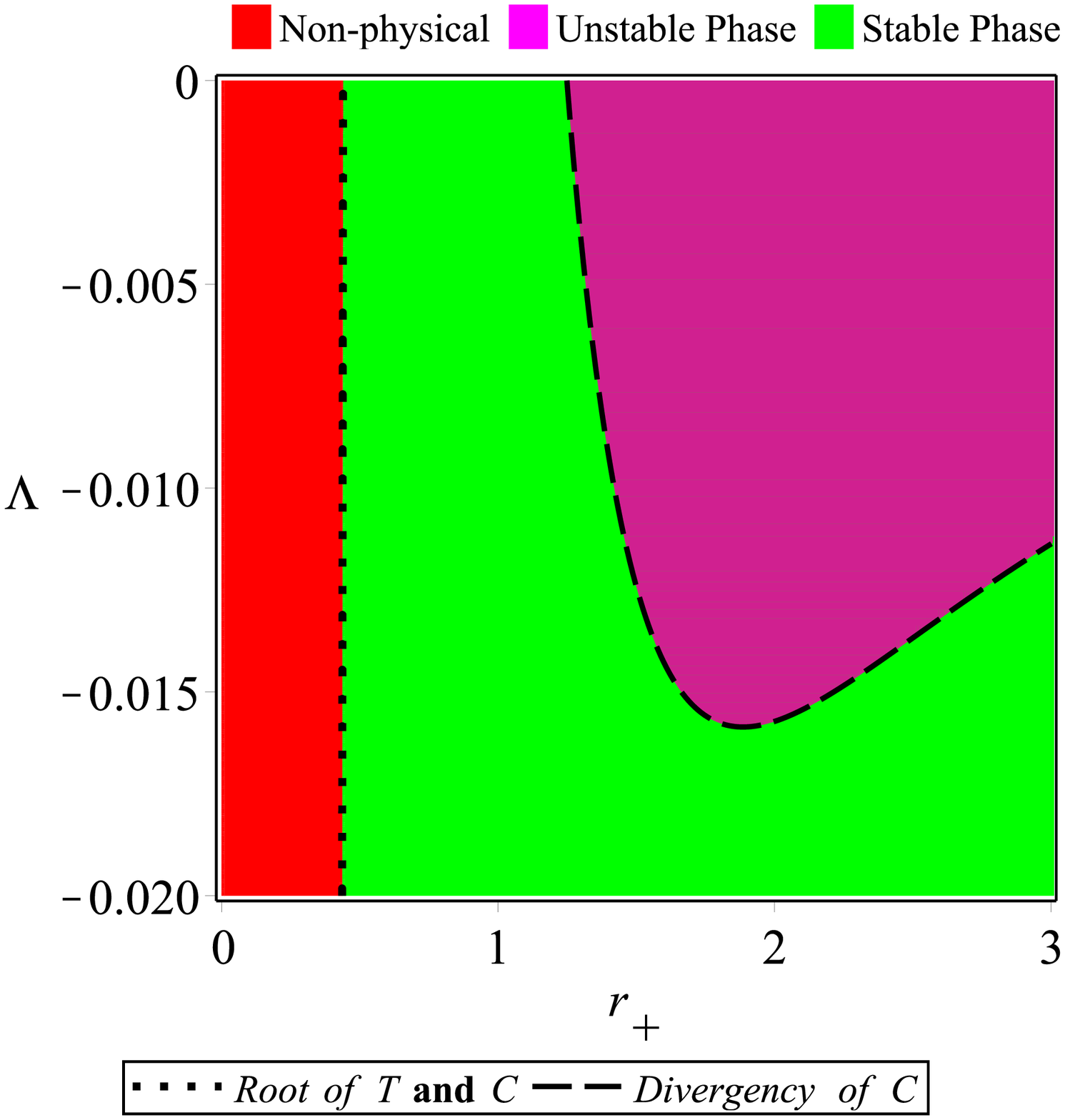}}\newline
\caption{Thermally stable and/or unstable regions of the black
holes.} \label{FigStability}
\end{figure}

Employing Eqs. (\ref{S}), (\ref{temp3F(R)}) and (\ref{C}), one can find
\begin{eqnarray}
C_{P,Q,J} &=&\frac{\pi r_{e}(\frac{r_{e}}{r_{0}})^{\gamma -\frac{z}{2}}}{%
B_{1} +B_{2} -B_{3} }((1+\gamma -\frac{z}{2})(\gamma -2\delta -4)B_{4} ),
\label{C2}
\end{eqnarray}
in which
\begin{eqnarray}
B_{1} &= &-3A_{1}r_{e}^{1+\delta } \left( w-2\right) ^{2}(4w-z-6)\left(
\frac{r_{e}}{r_{0}}\right) ^{2w-z}J^{2}  \nonumber \\
B_{2} &= &768\pi PA_{2}(2+z)(\gamma +2)r_{e}^{5+\delta }  \nonumber \\
B_{3} &= &-48A_{2} (A_{1}+8w)(\delta -\gamma )2^{-\frac{1}{4}}q^{ \frac{3}{2}%
} r_{+}^{3}  \nonumber \\
B_{4} &= &\frac{A_{1}J^{2}\left( w-2\right) ^{2}r_{e}^{1+\delta
}}{2}\left( \frac{r_{e}}{r_{0}}\right)
^{2w-z}-512r_{e}^{3}\left((\gamma +2)\pi Pr_{e}^{2+\delta
}-\frac{2^{-\frac{1 }{4}}q^{\frac{3}{2}}A_{2}\left( \gamma
+2\right) }{32}\right).  \nonumber
\end{eqnarray}

The behavior of the heat capacity with respect to the horizon radius is
addressed in Fig. \ref{FigHeat}. According to this figure, there are three
possibilities for the heat capacity:

Case I) One root: there are two phases of small and large black
holes. The small one is not physical due to the negativity of the
temperature. Whereas, the large black hole phase is thermally
stable.

Case II) One root and one divergency: in this case, there are
three phases small, medium, and large black holes. The small black
holes have negative temperatures and therefore, are not physical.
The medium and large phases are separated by a divergence point.
At this point, two phases of medium and large black holes are in
equilibrium and go from one to the other via a critical process.

Case III) One root and two divergencies: In this case, four
distinguishable phases can be observed for black holes; very
small, small, medium, and large black holes. For the very small
black hole phase, the temperature is negative and so this phase is
not a physical one. For small and large black hole phases, heat
capacity is positive and these two phases are thermally stable.
Medium black hole phases which are located between two
divergencies have negative heat capacity. Therefore, this phase is
not physical and accessible to the black holes.

Figure \ref{FigHeat} also displays the effects of different
parameters on the heat capacity. In general, we can highlight the
following effects of variation of different parameters on the heat
capacity.

i) According to Fig. \ref{FigHeat}(a), there is a critical value
for the electric charge where for values smaller than it, two
divergencies exist for the heat capacity. These two divergencies
coincide with each other for this critical value of the electric
charge. For the electric charges larger than this critical value,
no divergency appears in the structure of the heat capacity.

ii) Figure \ref{FigHeat}(b) shows that there is a critical value
for the angular momentum as well. The only difference is that for
angular momentums smaller than this critical value, no divergence
point is observed.

iii) The effect of parameter $r_{0}$ is depicted in Fig.
\ref{FigHeat}(c), indicating that its contribution to the heat
capacity is the same as the effect of the electric charge. In
other words, for the values of $r_{0}$ smaller than its critical
value, two divergence points appear. While, no divergency observe
for larger than the critical value.

iv) Figure \ref{FigHeat}(d) illustrates the effect of the exponent
$w$ on the heat capacity. Taking a closer look at this figure, one
can find that its effect is similar to the angular momentum. The
difference is that for fixed parameters $q$, $J$, $z$, $r_{0}$ and
$\Lambda$, there is a specific value of $w$ for which the heat
capacity has only one divergency without any root (see the dashed
curve of Fig. \ref{FigHeat}(d)). For this specific value, two
phases exist small and large black holes. Small black holes have a
negative heat capacity and are thermally unstable. Whereas, large
black holes are in a stable state due to the positivity of heat
capacity. For values of $w$ between this specific value and the
critical value, there are two divergencies for the heat capacity.

v) To study the effect of exponent $z$ and the cosmological
constant, we plot Figs. \ref{FigHeat}(e) and \ref{FigHeat}(f),
indicating that their effects are similar to the electric charge.

To have a more precise picture regarding the effects of different
parameters on thermal stability/instability of the solutions, we
have plotted Fig. \ref{FigStability}. As we see, by decreasing
(increasing) of the electric charge, cosmological constant and
parameter $r_{0}$ (angular momentum), the stability region of the
system decreases. In the case of exponents, the effect of each one
on the stability of the system depends on the value of the other.
In fact, the value of $w$ for which the system is thermally
unstable are quite dependent on the value of $z$ and vice versa.
\begin{figure}[!htb]
\centering
\subfloat[$ z=0.5 $, $ w=1.5 $, $ J=0.6 $, $ r_{0} =0.2$]{
        \includegraphics[width=0.31\textwidth]{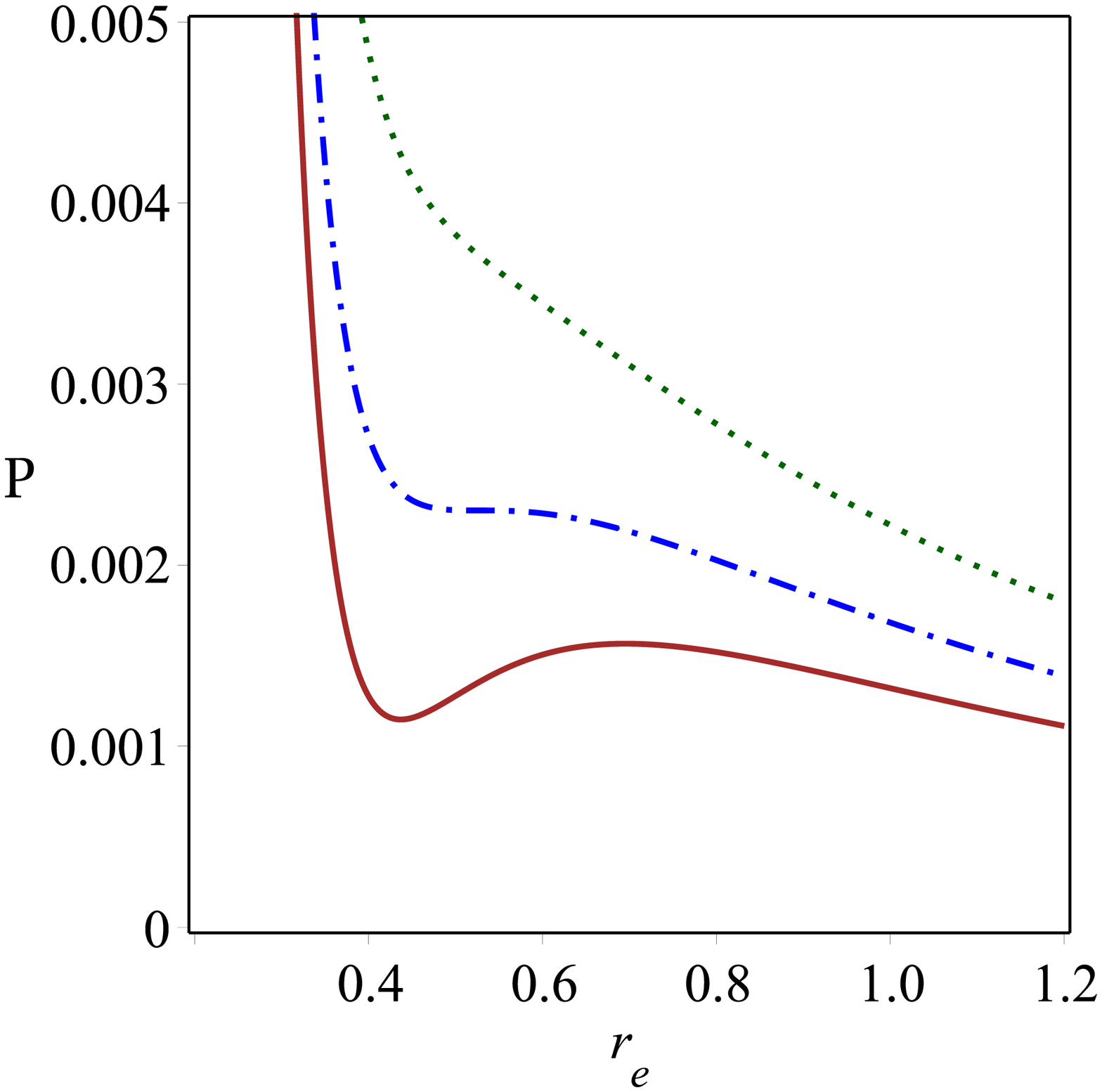}}
\subfloat[$ z=0.5 $, $ w=1.5 $, $ J=0.6 $, $ r_{0} =0.2$]{
        \includegraphics[width=0.31\textwidth]{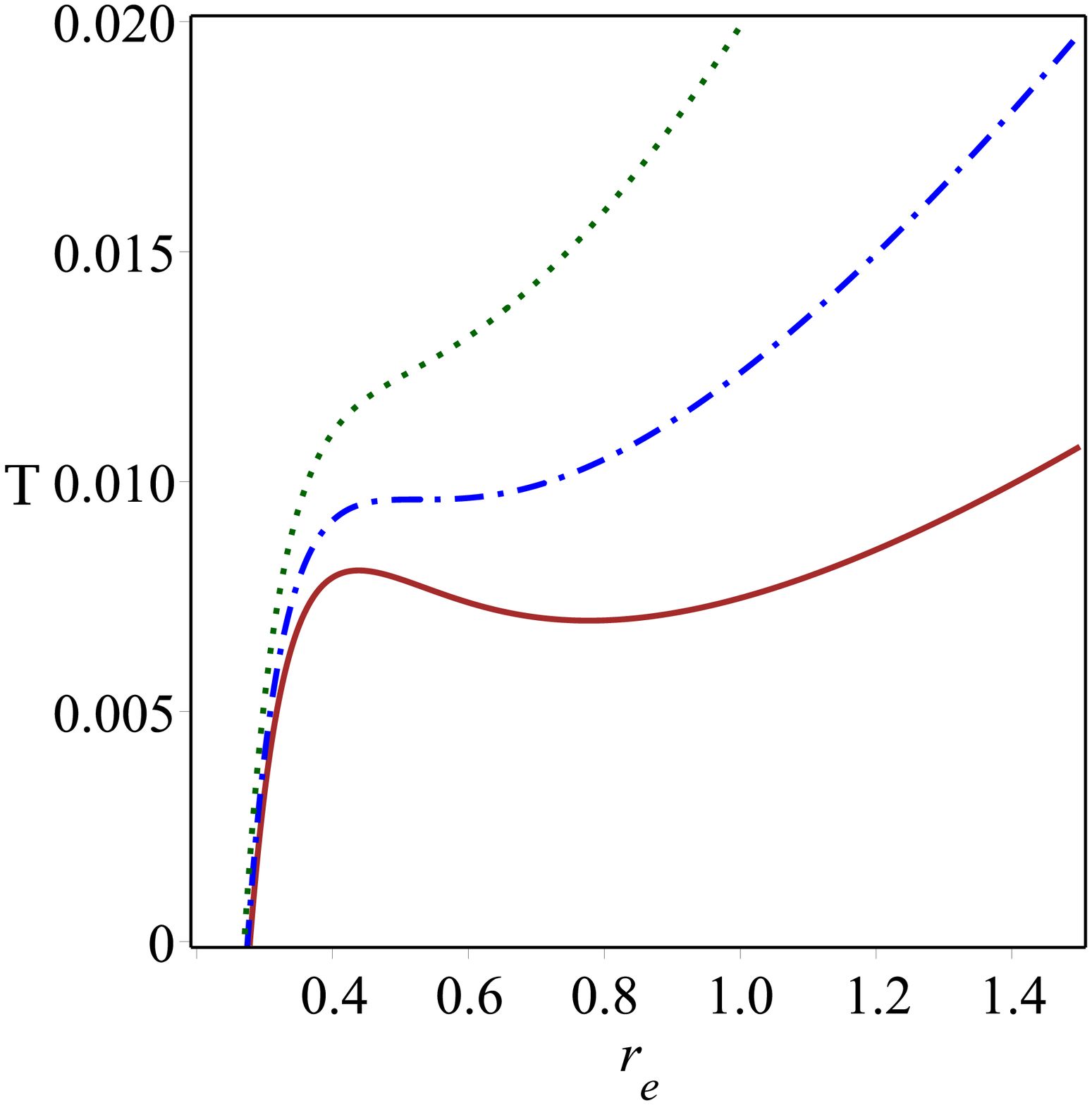}}
\subfloat[$ z=0.5 $, $ w=1.5 $, $ J=0.6 $, $ r_{0} =0.2$]{
        \includegraphics[width=0.32\textwidth]{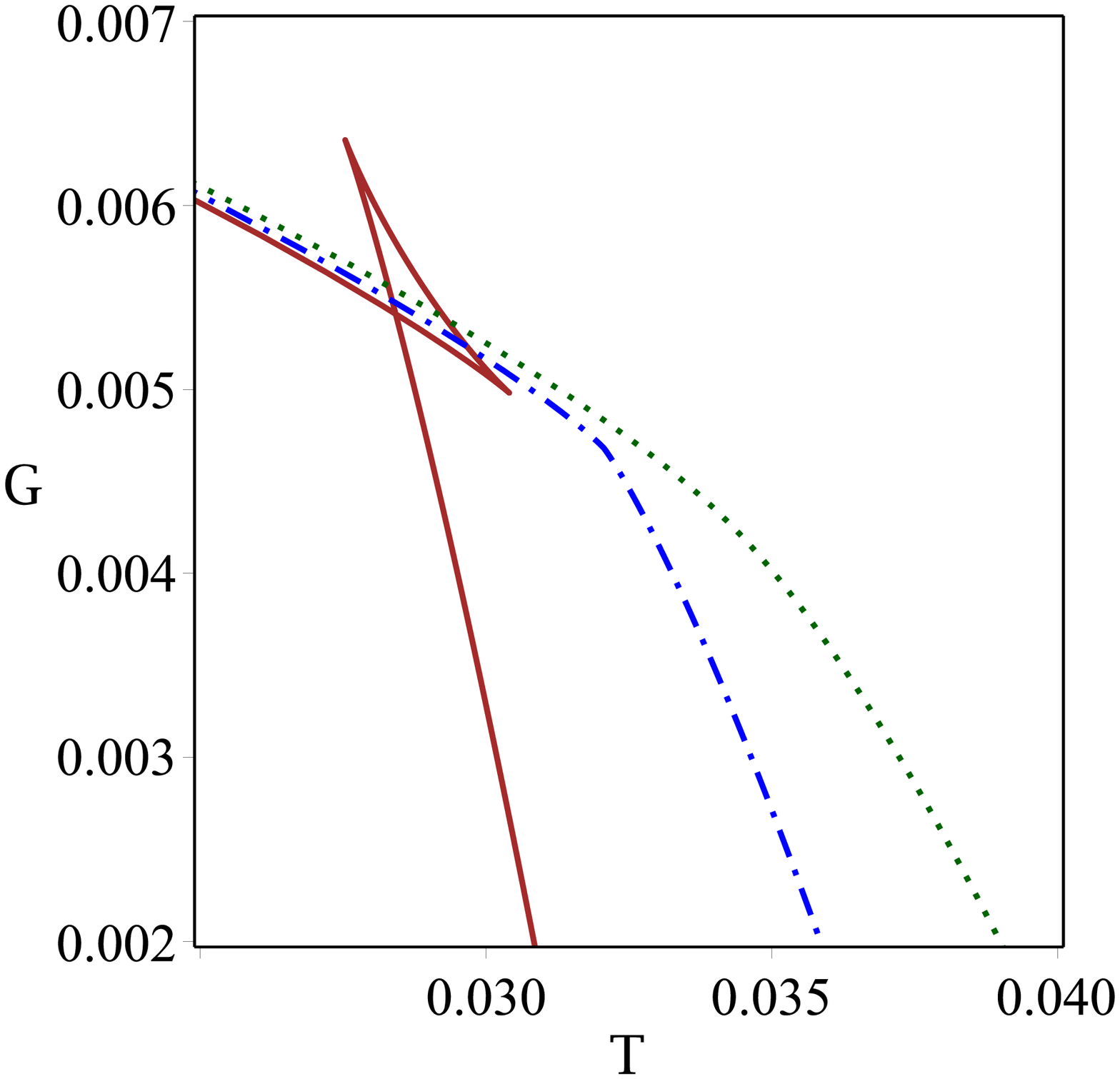}}\newline
\subfloat[$ z=1 $, $ w=0.5 $, $ J=0.3 $, $ r_{0} =1$]{
        \includegraphics[width=0.315\textwidth]{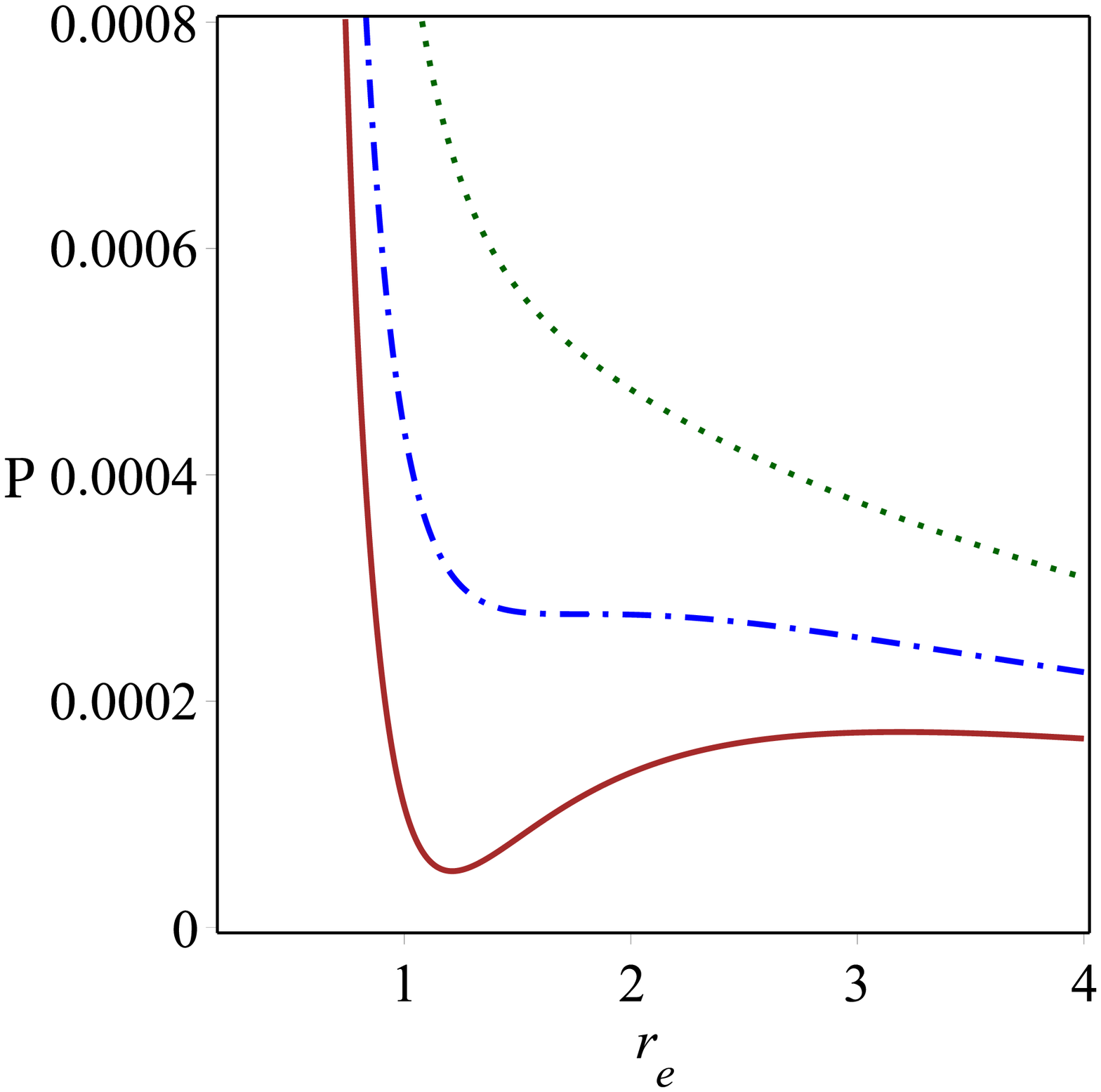}}
\subfloat[$ z=1 $, $ w=0.5 $, $ J=0.3 $, $ r_{0} =1$]{
        \includegraphics[width=0.31\textwidth]{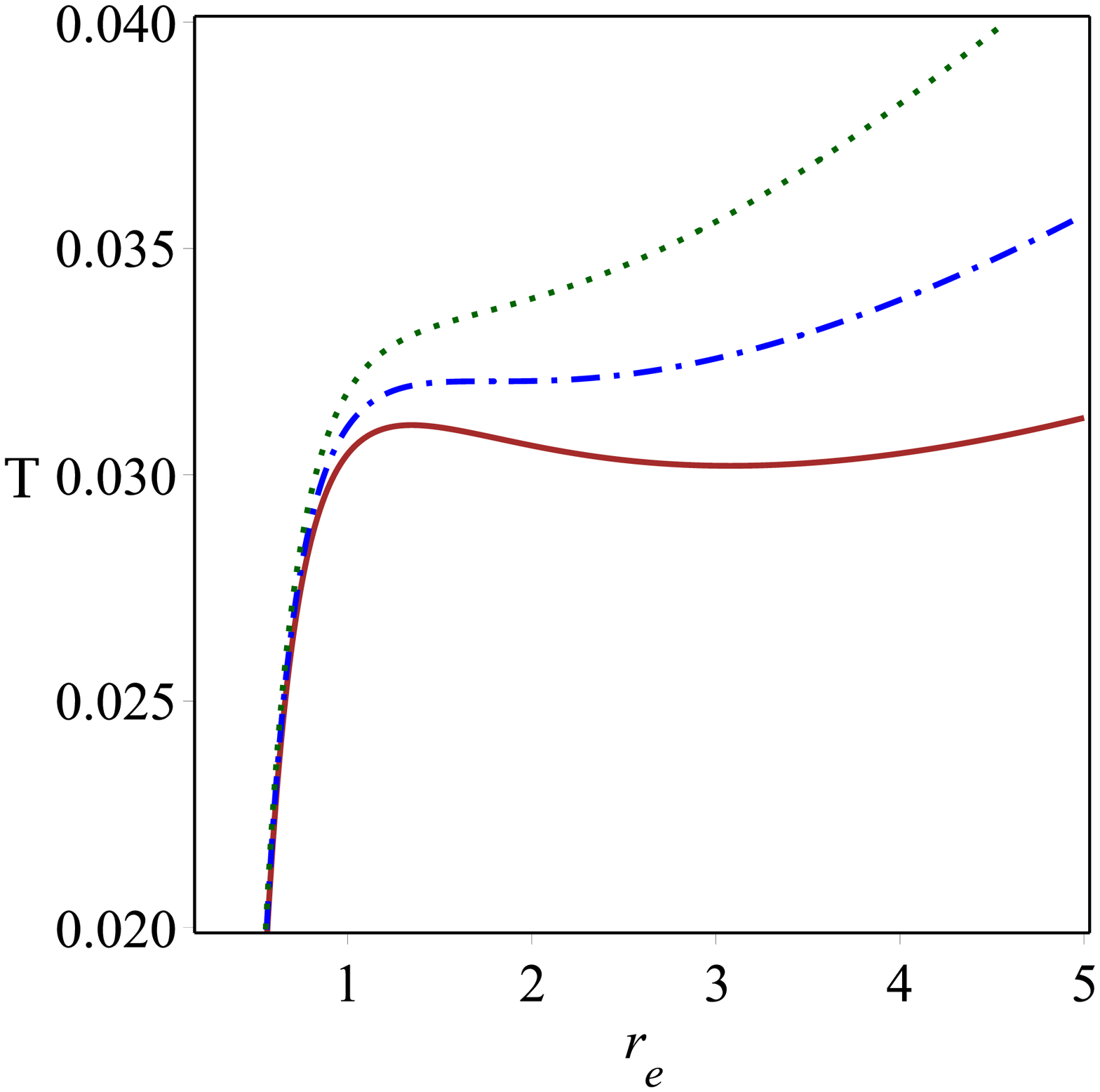}}
\subfloat[$ z=1 $, $ w=0.5 $, $ J=0.3 $, $ r_{0} =1$]{
        \includegraphics[width=0.33\textwidth]{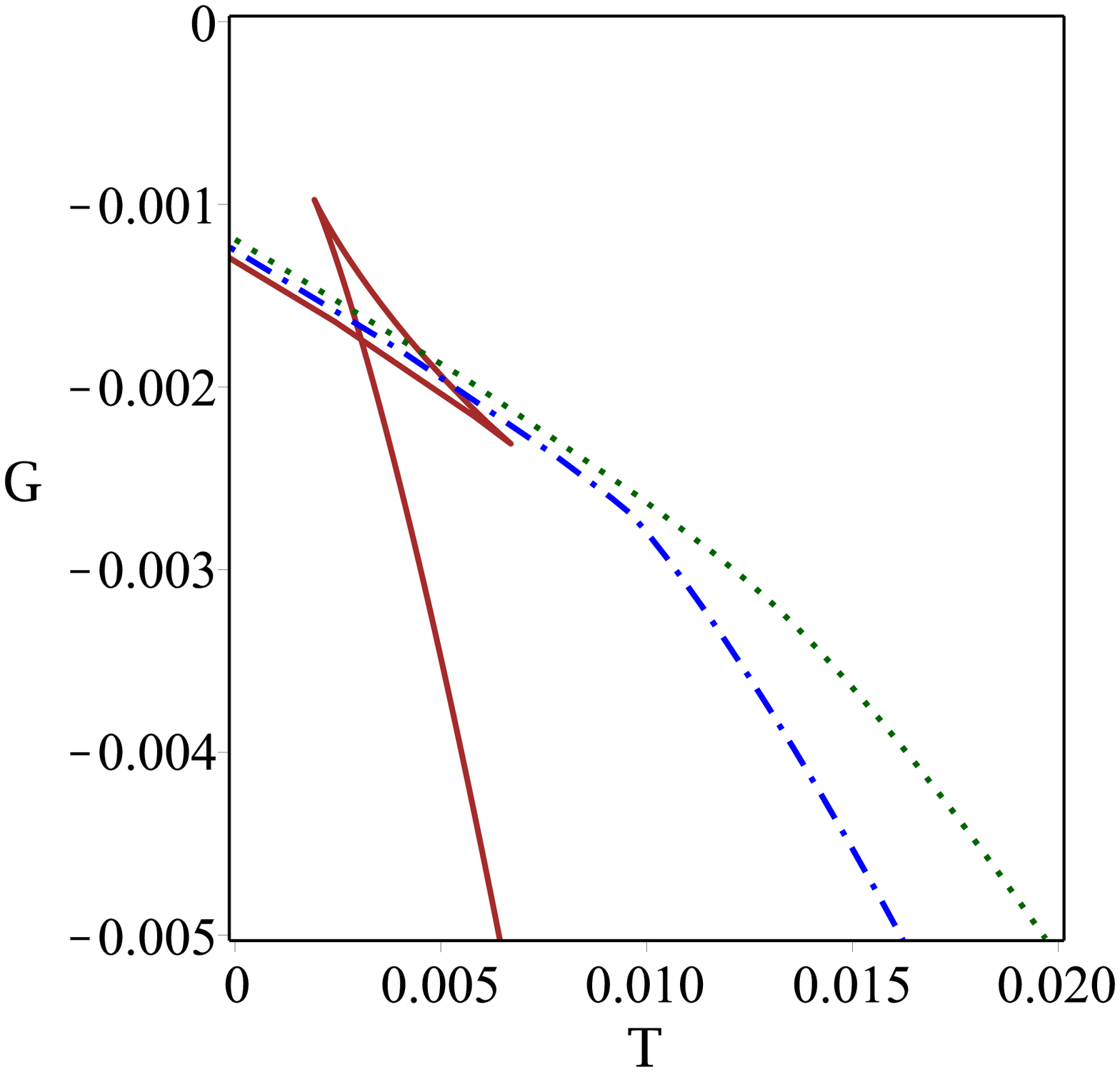}}\newline
\caption{van der Waals like phase diagrams for $q=0.1$. Left panels: $P -
r_{e}$ diagram for $T<T_{c} $ (continuous line), $T=T_{c} $ (dash-dotted
line), and $T>T_{c} $ (dotted line). Middle panels: $T - r_{e}$ diagram for $%
P<P_{c} $ (continuous line), $P=P_{c} $ (dash-dotted line), and $P>P_{c} $
(dotted line). Right panels: $G - T$ diagram for $P<P_{c} $ (continuous
line), $P=P_{c}$ (dash-dotted line) and $P>P_{c} $ (dotted line).}
\label{FigWaals1}
\end{figure}

\subsection{van der Waals like behavior}\label{SubSecIIIC}

Here, we look for the possibility of the existence of van der
Waals-like phase transition for the black holes. We also extract
critical thermodynamic quantities and analyze the effects of black
hole parameters on critical values. To do so, we require to
determine the equation of state which is obtained by writing down
the pressure as a function of the temperature and thermodynamic
volume. Since there is a direct relationship between the
thermodynamic volume and the horizon radius, we use the horizon
radius instead of using volume in the equation of state. From Eq.
(\ref{temp3F(R)}), the equation of state is obtained as
\begin{equation}
P=\frac{A_{1}J^{2}\left( w-2\right) ^{2} \left( \frac{r_{e}}{r_{0}}\right)
^{2w-z}}{256 A_{2}\pi r_{e}^{4} }+\frac{2^{-\frac{1}{4}}q^{\frac{3}{2}%
}\left( \delta -\gamma \right) }{8\pi \left( \gamma +2\right) r_{e}^{\delta
+2}}+ \frac{T}{2r_{e}\left( \frac{r_{e}}{r_{0}}\right) ^{\frac{z}{2}}\left(
\gamma +2\right) }.  \label{P(T)}
\end{equation}

The behaviors of the pressure and temperature under variation of
the event horizon radius are depicted in Fig. \ref{FigWaals1}.
Evidently, a van der Waals-like phase transition can be observed
for these black holes by suitable choices of different parameters.
Indeed, the presence of subcritical isobars in $T - r_{e}$ and
isothermal diagrams in $P - r_{e}$ confirm the existence of van
der Waals-like phase transition. As we know, the van der Waals
fluid goes under a first order phase transition for temperatures
smaller than the critical temperature $(T < T_{c})$. Whereas, at
the critical temperature, its phase transition is a second-order
one \cite{Kubiznak:2012}. The formation of the swallow-tail shape
in the $G-T$ diagram is another evidence of the first-order
(small-large) phase transition. Figures \ref{FigWaals1}(c) and
\ref{FigWaals1}(f), confirm the first-order phase transition for
our black hole solutions.

Our analysis shows that a first-order phase transition occurs for
values of the exponents that satisfy the condition
$x_{1}<z+w<x_{2}$. It is worth mentioning that the values of
$x_{1}$ and $x_{2}$ are highly governed by the parameter $z$. Some
values of $x_{1}$ and $x_{2}$ for different values of $z$ are as
follows
\begin{eqnarray}  \label{firstorder}
z &=&0.5 ~~\longrightarrow ~~ 0.9<z+w<2.1 \\
z &=&1 ~~~~\longrightarrow ~~~ 1.35<z+w<2.75  \nonumber \\
z &=&1.5~~ \longrightarrow ~~ 1.8<z+w<3.35  \nonumber \\
z &=&2~~~~ \longrightarrow ~~~2.3<z+w<3.85 ,  \nonumber
\end{eqnarray}

\begin{figure}[!htb]
\centering
\subfloat[]{
        \includegraphics[width=0.31\textwidth]{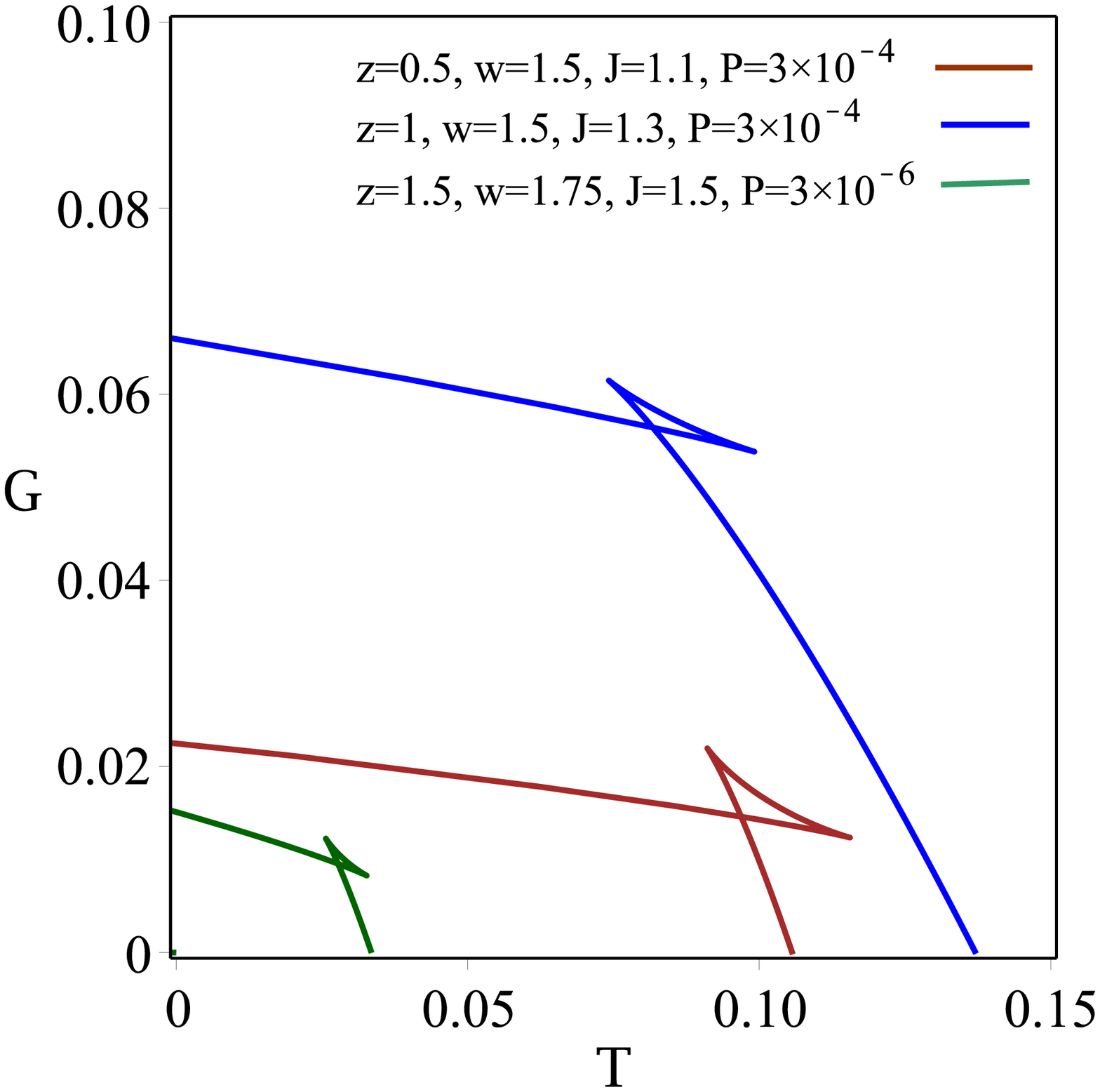}}
\subfloat[]{
        \includegraphics[width=0.33\textwidth]{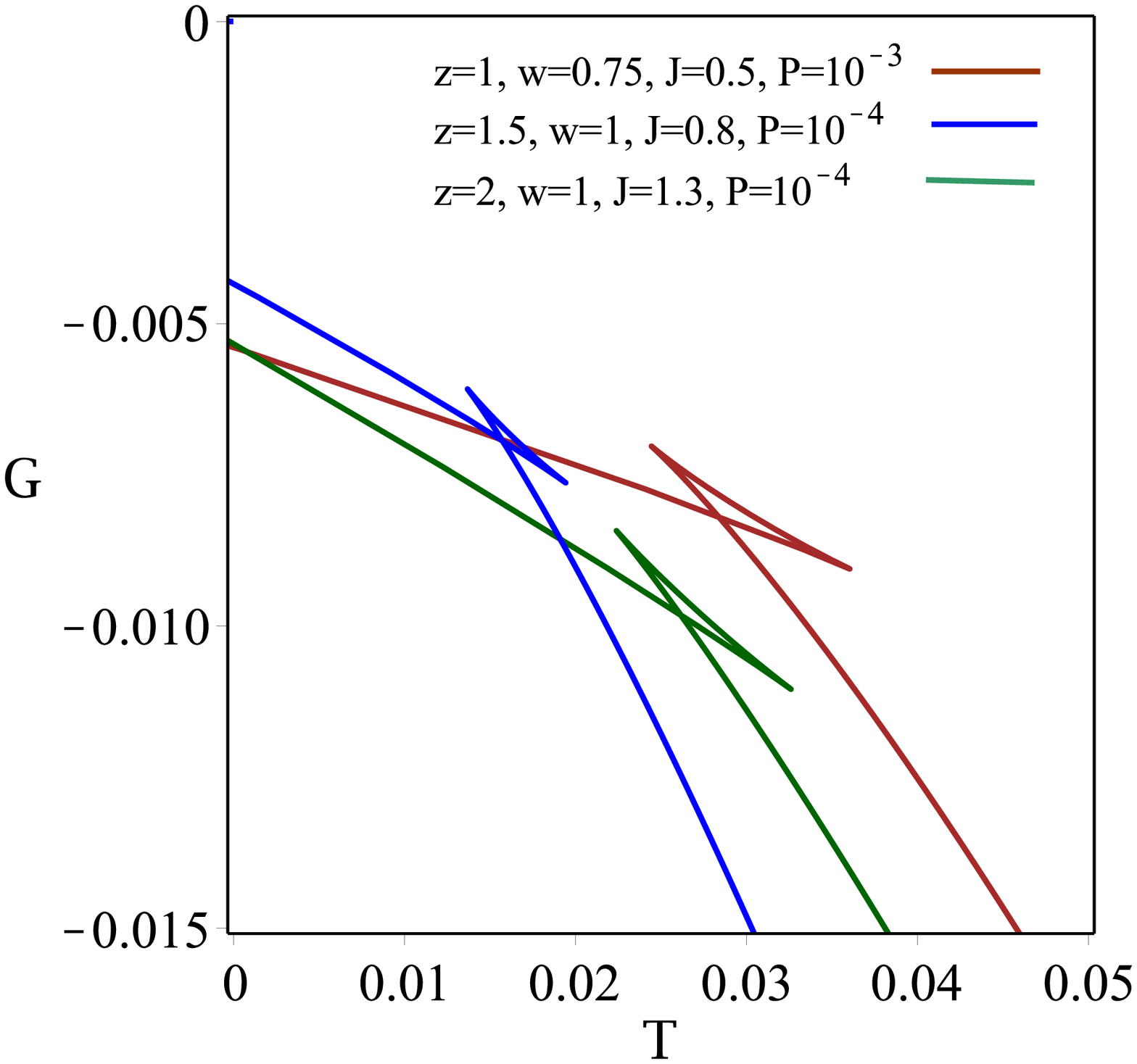}}\newline
\caption{$G - T$ diagram for $q=0.1 $ and $r_{0} =0.2$.}
\label{FigGibbs}
\end{figure}

A significant point here is that if $w>z$ in the mentioned region
(see the relation (\ref{firstorder})), the Gibbs free energy is
positive. Since the Gibbs free energy is obtained as $G=H-ST$,
this reveals the fact that the system is energy-dominated. But for
$z>w$, the Gibbs free energy is negative, indicating that the
system is entropy-dominated (see Fig. \ref{FigGibbs}).

To obtain the critical values of thermodynamic quantities, we use
the concept of the inflection point of isothermal $P-V$ diagram
given by
\begin{equation}
\left( \frac{\partial P}{\partial r_{e}}\right) _{T}=0,\ \ \ \left( \frac{%
\partial ^{2}P}{\partial r_{e}^{2}}\right) _{T}=0
\label{first and second critical}
\end{equation}

It is a matter of calculation to show that the critical horizon
radius (volume), temperature and pressure are given by
\begin{eqnarray}
r_{c}&=& \frac{32wA_{2}r_{0}^{2w-z}2^{-\frac{1}{4}}q^{\frac{3}{2}}\left(
\delta +2\right) \left( -2\delta +z-2\right) \left( \gamma -\delta \right) }{%
A_{1}J^{2}\left( w-2\right) ^{2} (A_{2}-2)(2w-z-4)} , \nonumber \\
T_{c}&=& \frac{2^{-\frac{1}{4}}q^{\frac{3}{2}}\left( \delta
+2\right) \left( \gamma -\delta \right)\left(
\frac{r_{c}}{r_{0}}\right) ^{\frac{z}{2}} r_{c}^{\delta
-2}}{2(2+z)\pi r_{c}^{3}}+ \frac{A_{1} \left( 2w-z-4\right)
\left( w-2\right) ^{2}J^{2}\left( \frac{r_{c}}{r_{0}}\right) ^{2w-\frac{z}{2}%
}}{32A_{2}(2+z)\pi r_{c}^{3}} ,  \nonumber \\
P_{c}&=&\frac{A_{1}J^{2}\left( w-2\right) ^{2}(A_{2}-2) \left( \frac{r_{c}}{%
r_{0}}\right) ^{2w-z}}{256 \pi wA_{2}r_{c}^{4}\left( \gamma +2\right)(2+z)}-%
\frac{2^{-\frac{1}{4}}q^{\frac{3}{2}}\left( \gamma -\delta \right)
(z-2-2\gamma )}{8(\gamma +2)(2+z)\pi r_{c}^{2+\delta }}.
\label{critical pressure}
\end{eqnarray}

\begin{table*}[htb!]
\caption{Critical values for the variation of $q$, $J$, $r_{0}$,
$w$ and $z$.} \label{table4}\centering
\begin{tabular}{||c|c|c|c|c||}
\hline
{\footnotesize $q$ \hspace{0.3cm}} & \hspace{0.3cm}$0.1$ \hspace{0.3cm} &
\hspace{0.3cm} $0.12$\hspace{0.3cm} & \hspace{0.3cm} $0.14$\hspace{0.3cm} &
\hspace{0.3cm}$0.16$\hspace{0.3cm} \\ \hline
$P_{c}$ ($z=0.5$, $w=1.5$, $J=0.6 $, $r_{0} =0.2$) & $27\times 10^{-5}$ & $%
22\times 10^{-5}$ & $19\times 10^{-5}$ & $17\times 10^{-5}$ \\ \hline
$T_{c}$ ($z=0.5$, $w=1.5$, $J=0.6 $, $r_{0} =0.2$) & $0.032 $ & $0.031$ & $%
0.03$ & $0.029$ \\ \hline
$r_{c}$ ($z=0.5$, $w=1.5$, $J=0.6 $, $r_{0} =0.2$) & $1.77 $ & $2.01$ & $2.24
$ & $2.45$ \\ \hline
$P_{c}r_{c}/T_{c}$ ($z=0.5$, $w=1.5$, $J=0.6 $, $r_{0} =0.2$) & $0.0153 $ & $%
0.0148$ & $0.0144$ & $0.0141$ \\ \hline\hline
&  &  &  &  \\
{\footnotesize $J$ \hspace{0.3cm}} & \hspace{0.3cm}$0.5$ \hspace{0.3cm} &
\hspace{0.3cm} $0.6$\hspace{0.3cm} & \hspace{0.3cm} $0.7$\hspace{0.3cm} &
\hspace{0.3cm}$0.8$\hspace{0.3cm} \\ \hline
$P_{c}$ ($z=0.5$, $w=1.5$, $q=0.1 $, $r_{0} =0.2$) & $14\times 10^{-5}$ & $%
27\times 10^{-5}$ & $46\times 10^{-5}$ & $73\times 10^{-5}$ \\ \hline
$T_{c}$ ($z=0.5$, $w=1.5$, $q=0.1 $, $r_{0} =0.2$) & $0.021 $ & $0.032$ & $%
0.045$ & $0.060$ \\ \hline
$r_{c}$ ($z=0.5$, $w=1.5$, $q=0.1 $, $r_{0} =0.2$) & $2.10 $ & $1.77$ & $1.54
$ & $2.45$ \\ \hline
$P_{c}r_{c}/T_{c}$ ($z=0.5$, $w=1.5$, $q=0.1 $, $r_{0} =0.2$) & $0.0147 $ & $%
0.0153$ & $0.0159$ & $0.0163$ \\ \hline\hline
&  &  &  &  \\
{\footnotesize $r_{0}$ \hspace{0.3cm}} & \hspace{0.3cm}$0.1$ \hspace{0.3cm}
& \hspace{0.3cm} $0.15$\hspace{0.3cm} & \hspace{0.3cm} $0.2$\hspace{0.3cm} &
\hspace{0.3cm}$0.25$\hspace{0.3cm} \\ \hline
$P_{c}$ ($z=0.5$, $w=1.5$, $J=0.6 $, $q =0.1$) & $51\times 10^{-4}$ & $%
93\times 10^{-5}$ & $27\times 10^{-5}$ & $11\times 10^{-5}$ \\ \hline
$T_{c}$ ($z=0.5$, $w=1.5$, $J=0.6 $, $q =0.1$) & $0.263 $ & $0.076$ & $0.032$
& $0.016$ \\ \hline
$r_{c}$ ($z=0.5$, $w=1.5$, $J=0.6 $, $q =0.1$) & $0.801 $ & $1.27$ & $1.77$
& $2.29$ \\ \hline
$P_{c}r_{c}/T_{c}$ ($z=0.5$, $w=1.5$, $J=0.6 $, $q =0.1$) & $0.0157 $ & $%
0.0155$ & $0.0153$ & $0.0152$ \\ \hline\hline
&  &  &  &  \\
{\footnotesize $w$ \hspace{0.3cm}} & \hspace{0.3cm}$1$ \hspace{0.3cm} &
\hspace{0.3cm} $1.25$\hspace{0.3cm} & \hspace{0.3cm} $1.5$\hspace{0.3cm} &
\hspace{0.3cm}$1.6$\hspace{0.3cm} \\ \hline
$P_{c}$ ($z=0.5$, $J=0.6$, $q=0.1 $, $r_{0} =0.2$) & $0.038$ & $31\times
10^{-4}$ & $27\times 10^{-5}$ & $19\times 10^{-6}$ \\ \hline
$T_{c}$ ($z=0.5$, $J=0.6$, $q=0.1 $, $r_{0} =0.2$) & $0.188 $ & $0.059$ & $%
0.032$ & $0.026$ \\ \hline
$r_{c}$ ($z=0.5$, $J=0.6$, $q=0.1 $, $r_{0} =0.2$) & $0.342 $ & $0.787$ & $%
1.77$ & $3.87$ \\ \hline
$P_{c}r_{c}/T_{c}$ ($z=0.5$, $J=0.6$, $q=0.1 $, $r_{0} =0.2$) & $0.0695 $ & $%
0.0423$ & $0.0153$ & $0.0029$ \\ \hline\hline
&  &  &  &  \\
{\footnotesize $z$ \hspace{0.3cm}} & \hspace{0.3cm}$0.5$ \hspace{0.3cm} &
\hspace{0.3cm} $1$\hspace{0.3cm} & \hspace{0.3cm} $1.5$\hspace{0.3cm} &
\hspace{0.3cm}$2$\hspace{0.3cm} \\ \hline
$P_{c}$ ($w=1.5$, $J=0.6$, $q=0.1 $, $r_{0} =0.2$) & $27\times 10^{-5}$ & $%
91\times 10^{-6}$ & $16\times 10^{-6}$ & $19\times 10^{-7}$ \\ \hline
$T_{c}$ ($w=1.5$, $J=0.6$, $q=0.1 $, $r_{0} =0.2$) & $0.032$ & $0.016$ & $%
0.007$ & $0.002$ \\ \hline
$r_{c}$ ($w=1.5$, $J=0.6$, $q=0.1 $, $r_{0} =0.2$) & $1.77 $ & $2.07$ & $2.64
$ & $3.45$ \\ \hline
$P_{c}r_{c}/T_{c}$ ($w=1.5$, $J=0.6$, $q=0.1 $, $r_{0} =0.2$) & $0.0153 $ & $%
0.0113$ & $0.0057$ & $0.0023$ \\ \hline\hline
\end{tabular}%
\end{table*}

Table \ref{table4} shows how critical quantities and universal
critical ratio $\left(\frac{P_{c}r_{c}}{T_{c}}\right)$ change
under variation of black hole parameters. From this table, one can
find that as the electric charge increases, the critical pressure,
temperature, and universal critical ratio decrease, whereas the
critical horizon radius (volume) increases. Regarding the effect
of angular momentum on the critical quantities, one can see that
its effect is opposite of that of the electric charge. Studying
the effects of exponents and parameter $r_{0}$ indicates that
their contribution to critical values is the same as the electric
charge. In other words, the critical volume is an increasing
function of these three parameters, whereas the critical pressure,
temperature, and universal critical ratio are decreasing functions
of them.

\subsection{Heat Engine}\label{SubSecIIID}

As the final step, we would like to consider the Lifshitz rotating
black hole as a heat engine and discuss its efficiency.

A heat engine is a physical system that works between two hot and
cold reservoirs and its main role is transferring heat from the
hot reservoir to the cold one. The total mechanical work done, by
the First Law, is $W = Q_{H}$ - $Q_{C}$. So, the efficiency of the
heat engine is
\begin{equation}
\eta = \frac{W}{Q_{H}}=1-\frac{Q_{C}}{Q_{H}}.
\end{equation}

\begin{figure}[!htb]
\centering
\includegraphics[width=0.36\textwidth]{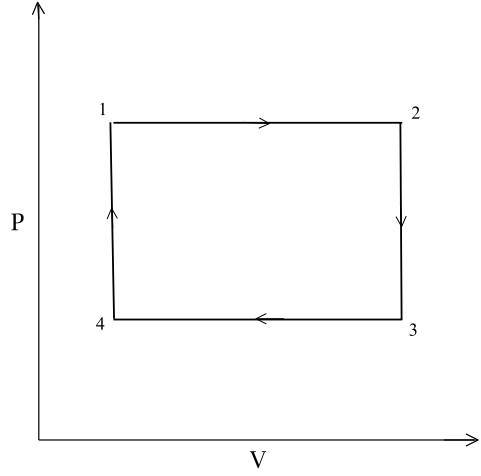}
\caption{Our engine cycle} \label{FigNCa}
\end{figure}

In order to calculate the efficiency, one may use the heat
capacity. According to Eqs. (\ref{S}) and (\ref{Volume}), the
entropy and thermodynamic volume are related to the horizon
radius. So, these two quantities are dependent to each other for
this kind of solution. This shows that the specific heat at
constant volume vanishes $C_{V}=0$ which is the "isochore equals
adiabat" result \cite{CVJohnson}. In this case, the specific heat
at constant pressure is not zero. An explicit expression for
$C_{P}$ would suggest that we can consider a rectangular cycle
such as Fig. \ref{FigNCa}, involving two isobars (paths of $1
\rightarrow 2$ and $3 \rightarrow 4$) and two isochores/adiabats
(paths of $2 \rightarrow 3$ and $4 \rightarrow 1$). Figure
\ref{FigNCa} shows a schematic of the proposed cycle. We can
calculate the work done along the heat cycle as
\begin{eqnarray}
W &=&\oint PdV=W_{1\longrightarrow 2}+W_{2\longrightarrow
3}+W_{3\longrightarrow 4}+W_{4\longrightarrow 1}  \nonumber \\
&&  \nonumber \\
&=&W_{1\longrightarrow 2}+W_{3\longrightarrow 4}=P_{1}\left(
V_{2}-V_{1}\right) +P_{4}\left( V_{4}-V_{3}\right).  \label{EqWo}
\end{eqnarray}

The upper isobar will give the net inflow of heat ($Q_{H}$) as
follows
\begin{equation}
Q_{H}=\int_{T_{1}}^{T_{2}}C_{p}\left( P_{1},T\right)
dT=\int\limits_{r_{+1}}^{r_{+2}}C_{p}\left( P_{1},T\right) \frac{\partial T}{%
\partial r}dr=Q_{H2}-Q_{H1}.  \label{P-V cycle}
\end{equation}

Taking advantage of Eqs. (\ref{temp3F(R)}) and (\ref{C2}), the
input heat flow to the cycle is
\begin{eqnarray}
Q_{H} &=&\frac{A_{1} (w-2)^{2}J^{2}(1+\gamma -\frac{z}{2})\left( 2\delta
+4-\gamma \right) (\frac{r_{e}}{r_{0}})^{2w-z+\gamma }}{3072\pi A_{2}\left(
2w-z+\gamma -2\right) r_{e}^{2}} \Bigg\vert_{r_{e1}}^{r_{e2}}  \nonumber \\
&&+\frac{\left( 1+\gamma -\frac{z}{2}\right) \left( \gamma -4-2\delta
\right)(2^{\frac{4}{3}}q^{\frac{3}{2}}+16P_{1}\pi r_{+}^{\delta +2}) (\frac{%
r_{e}}{r_{0}})^{\gamma }}{192r_{e}^{\delta }} \Bigg\vert_{r_{e1}}^{r_{e2}}.
\label{QH2}
\end{eqnarray}

Using Eq. (\ref{P-V cycle}) and (\ref{QH2}), one can obtain the engine
efficiency as
\begin{eqnarray}
\eta &=&\frac{W}{Q_{H}}=\frac{64\pi D_{1}A_{2} r_{e2}^{2}r_{e1}^{2}}{%
-(D_{2}+D_{3})\left( (6+z)^{2}-(2\delta -2\gamma)^{2}\right)},
\end{eqnarray}
in which
\begin{eqnarray}
D_{1} &= &
\frac{r_{e1}^{2+\gamma}-r_{e2}^{2+\gamma}}{r_{0}^{\gamma}} \nonumber \\
D_{2} &= &\left( 4A_{2}P_{1}\pi r_{e1}^{\delta +2}+\frac{%
J^{2}(w-2)^{2}r_{e1}^{2w-z+\gamma -1}}{32r_{0}^{2w-z}}+\frac{1}{4}A_{2}2^{-%
\frac{1}{4}}q^{\frac{3}{2}}\right) r_{e2}^{\delta }\left( \frac{r_{e1}}{r_{0}%
}\right) ^{\gamma }  \nonumber \\
D_{3} &= &\left( 4A_{2}P_{1}\pi r_{e2}^{\delta +2}+\frac{%
J^{2}(w-2)^{2}r_{e2}^{2w-z+\gamma -1}}{32r_{0}^{2w-z}}+\frac{1}{4}A_{2}2^{-%
\frac{1}{4}}q^{\frac{3}{2}}\right) r_{e1}^{\delta }\left( \frac{r_{e2}}{r_{0}%
}\right) ^{\gamma } .  \nonumber
\end{eqnarray}

Among different classical cycles, the Carnot cycle is one of the
interesting simplest cycle that can be considered. The efficiency
of this cycle is the maximum efficiency of the heat engines in
such a way that any higher efficiency would violate the second law
of thermodynamics. To calculate the Carnot efficiency, we consider
the $T_{H}$ and $T_{C}$ in our cycle to correspond to $T_{2} $ and
$T_{4}$, respectively. So, this efficiency is
\begin{eqnarray}
\eta_{c} &=&1-\frac{T_{C}}{T_{H}}=1-\frac{X_{1} (\frac{r_{e1}}{r_{0}})^{%
\frac{z }{2}}}{X_{2} (\frac{r_{e2}}{r_{0}})^{\frac{z}{2}}}  \label{Carnot}
\end{eqnarray}
where
\begin{eqnarray}
X_{1} &=&-\frac{A_{1}J^{2}(w-2)^{2}(\frac{r_{e1}}{r_{0}} )^{2w-z}}{%
32A_{2}r_{e1}^{3}}+8\pi P_{4}r_{e1}\left( \gamma +2\right) -\frac{2^{-\frac{1%
}{4}}q^{\frac{3}{2}}(\delta -\gamma )}{ r_{e1}^{\delta +1}}  \nonumber \\
X_{2} &=&-\frac{A_{1}J^{2}(w-2)^{2}(\frac{r_{e2}}{r_{0}} )^{2w-z}}{%
32A_{2}r_{e2}^{3}}+8\pi P_{4}r_{e2}\left( \gamma +2\right) -\frac{2^{-\frac{1%
}{4}}q^{\frac{3}{2}}(\delta -\gamma )}{ r_{e2}^{\delta +1}},
\nonumber
\end{eqnarray}
\begin{figure}[!htb]
\centering
\subfloat[]{
        \includegraphics[width=0.32\textwidth]{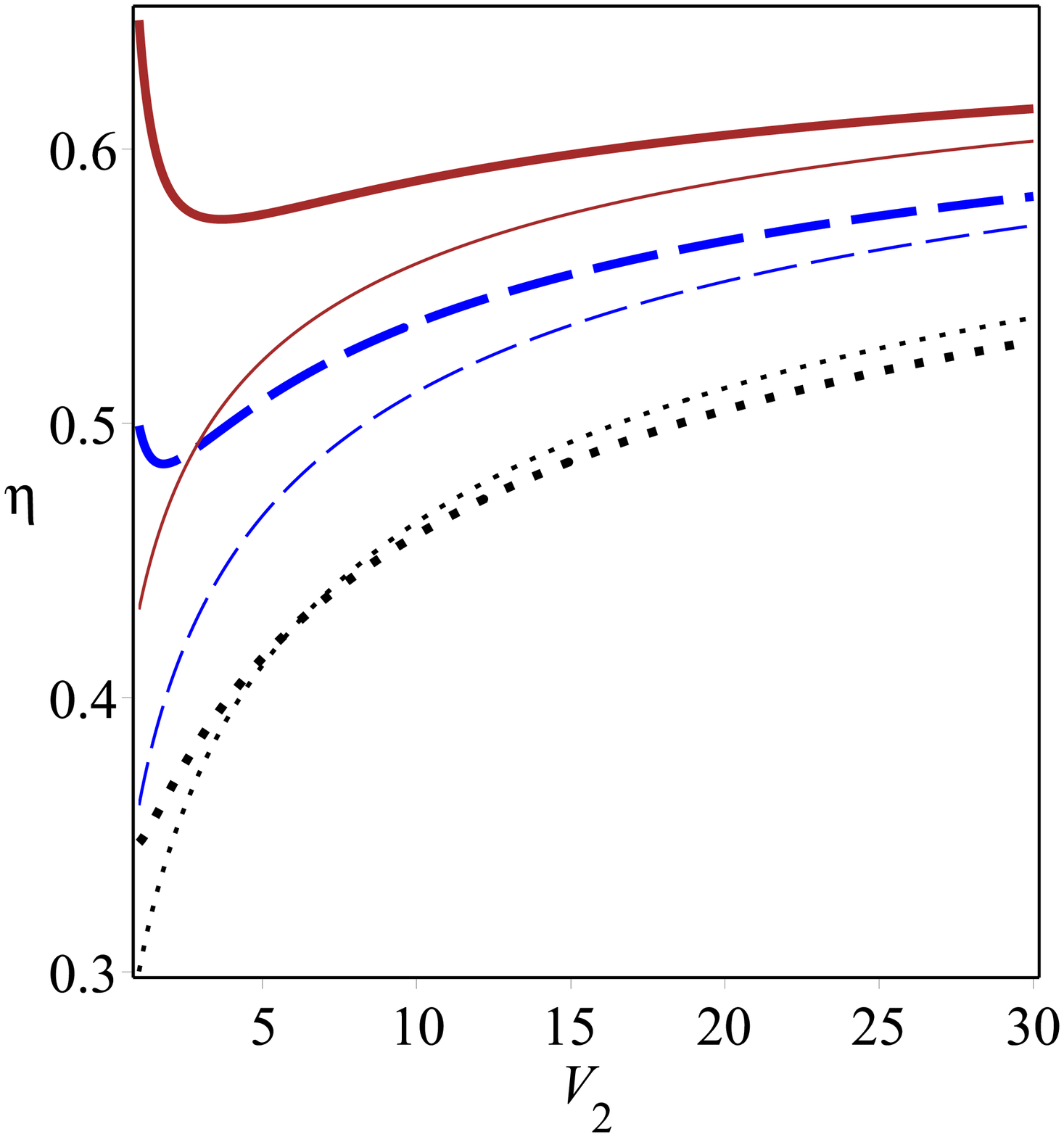}}
\subfloat[]{
        \includegraphics[width=0.335\textwidth]{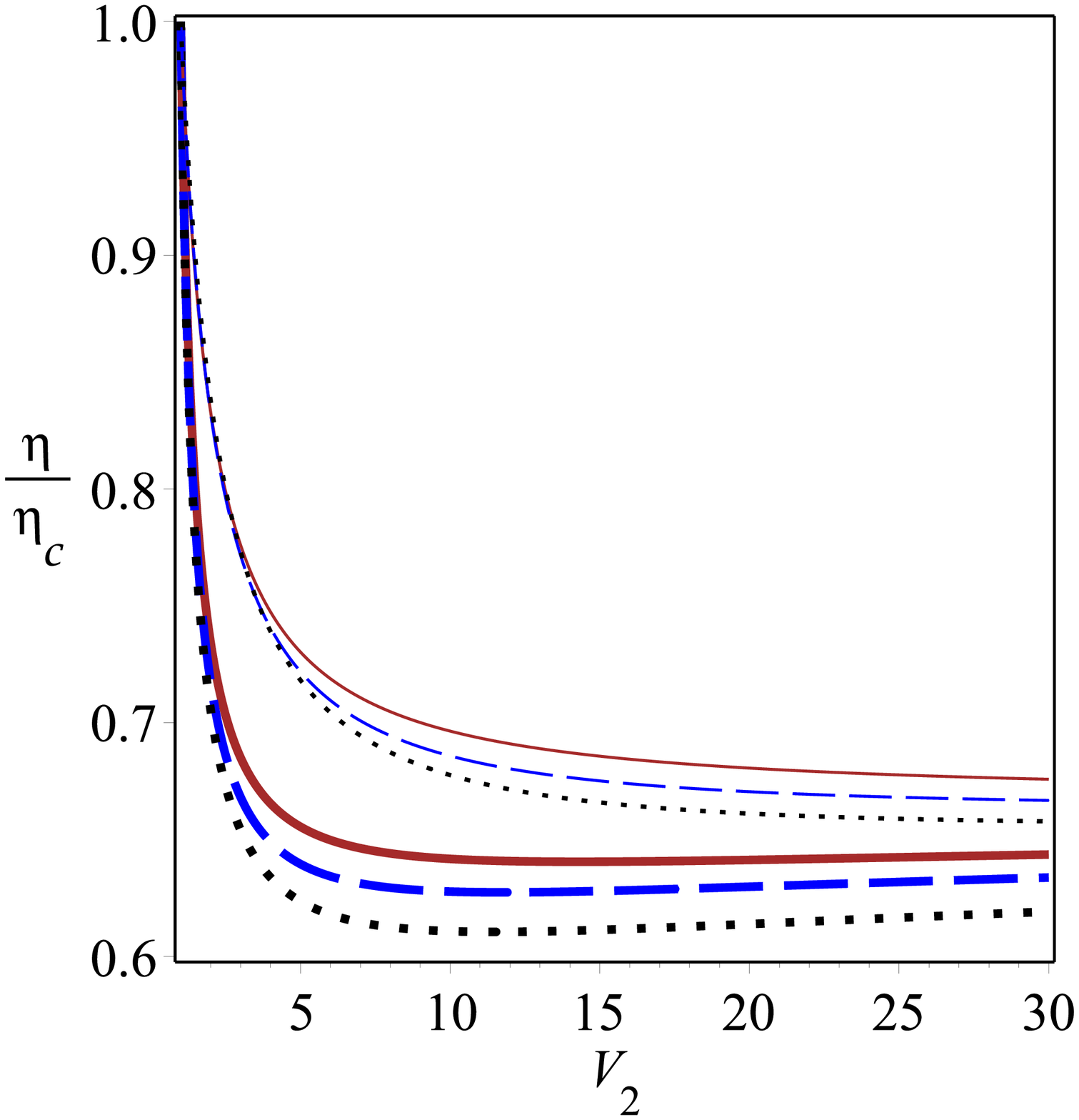}} \newline
\subfloat[]{
        \includegraphics[width=0.32\textwidth]{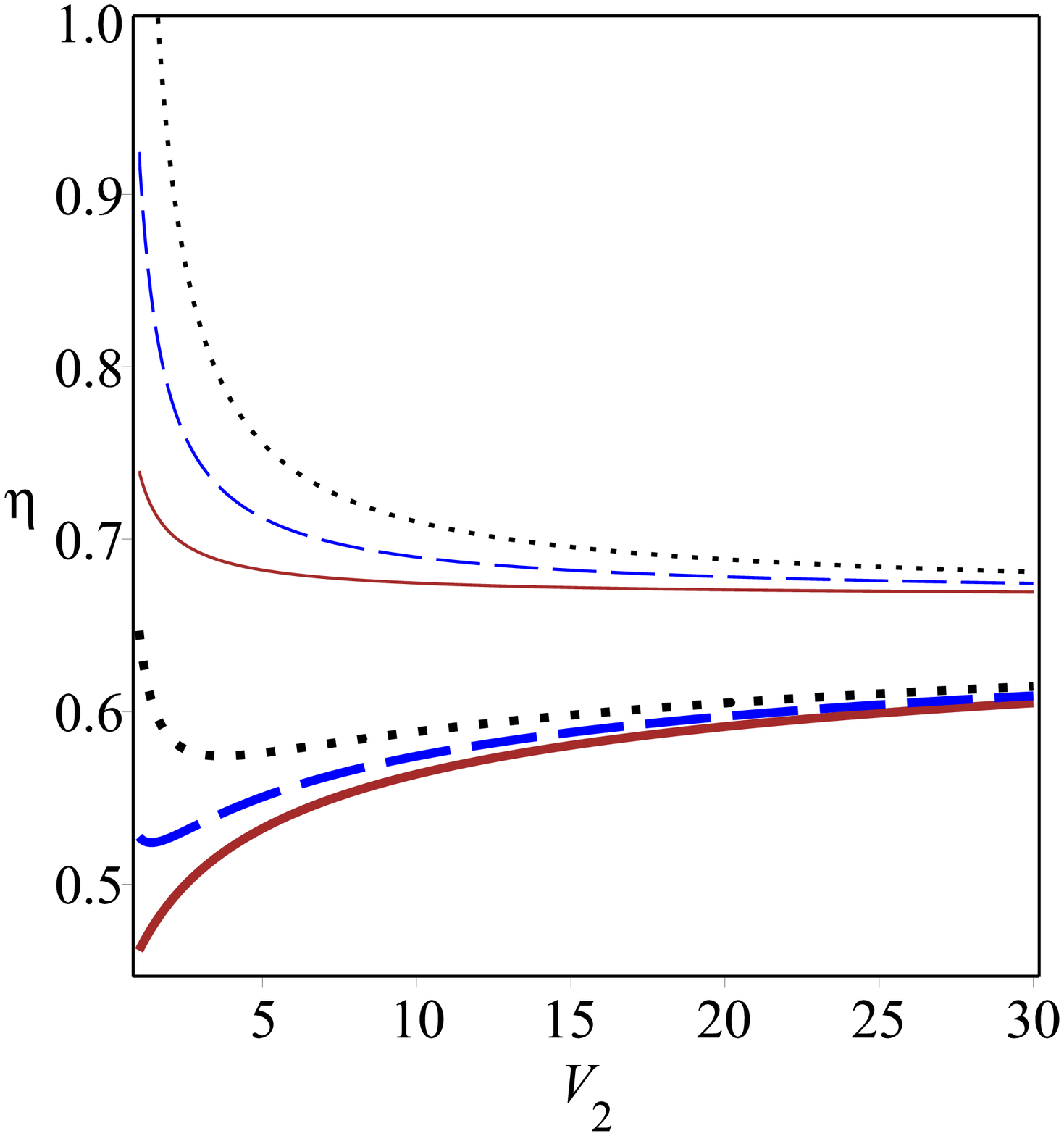}}
\subfloat[]{
        \includegraphics[width=0.34\textwidth]{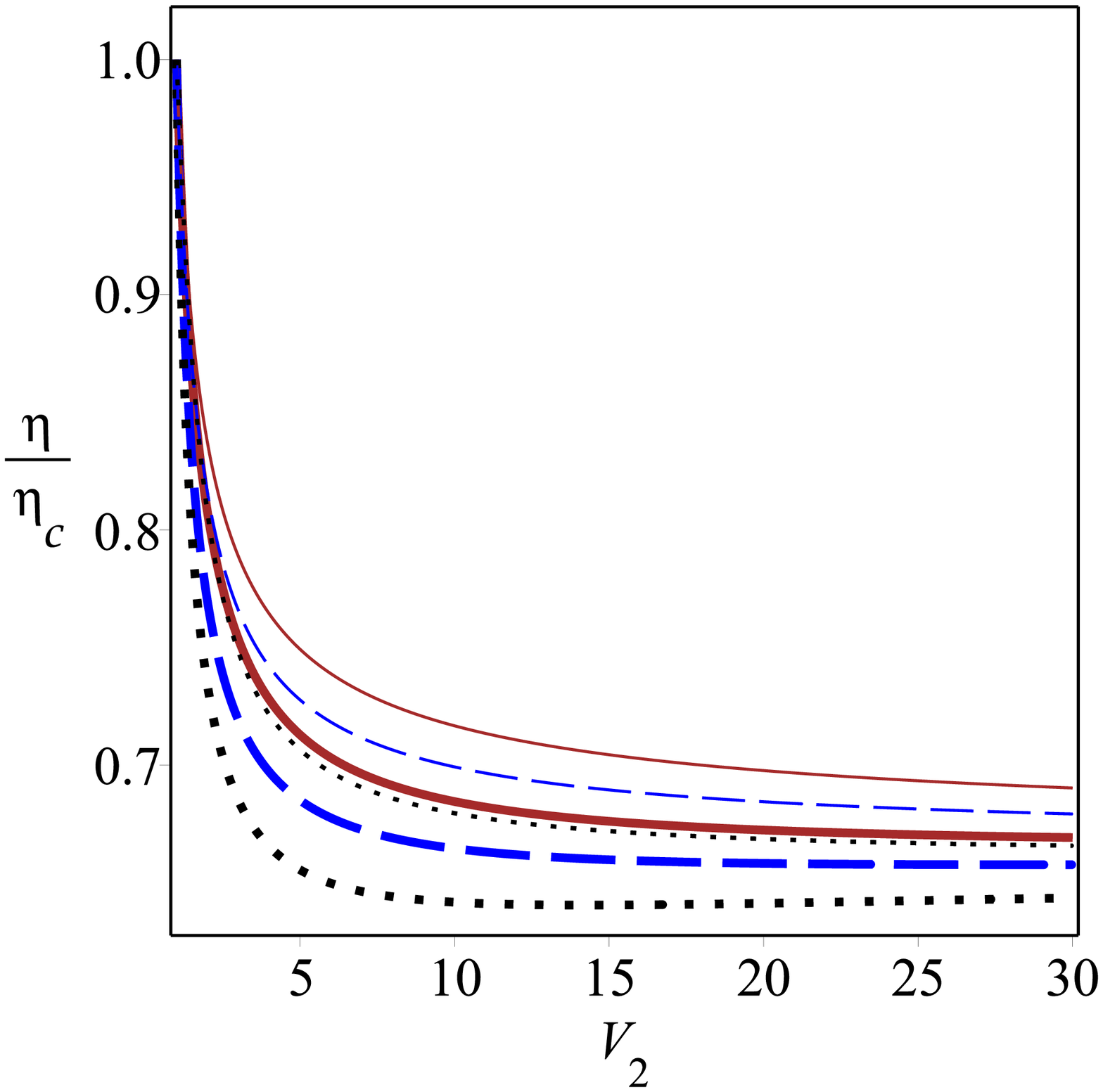}} \newline
\caption{Variation of $\protect\eta $ and $\frac{\protect\eta}{\protect\eta %
_{C}}$ versus $V_{2}$ for $z=0.5 $, $w=1.5 $, $r_{0}=0.2 $, $P_{1}=0.003$, $%
P_{4}=0.001$ and $V_{1}=1$. Up panels: bold lines for $q=0.3$ and thin lines
for $q=0$; $J=0.4$ (continues line), $J=0.5$ (dashed line) and $J=0.65$
(dotted line). Down panels: bold lines for $J=0.4$ and thin lines for $J=0$;
$q=0.1$ (continues line), $q=0.2$ (dashed line) and $q=0.3$ (dotted line).}
\label{FigEf1}
\end{figure}
The behavior of the heat engine efficiency $\eta $ and the ratio $\frac{\eta
}{\eta _{C}}$ under variation of black hole parameters is depicted in Figs. (%
\ref{FigEf1})-(\ref{FigEf3}). In Fig. \ref{FigEf1}, we examine the
influence of electric charge and angular momentum on $\eta $ and
the ratio $\frac{\eta }{\eta _{C}}$ for the fixed exponents,
parameter $r_{0}$ and pressures $P_{1}$, $P_{4}$. As one can
check, from the up panels of Fig. \ref{FigEf1}, both $\eta$ and
the ratio $\frac{\eta }{\eta _{C}}$ are decreasing functions of
the angular momentum. For large values of $J$, the efficiency
monotonically
increases as the volume $V_{2}$ grows (see the bold dotted line of Fig. \ref%
{FigEf1}(a)). This means that for rapidly charged rotating black
holes, the increase of volume difference between the small black
hole ($V_{1}$) and larger black hole ($V_{2}$) will make the heat
engine more efficient. For small angular momentum, the efficiency
curve has a local minimum value, indicating that there exists a
specific value of the volume $V_{2}$ at which the black hole heat
engine works at the lowest efficiency (see the bold continuous
line of Fig. \ref{FigEf1}(a)). In the absence of the electric
charge, for all values of the angular momentum the heat engine
efficiency monotonously increases with the growth of $V_{2}$ and
then tends to a constant value (see thin lines of Fig.
\ref{FigEf1}(a)). Taking a close look at Fig. \ref{FigEf1}(a), one
can find that charged rotating black holes have a bigger
efficiency than their uncharged counterparts (compare bold and
thin lines). Just for large volume difference $\Delta
V=V_{2}-V_{1}$, their efficiency becomes smaller compared to that
of a rapidly rotating black hole (compare bold-dotted and
thin-dotted lines in Fig. \ref{FigEf1}(a)).
\begin{figure}[!htb]
\centering
\subfloat[$ w=1.5 $]{
        \includegraphics[width=0.32\textwidth]{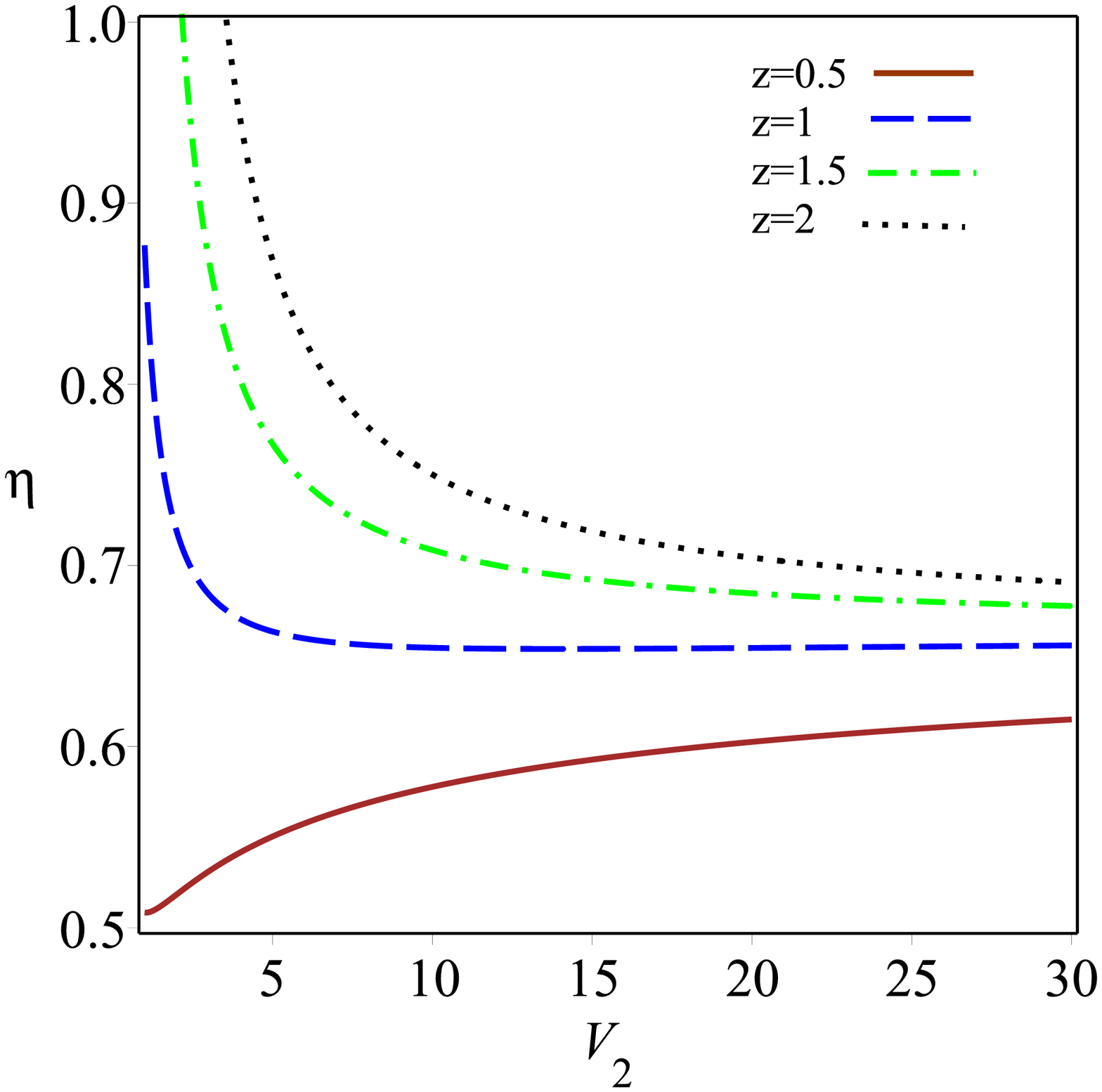}}
\subfloat[$ w=1.5 $]{
        \includegraphics[width=0.305\textwidth]{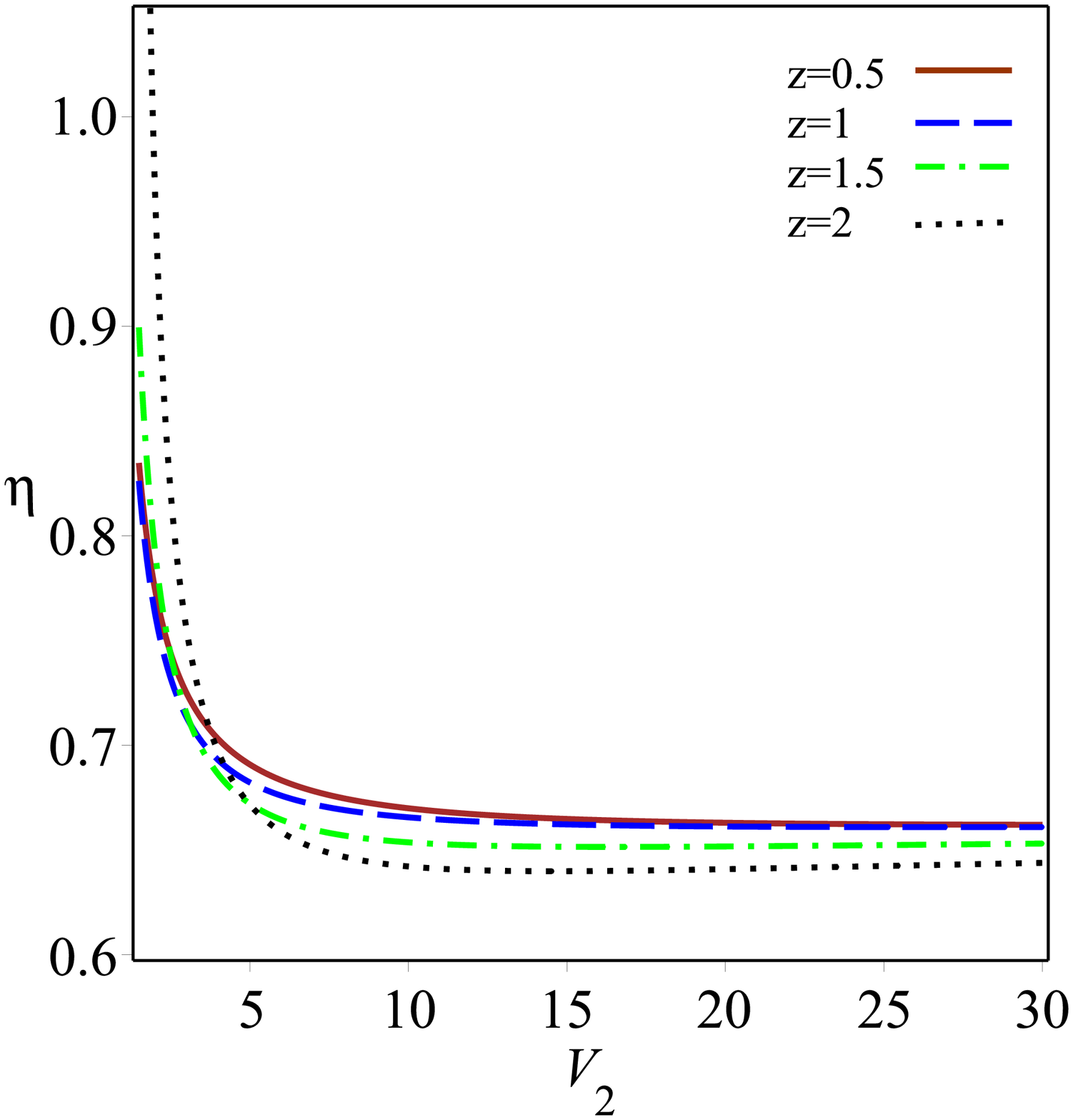}} \newline
\subfloat[$ z=0.5 $]{
        \includegraphics[width=0.32\textwidth]{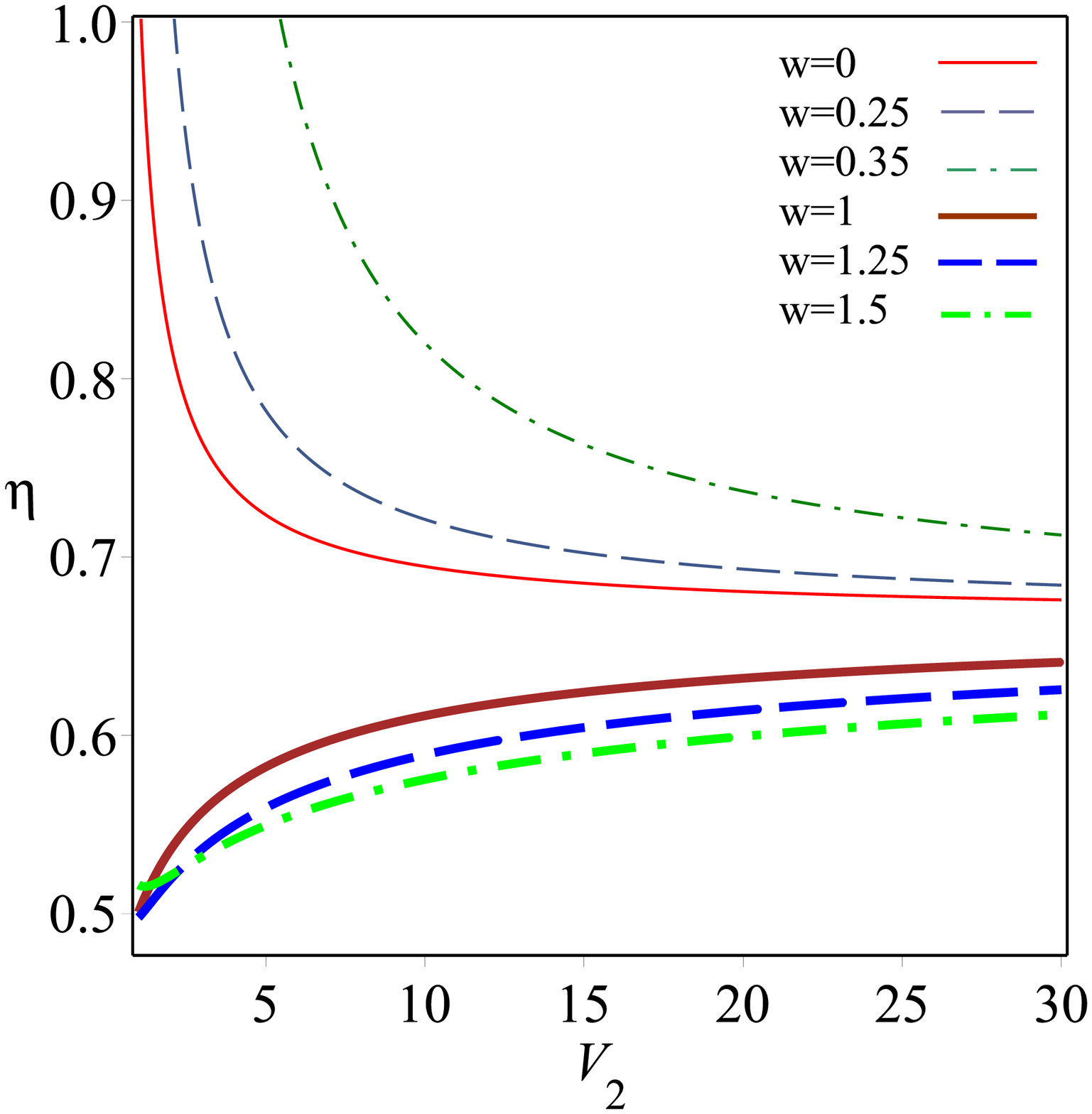}}
\subfloat[$ z=0.5 $]{
        \includegraphics[width=0.325\textwidth]{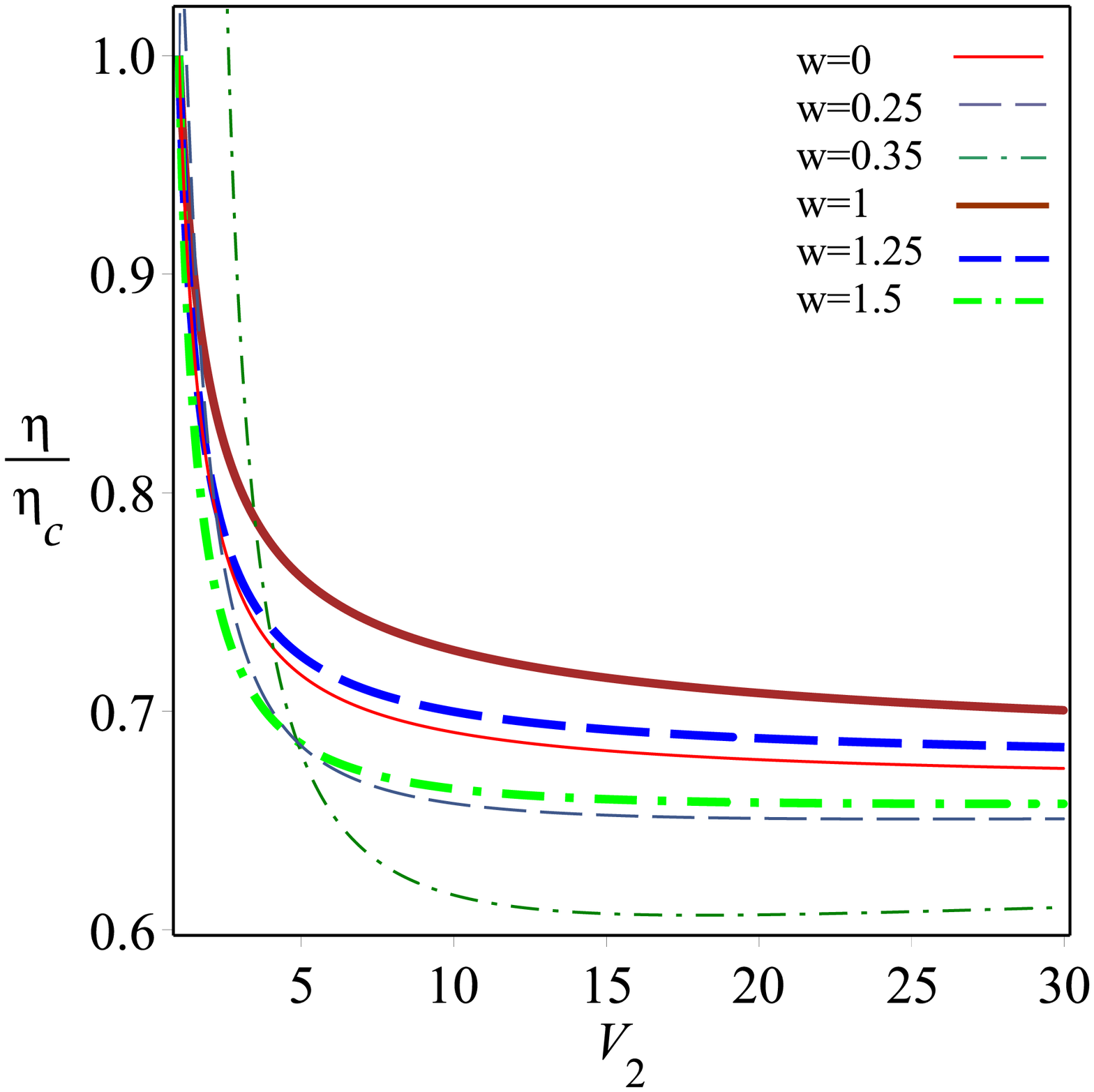}} \newline
\caption{Variation of $\protect\eta $ and $\frac{\protect\eta}{\protect\eta %
_{C}}$ versus $V_{2}$ for $J=0.4 $, $q=0.2 $, $r_{0}=0.2 $, $P_{1}=0.003$, $%
P_{4}=0.001$ and $V_{1}=1$. Up panels: for different values of $z $
parameter. Down panels: for different values of $w $ parameter.}
\label{FigEf2}
\end{figure}

Down panels of Fig. \ref{FigEf1} displays the effect of the electric charge
on $\eta $ and the ratio $\frac{\eta }{\eta _{C}}$. As we see, although this
parameter has an increasing contribution to the efficiency, its effect is to
decrease the ratio $\frac{\eta }{\eta _{C}}$. For non-rotating black holes,
all curves monotonic reduce rapidly firstly, then the efficiency reaches a
constant value after the certain value of volume $V_{2}$ (see thin lines of
Fig. \ref{FigEf1}(c)). For slowly charged rotating black holes, the
efficiency gradually increases as the volume $V_{2}$ increases and then
tends to a constant value (see the bold continuous line of Fig. \ref{FigEf1}%
(c)). While for rapidly charged rotating black holes, the efficiency
decreases to a minimum value with an increase of $V_{2}$ and then gradually
grows as the volume increases more, and finally reaches a constant value in
the limit of that $V_{2}$ goes to the infinity (see the bold dotted line of
Fig. \ref{FigEf1}(c)). Fig. \ref{FigEf1}(c) also shows that non-rotating
black holes have a bigger efficiency compared to the charged rotating black
holes (compare bold lines to thin lines). Comparing Fig. \ref{FigEf1}(a) to
Fig. \ref{FigEf1}(c), one can find that variation of $J$ has a stronger
effect on the efficiency than the electric charge.

Figure \ref{FigEf2} shows how $\eta $ and the ratio $\frac{\eta
}{\eta _{C}}$ are affected by the exponents. According to the up
panels of this figure, the exponent $z$ has an increasing (a
decreasing) effect on the efficiency (the ratio $\frac{\eta }{\eta
_{C}}$). For small values of $z$, the efficiency monotonically
increases as the volume $V_{2}$ grows and
then tends to the saturation value (see the continuous line of Fig. \ref%
{FigEf2}(a)). While for large values, the opposite behavior will
be observed. Figure \ref{FigEf2}(c) displays the influence of the
exponent $w$ on the efficiency. Our findings indicate that for
$w<0.5$, the efficiency increases with the increase of $w$. While
for $w>0.5$, increasing $w$ leads to the decreasing of the heat
engine efficiency. Regarding the effect of this parameter on the
ratio $\frac{\eta }{\eta _{C}}$, in both regions ($w<0.5$ or
$w>0.5$), one finds that increasing the this parameter makes the
increasing of the ratio $\frac{\eta }{\eta _{C}}$ (see Fig. \ref{FigEf2}%
(d)). Taking a close look at the right panels of Fig. \ref{FigEf2}, one can
notice that in the region of volume $V_{2}$ near $V_{1}$ and for small
(large) values of $w$ ($z$), the efficiency becomes larger than Carnot
efficiency which violates the second law of thermodynamics. This shows that
small (large) values of $z$ ($w$) should be considered to observe an
acceptable efficiency of the system.
\begin{figure}[!htb]
\centering
\subfloat[$ \Delta P=0.002 $]{
        \includegraphics[width=0.32\textwidth]{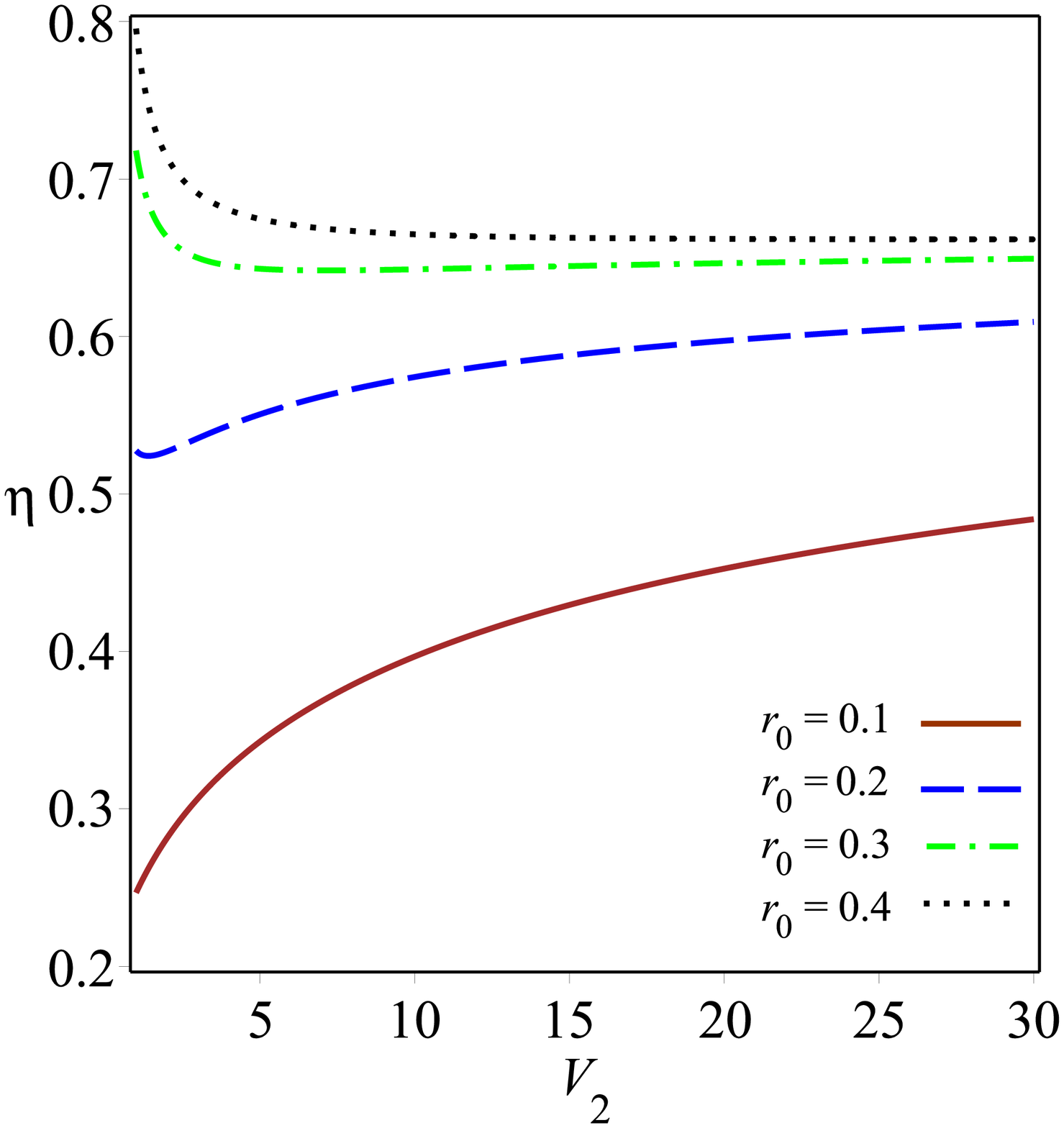}}
\subfloat[$ \Delta P=0.002 $]{
        \includegraphics[width=0.32\textwidth]{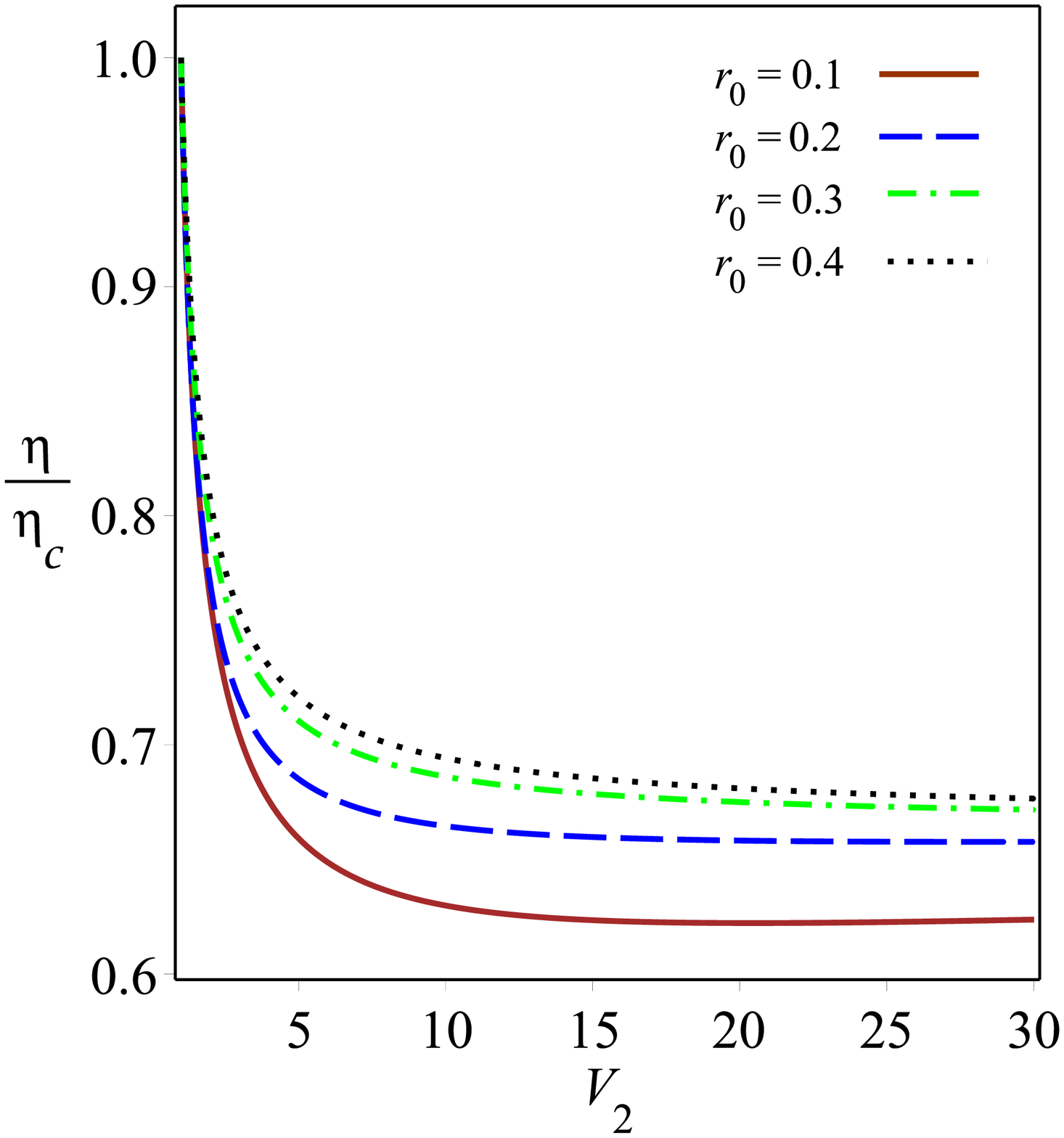}} \newline
\subfloat[$ r_{0}=0.2 $]{
        \includegraphics[width=0.32\textwidth]{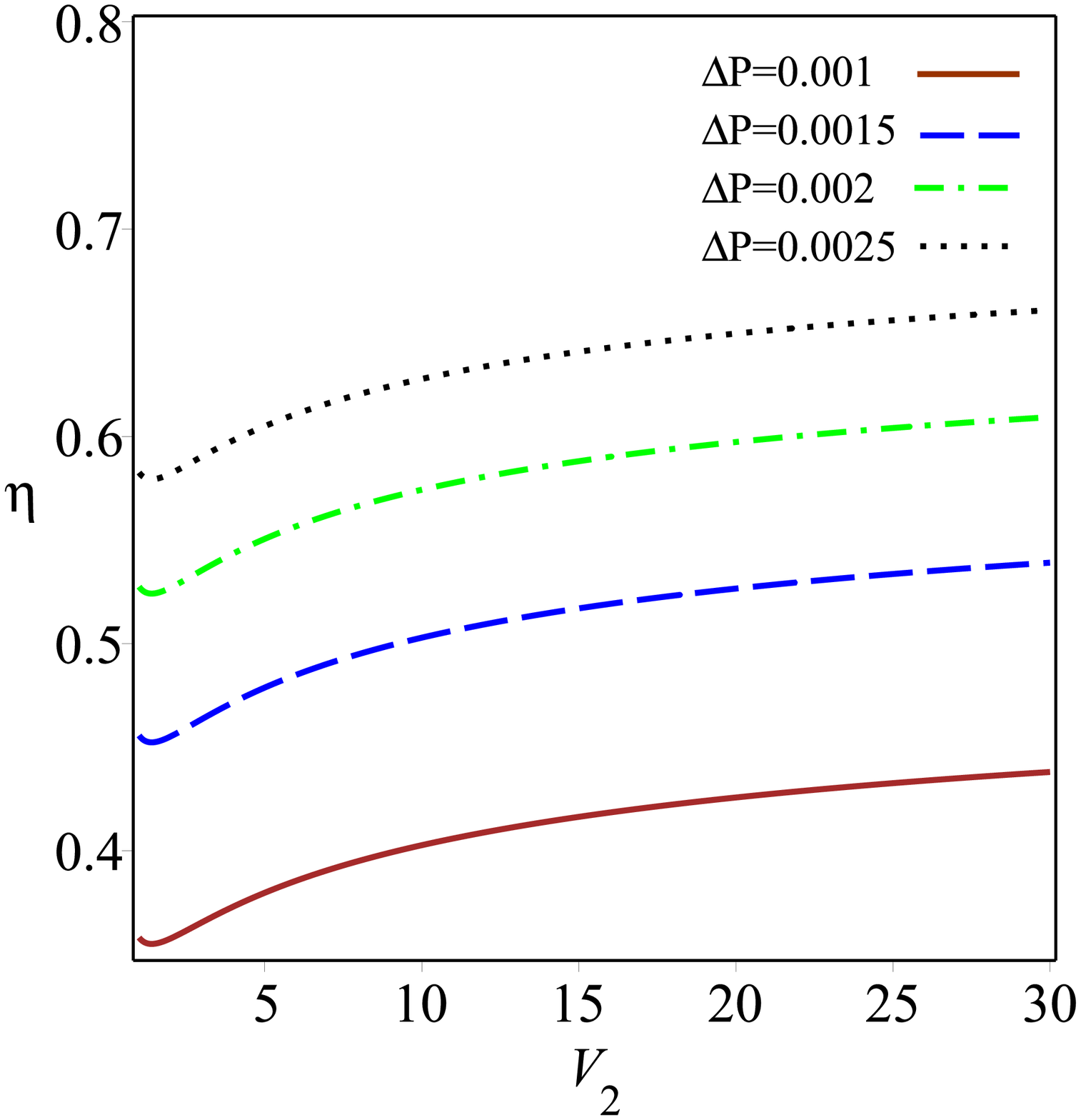}}
\subfloat[$ r_{0}=0.2 $]{
        \includegraphics[width=0.32\textwidth]{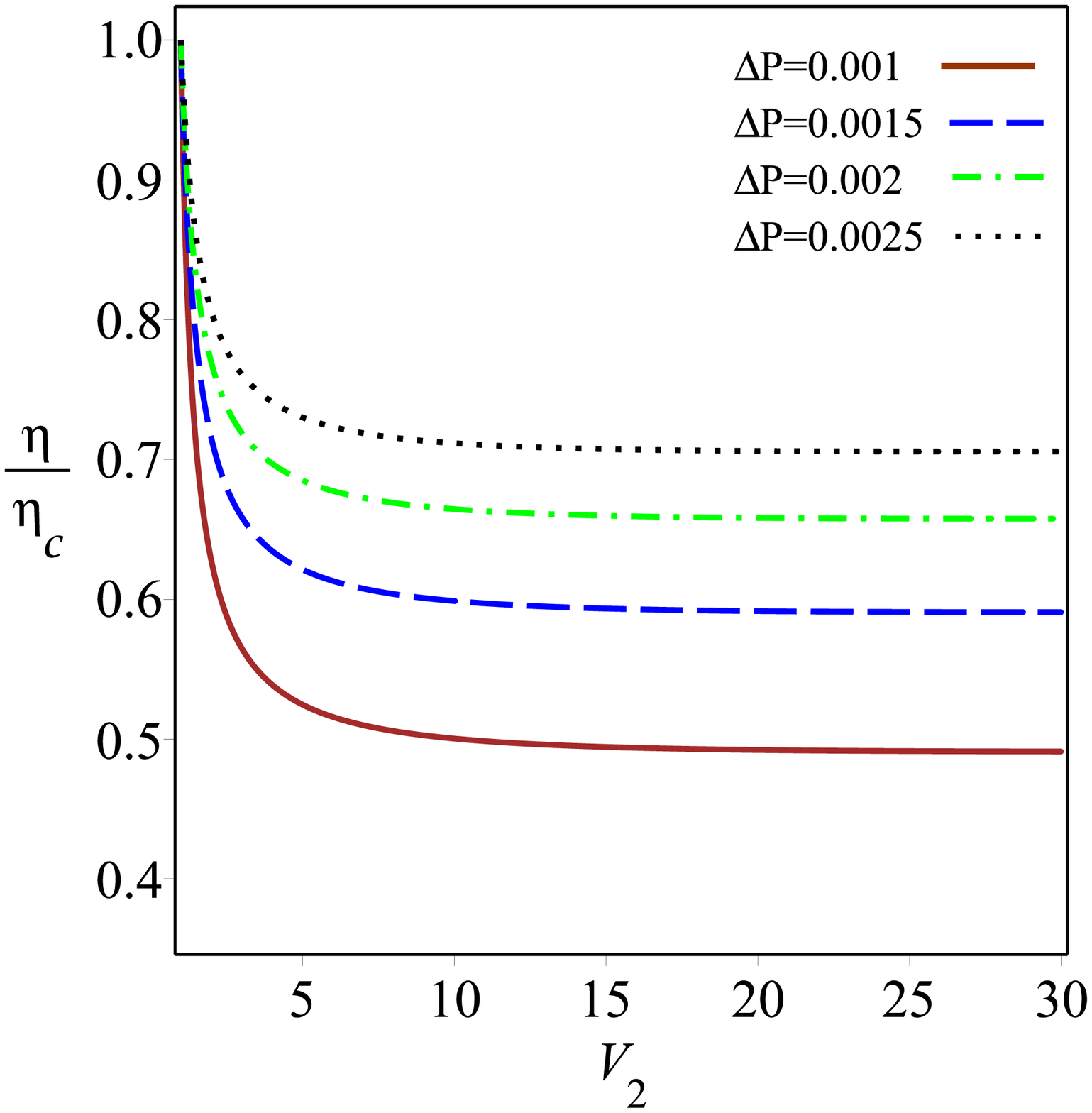}} \newline
\caption{Variation of $\protect\eta $ and $\frac{\protect\eta}{\protect\eta %
_{C}}$ versus $V_{2}$ for $J=0.4 $, $q=0.2 $, $z=0.5 $, $w=1.5$ and $V_{1}=1$%
. Up panels: for different values of $r_{0} $ parameter. Down panels: for
different values of the pressure difference. }
\label{FigEf3}
\end{figure}

The effects of the parameter $r_{0}$ and pressure on $\eta $ and
the ratio $\frac{\eta }{\eta _{C}}$ are reflected in Fig.
\ref{FigEf3}. We find that the parameter $r_{0}$ has an increasing
effect on both $\eta $ and $\frac{\eta }{\eta _{C}}$. For small
(large) values of $r_{0}$, the efficiency gradually grows
(reduces) with the increase of $V_{2}$ (see Fig. \ref{FigEf3}(a)).
Regarding the pressure, down panels of Fig. \ref{FigEf3} show that
increasing the difference of pressure $\Delta P$ makes the
increasing of both $\eta $ and $\frac{\eta }{\eta _{C}}$. Taking a
look at right panels of Fig. \ref{FigEf3}, one can find that the
second law of thermodynamics is satisfied for all values of these
two parameters.

\section{conclusion}

In this paper, we have obtained a new Lifshitz-like rotating black
hole solution in three dimensional $F(R)$ gravity. Investigating
geometrical properties of the solution, we have found that this
solution reduces to the charged rotating BTZ-like black hole in
special limits. We also studied the optical features of the black
hole and noticed that some constraints should be imposed on the
exponents to have an acceptable optical behavior. Studying the
impact of parameters of the model on the photon
orbit radius illustrated that as the electric charge, angular
momentum and absolute
value of the cosmological constant increase, both event horizon and photon orbit radii decrease. Regarding the effect of
exponent $z$, our analysis showed that this parameter has an
increasing contribution to the horizon radii and photon orbit.

After that, we continued our analysis by investigating the energy
emission rate and examining the influence of parameters on the
radiation process. The results indicated that the angular momentum
and exponent $z$ have an increasing contribution to the emission
rate, namely, the emission of particles around the BH increases by
increasing these two parameters. Regarding the role of the
electric charge, cosmological constant and exponent $w$, we have
found that as the effects of these parameters get stronger, the
evaporation process gets slower. In other words, the lifetime of a
black hole would be longer under such conditions.

As the next step, we have studied the thermodynamic properties of
the system in the extended phase space thermodynamics. We
calculated thermodynamic quantities of black holes and showed that
these quantities satisfy the first law of thermodynamics. We also
obtained the modified Smarr relation and found that regardless of
the cosmological constant term, scaling other thermodynamic
quantities is modified. Using heat capacity, we investigated the
thermal stability of the system and showed how the parameters of
the model affect the region of the stability. Moreover, we look
for possible phase transitions and found that three-dimensional
Lifshitz-like rotating black hole experiences the
first-order/second-order phase transitions with a suitable choice
of parameters.

Finally, we have considered this kind of black hole as a working
substance and studied the holographic heat engine by taking a
rectangle heat cycle in the $P-V$ plot. Investigating the black
hole heat engine efficiency and comparing obtained results with
the Carnot efficiency led to the following interesting results:

I) The angular momentum (electric charge) has a decreasing (an
increasing) contribution to the efficiency of the system. For all
values of these two parameters, the efficiency is always smaller
than the Carnot efficiency which is consistent with the second law
of thermodynamics.

II) The charged rotating black hole has a bigger (smaller)
efficiency than its uncharged (non-rotating) counterpart. It is
worth pointing out that for very large volume difference, the
efficiency of rapidly rotating black hole becomes bigger than
them.

III) The heat engine efficiency is an increasing function of the
exponent $z$. For small values of this parameter, the condition
$\frac{\eta }{\eta _{C}}<1$ is satisfied all the time. But for
large values of $z$, this condition is violated in the region of
volume $V_{2}$ near $V_{1}$. The contribution of the exponent $w$
on the efficiency is a little different. For $w < 0.5$, the
efficiency increases with the increase of $w$, whereas for $w >
0.5$, increasing this parameter leads to the decrease of the heat
engine efficiency. For $w < 0.5$, the second law of thermodynamics
violates for a very small volume difference, whereas for $w > 0.5$
it is always preserved.

IV) Increasing the pressure difference makes the increasing the
efficiency of the system. For all values of pressure, the
efficiency is always smaller than the Carnot efficiency which is
consistent with the second law of thermodynamics.

\begin{acknowledgements}

The authors thank Shiraz University Research Council. KhJ is
grateful to the Iran Science Elites Federation for the financial
support.

\end{acknowledgements}

\end{document}